\documentclass[aps,twocolumn,showpacs,preprintnumbers,nofootinbib,prd,10pt,superscriptaddress]{revtex4-1}

\makeatletter
\def\l@subsubsection#1#2{}
\def\l@subsubsubsection#1#2{}
\makeatother

\usepackage{graphicx,amssymb,amsmath,amsthm,amsfonts,epsfig,epsf}
\usepackage{mathrsfs}
\usepackage[usenames]{color}
\usepackage{epstopdf}
\usepackage{epstopdf}
\usepackage{aas_macros}
\usepackage{mathtools}
\usepackage[thinc]{esdiff}
\usepackage[italicdiff]{physics}
\usepackage{bm}
\usepackage{dcolumn}
\usepackage{latexsym}
\usepackage{rotating}

\usepackage{enumerate}
\usepackage{tensor}
\usepackage{multirow}
\usepackage{url}
\usepackage{subcaption}
\usepackage{fontawesome5}

\makeatletter
\long\def\@makecaption#1#2{%
  \par
  \vskip\abovecaptionskip
  \setbox\@tempboxa\hbox{%
    \small\bfseries #1.\ \normalfont #2}%
  \ifdim \wd\@tempboxa >\hsize
    {\small\bfseries #1.\ \normalfont
    \justifying #2\par}%
  \else
    \hbox to\hsize{\hfil\box\@tempboxa\hfil}%
  \fi
  \vskip\belowcaptionskip}
\makeatother
\usepackage{ragged2e} 

\usepackage[dvipsnames]{xcolor}
\usepackage[unicode]{hyperref}
\usepackage{orcidlink}
\hypersetup{colorlinks=true, citecolor=MidnightBlue,
            linkcolor=MidnightBlue, urlcolor=MidnightBlue, linktocpage=true}

\newcommand{\tn}{\textnormal}

\renewcommand{\dd}{\mathrm{d}}

\newcommand{\eeq}{\end{equation}}
\newcommand{\ba}{\begin{align}}
\newcommand{\ea}{\end{align}}

\newcommand{\EN}{\mathcal{E}}
\newcommand{\ANG}{\mathcal{L}}
\newcommand{\CAR}{\mathcal{Q}}

\newcommand{\GSSI}{Gran Sasso Science Institute (GSSI), I-67100 L’Aquila, Italy}
\newcommand{\GranSasso}{INFN, Laboratori Nazionali del Gran Sasso, I-67100 Assergi, Italy}

\begin{document}
\title{Adiabatic evolution of asymmetric binaries on generic orbits with new fundamental fields I: characterization of gravitational wave fluxes}

\author{Sara Gliorio \orcidlink{0009-0007-7207-784X}}
 \email{sara.gliorio@gssi.it}
\affiliation{\GSSI}
\affiliation{\GranSasso}
\author{Matteo Della Rocca \orcidlink{0009-0001-4470-3694}}
\email{matteo.dellarocca@uniroma1.it}
\affiliation{Dipartimento di Fisica, Sapienza Universit\`a di Roma, Piazzale Aldo Moro 5, 00185, Roma, Italy}
\affiliation{Dipartimento di Fisica, Universit\`a di Pisa, Largo B. Pontecorvo 3, 56127 Pisa, Italy}
\affiliation{INFN, Sezione di Roma, Piazzale Aldo Moro 2, 00185, Roma, Italy}
\author{Susanna Barsanti\,\orcidlink{0000-0001-7321-2512}}
\email{susanna.barsanti@ucd.ie}
\affiliation{School of Mathematics and Statistics, University College Dublin, Belfield, Dublin 4, Ireland}
\author{Leonardo Gualtieri \orcidlink{0000-0002-1097-3266}}
 \email{leonardo.gualtieri@unipi.it}
\affiliation{Dipartimento di Fisica, Universit\`a di Pisa, Largo B. Pontecorvo 3, 56127 Pisa, Italy}
\affiliation{INFN, Sezione di Pisa, Largo B. Pontecorvo 3, 56127 Pisa, Italy}
\author{Andrea Maselli \orcidlink{0000-0001-8515-8525}}
 \email{andrea.maselli@gssi.it}
\affiliation{\GSSI}
\affiliation{\GranSasso}
\author{Thomas P. Sotiriou \orcidlink{0000-0002-9089-4866}}
\email{Thomas.Sotiriou@nottingham.ac.uk}
\affiliation{School of Mathematical Sciences \& School of Physics and Astronomy, University of Nottingham, University Park, Nottingham, NG7 2RD, UK}
\affiliation{Nottingham Centre of Gravity, University of Nottingham, University Park, Nottingham, NG7 2RD, UK}

\begin{abstract} 
We investigate the dynamics of asymmetric binaries in 
extensions of General Relativity featuring a massless scalar field non-minimally coupled to gravity, focusing on the interplay between eccentricity 
and inclination in fully generic bound orbits. Building on 
an effective field theory framework tailored to extreme- 
and intermediate-mass-ratio inspirals, we compute scalar-field 
perturbations using a new arbitrary-precision \texttt{C++} code capable of evolving 
perturbations along generic Kerr geodesics, \href{https://github.com/saragliorio/STORM}{\texttt{STORM }\faGithub}. We investigate 
the complete set of scalar fluxes at infinity and through 
the horizon across the relevant parameter space and analyze 
their harmonic structure as a function of orbital geometry 
and black-hole spin. Our results advance ongoing efforts to 
construct accurate waveform models for asymmetric binaries 
beyond GR and lay the groundwork for precision tests of 
fundamental physics with next-generation 
gravitational-wave detectors.
\end{abstract}

\maketitle

%
\section{Introduction}
%
We are now more than a decade into the gravitational-wave (GW) era. 
LIGO, Virgo, and KAGRA (LVK) have detected over two hundred events, reshaping our view of the Universe and revealing 
the dynamics of black holes (BHs) in the highly dynamical, 
strong-field regime of compact binary coalescences \cite{LIGOScientific:2025slb}. These observations have transformed GW sources into unparalleled laboratories 
for fundamental physics \cite{LIGOScientific:2021sio}. 
Ground-based interferometers have also enabled searches 
for new physics beyond General Relativity (GR) or the 
Standard Model using comparable-mass binaries with mass 
ratios up to $\epsilon = m_2/m_1 \sim 1/10$ \cite{LIGOScientific:2020stg,LIGOScientific:2020zkf}, 
and have already placed stringent constraints on deviations 
from GR \cite{Yunes:2013dva,Sanger:2024axs,Gao:2024rel}.

The coming decade — driven by the advent of next-generation 
detectors on the ground \cite{Punturo:2010zz,2015PhRvD..91h2001D} 
and in space \cite{LISA:2017pwj,TianQin:2020hid,Ajith:2024mie}, 
together with the ambitious science goals of facilities 
such as Einstein Telescope and LISA \cite{ET:2025xjr,LISA:2024hlh,Barack:2018yly,LISA:2022kgy,Sathyaprakash:2019yqt} — promises a transformative leap in GW 
astronomy. These instruments will detect sources across 
a far broader mass spectrum, including populations that 
remain dim to current interferometers.

Among the most compelling targets are asymmetric binaries 
with mass ratios $\epsilon \sim 10^{-7}$–$10^{-2}$, such 
as extreme- and intermediate-mass-ratio inspirals 
(EMRIs/IMRIs). Composed of stellar-mass compact objects 
inspiraling into massive BHs, these systems can be tracked 
for hundreds of thousands of orbital cycles — often on highly 
relativistic, inclined, and eccentric trajectories — within 
the LISA band. Their rich harmonic structure and extended 
signal duration enable sub-percent precision in parameter 
estimation \cite{Babak:2017tow,Chapman-Bird:2025xtd}.

While asymmetric binaries are exceptionally promising for 
fundamental-physics tests 
\cite{Cardenas-Avendano:2024mqp,Barack:2018yly,Barausse:2020rsu,Barausse:2016eii,Blazquez-Salcedo:2016enn,Glampedakis:2005cf,Barack:2006pq,Cardoso:2018zhm,Datta:2019epe,Pani:2019cyc,Maggio:2021uge,Destounis:2020kss,Piovano:2020ooe,Sago:2021iku,Piovano:2022ojl,Collodel:2021jwi}, 
modeling them 
remains a substantial challenge. 

Robust tests of gravity 
require accurate theoretical predictions of BH dynamics 
and their GW emission both within GR and in extensions 
beyond GR. The self-force (SF) framework provides the leading 
methodology for modeling asymmetric binaries in GR, leveraging 
the small mass ratio as an expansion parameter within 
relativistic perturbation theory \cite{Barack:2018yvs,Pound:2019lzj,Warburton:2021kwk,Wardell:2021fyy}. Extending these models beyond GR is considerably 
more demanding, owing to additional fields, new couplings, 
and the absence of a Kerr-like solution that could serve 
as the background spacetime for perturbative 
analyses. Progress in this direction has been made recently, with Refs.~\cite{Wagle:2023fwl,Li:2022pcy} 
developing a formalism that generalizes Teukolsky’s 
equation to modified theories of gravity.

Effective field theory (EFT) methods have emerged as a 
powerful toolkit for studying BH dynamics in extensions of GR. 
Recent work \cite{Maselli:2020zgv} has applied this approach
 to asymmetric binaries, exploiting their intrinsic 
mass hierarchy to model their evolution in a broad class of 
theories featuring massless scalar fields non-minimally 
coupled to gravity, and showing that deviations from 
the Kerr geometry of the primary are negligible at leading 
order in the mass ratio, while the scalar charge of the 
secondary fully determines the departure from GR. This 
enables the construction of adiabatic-order waveforms 
that remain modularly adaptable to existing GR templates 
and introduce only one new parameter, the scalar charge $d$ of the secondary BH. The same 
framework has been embedded within a genuine SF 
formalism to compute post-adiabatic corrections to the 
binary dynamics \cite{Spiers:2023cva}.
This analysis --~which for brevity we shall call EFT approach~--
has been extended to a variety of 
orbital configurations, including equatorial eccentric~\cite{Barsanti:2022ana,Zhang:2022rfr} and spherical inclined~\cite{DellaRocca:2024pnm} trajectories, as well 
as to fully generic bound inspirals in recent analyses 
focused on the evolution of EMRIs on arbitrary orbits~\cite{Zi:2025lio}. 
In parallel, a complementary line of work has applied 
SF methods to model not only the inspiral but also the 
merger and ringdown in theories beyond GR, under the 
small-mass-ratio approximation \cite{Roy:2025kra}. 
Additional theoretical progress 
has explored the impact of massive scalar fields using 
the EFT approach in the context of circular orbits \cite{Barsanti:2022vvl}, time-dependent scalar charges 
\cite{DellaRocca:2024sda}, and has further extended the 
framework to vector fields, for circular and eccentric \cite{Zhang:2023vok,Zi:2025jxy,Zi:2025qos} trajectories. 
The EFT approach has also been exploited to show that the 
scalar charge of the secondary can imprint strong, detectable 
signatures on EMRI waveforms \cite{Maselli:2021men,Speri:2024qak}.
Finally, scalar self-force studies were also carried 
out for equatorial circular~\cite{Warburton:2010eq}, 
equatorial eccentric~\cite{Warburton:2011hp},  spherical inclined~\cite{Warburton:2014bya} and generic 
orbits~\cite{Nasipak:2019hxh}.

In this paper, we take a step further by generalizing the 
framework introduced in Ref.~\cite{Maselli:2020zgv} to describe the 
dynamics of asymmetric binaries on fully generic orbits, 
simultaneously incorporating eccentricity and inclination 
in the presence of massless scalar fields. To achieve this, 
we integrate the scalar-field perturbation equations using \texttt{STORM} \cite{STORM}, a 
new \texttt{C++} code which computes perturbations along generic bound geodesics with arbitrary-precision numerics. We characterize, for the first time 
within this framework, the full spectrum of scalar fluxes 
emitted at the horizon and at infinity as a function of the 
orbital parameters and the BH spin. We examine in 
detail the structure and properties of their harmonic 
content, analyzing the relative weight of each mode in the 
overall flux emitted by the binary.

These developments complement ongoing efforts to model 
compact-binary dynamics and waveforms beyond GR and 
contribute to a broader program aimed at enabling precision 
tests of fundamental physics with asymmetric binaries. The 
present work is intended as a foundational element of this 
program, providing both the theoretical framework and numerical infrastructure upon which more complete beyond-GR waveform 
models can be built.

In Section\,\ref{sec:theoreticalsetup} we review the theoretical framework. In Sec.\,\ref{sec:resultsscalarspectra} we discuss our results. Finally, in Sec.\,\ref{sec:concl} we draw our conclusions. Throughout the paper we use geometric $G=c=1$ units.
%
\section{Theoretical setup}\label{sec:theoreticalsetup}
%

In this section we detail the theoretical formalism used 
to compute the scalar-led and gravitational-led fluxes along 
generic orbits. We begin by describing the classes of beyond-GR models we consider, discussing then how to determine the 
equations of perturbations at the adiabatic order. 
%

%
\subsection{Scalar fields and EMRIs}\label{subsec:scalarchargesEMRIs}
%

We consider a general action describing a non-minimally 
coupled, real, and massless scalar field $\psi$ \cite{Spiers:2023cva},
\begin{equation}
    S\left[\textbf{g}, \psi, \Psi \right] = S_0\left[\textbf{g}, \psi\right] + \alpha S_{\rm c} \left[\textbf{g}, \psi\right] + S_{\rm m}\left[\textbf{g}, \psi, \Psi\right]\ ,
    \label{action}
\end{equation}
where ${\bf g}$ denotes the spacetime metric, $S_0$ is the GR action 
\begin{equation}    
S_0 = \int \dd ^4 x \frac{\sqrt{-g}}{16 \pi} \left(R - \frac{1}{2} \partial_\mu \psi \partial^{\mu} \psi \right)\ ,
\end{equation}
and $R$ is the Ricci scalar. The non-minimal coupling between the 
scalar field and the metric is encoded in the term 
$\alpha S_{\rm c}$, where the coupling constant $\alpha$ has dimensions, in geometrized units, $({\rm mass})^{N}$.
Matter fields, denoted by $\Psi$, are 
described by the action $S_{\rm m}$.
In our setup, we assume that either $\alpha=0$, i.e. the contribution $\alpha S_{\rm c}$ vanishes, or $\alpha$ has dimensions $N>1$, defining in physical units 
a characteristic energy scale above which interactions 
are suppressed. 
This choice includes the vast majority of scalar-tensor gravity theories  that are currently of interest, including Bergmann-Wagoner theories (for which $\alpha=0$), all shift-symmetric Horndeski theories that admit Minkowski as a solution \cite{Saravani:2019xwx}, as well as all known scalarization models \cite{Damour:1993hw,Doneva:2017bvd,Silva:2017uqg,Dima:2020yac,Herdeiro:2020wei,Doneva:2022ewd}.\footnote{Chern-Simons gravity \cite{Alexander:2009tp} is also formally covered by our action if one takes $\phi$ to be a pseudoscalar. However, in that case the scalar charge is dipolar \cite{Yunes:2011we}, which will not be covered by our analysis of the particle action below.} 

Varying the action \eqref{action} with respect to the 
field content yields the equations of motion:
\begin{equation}
    G_{\mu\nu} = 8 \pi T^{\rm scal}_{\mu \nu}+\alpha T^{\rm c}_{\mu\nu}+ T^{\rm m}_{\mu\nu}\quad ,\quad \Box \psi  = \alpha T^{\rm c}+ T^{\rm m} \ ,\label{eq:fieldseq}
\end{equation}
where $\square=\nabla_\mu\nabla^\mu$,
$T^{\rm scal}_{\mu \nu} = \frac{1}{16 \pi } \left[\partial_{\mu} \psi\partial_{\nu} \psi - \frac{1}{2}g_{\mu\nu}(\partial \psi)^2\right]$, and 
\begin{align}
T^{\rm c}_{\mu\nu} =- \frac{16 \pi}{\sqrt{-g}}\frac{\delta S_{\rm c}}{\delta g^{\mu \nu}}   \quad\ &,\quad
T^{\rm m} =- \frac{16 \pi}{\sqrt{-g}}\frac{\delta S_{\rm m}}{\delta \psi}\ . 
\end{align}
Solving for the metric and scalar field requires 
addressing the coupled system \eqref{eq:fieldseq}. 
However, the mass asymmetry characteristic of EMRIs 
naturally simplifies the problem, allowing tensor and 
scalar perturbations to decouple at leading dissipative 
order in the mass ratio. A rigorous treatment of this 
decoupling within a SF framework, including 
post-adiabatic corrections, was developed in 
Ref.~\cite{Spiers:2023cva}, to which we refer for 
details. Here we present the fundamental equations 
governing the leading-order perturbations and 
their application to generic orbits.

We consider binaries in which the primary is a BH of 
mass $M$, and the secondary a compact object of mass $m_{\rm p}\ll M$. Thus, the field equations\,\eqref{eq:fieldseq} can be solved perturbative in the mass ratio $\epsilon=m_{\rm p}/M\ll1$. Moreover, this choice leads to a hierarchy of scales, in which the secondary body, with a typical length-scale $\sim m_{\rm p}$, moves in the ``exterior'' spacetime of the primary, with the larger length-scale $\sim M$. We can then use an EFT (or ``skeletonized'') approach\,\cite{Eardley:1975fgi,Damour:1992we,Julie:2017ucp}, in which the matter action is replaced by the particle action 
\begin{equation}
    S_{\rm p} = - 
    \int_\gamma m \left(\psi \right) \sqrt{g_{\mu \nu} \frac{\dd y^\mu_{\rm p}}{\dd\tau}\frac{\dd y^\nu_{\rm p}}{\dd\tau}} 
    \dd\tau\ ,\label{eq:partac}
\end{equation}
where $\gamma$: $\tau\to y^\mu_{\rm p}(\tau)$ is the worldline of the secondary body, in a coordinate system $\{y^\mu\}$. In the length-scale of the exterior spacetime, the secondary body is treated as a point-like object, and the function $m(\psi)$ -- encoding the coupling of the body with the scalar field -- is evaluated at its location.

When $\alpha=0$, the exterior spacetime is solution to the Einstein-Klein-Gordon field equations and thus, due to no-hair theorems\,\cite{Chase:1970omy,hawking_black_1972,bekenstein_novel_1995,Sotiriou:2011dz,Hui:2012qt} is described by the Kerr metric with a constant scalar field. When $\alpha\neq0$, assuming that for 
$\alpha\rightarrow 0$ 
the  solutions of Eqs.~\eqref{eq:fieldseq} 
are continuously connected to GR, any corrections to the Kerr metric must depend on the dimensionless parameter
\begin{equation}
\zeta=\frac{\alpha}{M^{N}}=\epsilon^{N}\frac{\alpha}{m^N_{\rm p}}\, , \label{eq:zetapar}
\end{equation}
where, as discussed above, we are assuming $N>1$. Astrophysical observations in both GW 
and electromagnetic spectra for stellar-mass 
sources constrain 
$\alpha/m_{\rm p}^{N} \lesssim \mathcal{O}(1)$ 
\cite{Nair:2019iur,Silva:2020acr,Saffer:2021gak,Lyu:2022gdr,Gao:2024rel,Sanger:2024axs}. 
Therefore, $\zeta\lesssim \epsilon^{N}\ll 1$, implying that 
deviations from the Kerr background can be safely 
neglected at leading adiabatic order, and the background scalar field is constant due to no-hair theorems.
Crucially, 
Eq.~\eqref{eq:zetapar} also 
allows us to use $\epsilon$ as a single 
bookkeeping parameter to describe both metric and 
scalar perturbations, as in the gravitational SF formalism in 
GR:
\begin{equation}
g_{\mu\nu}=g_{\mu\nu}^{(0)}+\epsilon h^{(1)}_{\mu\nu}+\ldots\ \ ,\  \psi=\psi^{(0)}+\epsilon \psi^{(1)}+\ldots\ .
\end{equation} 
At zeroth order in the mass ratio, the background spacetime $g^{(0)}_{\mu\nu}$
is the Kerr metric, and
the scalar field $\psi^{(0)}$, arising from 
$S_0$, is constant and can be set to zero without loss of generality. 

The first-order perturbations, $h_{\mu\nu}^{(1)}$ and 
$\psi^{(1)}$, corresponding to the leading dissipative 
SF contribution, are sourced by the secondary object. 
In the field equations\,\eqref{eq:fieldseq}, the contributions of $T^{\rm scal}_{\mu\nu}$ and $\delta S_{\rm c}/\delta\psi$, which are at least quadratic in $\epsilon$, can be neglected.
Since $m_{\rm p}\ll M$, there exists a buffer region inside the world tube of the secondary object where the scalar field admits a multipolar decomposition. In a local coordinate frame $\{\tilde{x}_{\mu}\}$ centered on 
the secondary, this region is determined by the condition $m_{\rm{p}}\ll \tilde{r}\ll M$, where $\tilde r$ is the spatial distance from the 
worldline of the secondary. 
Within this region, the scalar field solution is
\begin{equation}
    \psi^{(1)}=\frac{m_{\rm p} d}{{\tilde r}}
    +\mathcal{O}\left(m^2_{\rm p}/{\tilde r}^2\right)\ ,
    \label{exp_sc}
\end{equation}
where the dimensionless constant $d$ is called {\it scalar charge} of the secondary body, and measures the leading monopole term in the scalar multipolar expansion. 
Note that $d$ is not a conserved, Noether charge. 

Substituting the expansion~\eqref{exp_sc} into the scalar 
field equation yields the matching conditions:
\begin{equation}
m_{\rm p}=m(\psi)\big\vert_{\psi=0}\ ,\quad 
d=-4\frac{m'(\psi)}{m_{\rm p}}\big\vert_{\psi=0}\ .  
\end{equation}
Expanding the field equations~\eqref{eq:fieldseq} at 
linear order in $\epsilon$, with the source term given by 
$S_{\rm p}$, yields decoupled equations for the metric and 
scalar perturbations:
\begin{equation}
    G^{\alpha \beta}[h^{(1)}_{\alpha\beta}] = 8 \pi m_{\rm p}  
    \int \frac{\delta^{(4)}\left(x-y_{\rm p}(\tau)\right)}{\sqrt{-g}}
    \frac{\dd y^{\alpha}_{\rm p}}{\dd\tau} 
    \frac{\dd y^{\beta}_{\rm p}}{\dd\tau} \dd\tau \ ,\label{eq:einstein}
\end{equation}

\begin{equation}
    \Box \psi^{(1)} = - 4 \pi d m_{\rm p} 
    \int \frac{\delta^{(4)} \left(x - y_{\rm p}(\tau)\right)}{\sqrt{-g}}\dd\tau\ . 
    \label{eq:scalar}
\end{equation}
where 
\begin{equation}
G^{\mu\nu}[h^{(1)}_{\alpha\beta}] = 
\left.\frac{\dd}{\dd\lambda}\,
G_{\mu\nu}\!\big[g^{(0)}_{\alpha\beta}+\lambda\,h^{(1)}_{\alpha\beta}\,\big]\right|_{\lambda=0}
\end{equation}
is the linearized Einstein operator acting on 
$h_{\alpha\beta}$. Equations~\eqref{eq:einstein} 
coincide with those of GR, leading to the Teukolsky equations for Kerr perturbations. The scalar 
field equation, by contrast, contains an 
integral encoding the orbital dynamics of the 
secondary and depends on the scalar charge $d$, 
which sets the overall amplitude of $\psi^{(1)}$.

The field equations~\eqref{eq:einstein}--\eqref{eq:scalar} show that, at leading order in the mass ratio $\epsilon$, the dynamics depends on the underlying modified theory of gravity only through the scalar charge $d$ appearing in the source term of Eq.~\eqref{eq:scalar}. Any dependence on the fundamental coupling $\alpha$ entering the interaction sector $S_{\rm c}$ is therefore encoded implicitly in the relation between the coupling $\alpha$ and the scalar charge $d$. As discussed in Ref. \cite{Maselli:2020zgv}, this property makes the framework completely agnostic with respect to the expression of the interaction terms included in $S_{\rm c}$.

This observation has important practical consequences. At leading order in the mass ratio, waveform modelling and data analysis can be performed directly in terms of the phenomenological parameter $d$, without specifying a particular modified theory of gravity. In the remainder of this work, we therefore treat $d$ as the fundamental parameter controlling beyond-GR effects.

Once a specific theory is selected, measurements or constraints on $d$ can be translated into measurements or constraints on the corresponding fundamental coupling through the theory-dependent relation $d=d(\alpha)$. For example, in shift-symmetric scalar Gauss-Bonnet gravity, for BHs regularity
conditions at the horizon fix $d$ to depend on the BH mass and on the fundamental coupling as 
$d \propto \alpha/{M_{\rm BH}^2}$, with a proportionality coefficient of order unity~\cite{Sotiriou:2014pfa,Sotiriou:2013qea,Thaalba:2022bnt}. Therefore, any observational bound on $d$ can be immediately reinterpreted as a bound on $\alpha$. Analogous mappings can be constructed for other theories once the corresponding relation between the scalar charge and the fundamental couplings is known.

\subsection{Orbital motion}\label{subsec:orbitalmotiton}
%
At adiabatic order, the trajectory of a small compact 
object during inspiral follows a sequence of geodesics 
around the primary, described by the Kerr metric. In 
Boyer–Lindquist coordinates $x^\mu=(t, r, z = \cos\theta, \phi)$, 
the line element reads
\begin{align}
\dd s^2 =& - \left(1 - \frac{2Mr}{\Sigma}\right)\dd t^2 - \frac{4 M a r (1-z^2)}{\Sigma} \dd t \dd\phi+\nonumber\\ 
&\frac{\Sigma}{\Delta}\dd r^2 + \frac{\Sigma}{1-z^2} \dd z^2 +\frac{1-z^2}{\Sigma}\left[2 M a^2 r  (1-z^2) + \right.\nonumber\\
&\left.(a^2+r^2)\Sigma\right] \dd\phi^2\; ,
\label{eq:metric}
\end{align}
where $j = aM$ is the BH angular momentum, and 
\begin{align*}
\Delta &= r^2 - 2 M r + a^2 = (r-r_+)(r-r_-)\ , \\
\Sigma &= r^2 + a^2 z^2 \ ,
\end{align*}
with $r_\pm = M\pm \sqrt{M^2-a^2}$, and 
$\pm$ denoting the inner and outer horizons.

Geodesics in Kerr spacetime admit four conserved quantities: the normalization of the four–velocity $u^\mu = \dd x^\mu/\dd\tau$ ($u_\mu u^\mu = - 1$), the specific energy $\EN$, the azimuthal angular momentum $\ANG$, and the Carter constant $\CAR$, the latter arising from the Killing tensor $\mathcal{K}_{\mu\nu}$ \cite{Carter:1968rr}:
\begin{align}
    \EN &= -g_{tt} \dot t - g_{t\phi}\dot\phi\; ,\\
    \ANG &= g_{\phi t}\dot t + g_{\phi\phi}\dot \phi \; , \\
    \CAR &= u^\mu \mathcal{K}_{\mu\nu}u^\nu -(\ANG - a \EN)^2\; .
\end{align}
With these conserved quantities, the geodesic equations take the form \cite{Ferrari:2020nzo}
\begin{align}
    \left(\frac{\dd r}{\dd \lambda}\right)^2 =& R(r)\nonumber\\
    =&\left[\EN(r^2+a^2)-a\ANG\right]^2\\
    &-\Delta\left[r^2+(\ANG-a \EN)^2+\CAR \right]\; ,\label{eq:r_KerrGeo}\\
    \left(\frac{\dd z}{\dd\lambda}\right)^2=&\varTheta(z)\nonumber \\
    =&\CAR - z^2 [a^2(1-\EN^2)(1-z^2)+ \ANG^2+ \CAR]\; ,\label{eq:z_KerrGeo}\\
    \frac{\dd t}{\dd\lambda} =& T(r,z)\nonumber\\ 
    =&\EN\left[\frac{(r^2+a^2)^2}{\Delta}-a^2 (1-z^2)\right]+\nonumber\\   
    &a \ANG \left( 1-\frac{r^2+a^2}{\Delta}\right)\; ,\label{eq:t_KerrGeo} \\
    \frac{\dd\phi}{\dd\lambda} =&  \Phi(r,z)\nonumber\\
    =&\frac{\ANG}{1-z^2}+a\EN\left( \frac{r^2+a^2}{\Delta}-1\right)-\frac{a^2 \ANG}{\Delta}\; ,
    \label{eq:phi_KerrGeo}
\end{align}
where $\tau$ is the proper time and $\lambda$ the Mino time~\cite{Mino:1997bx}, defined by $\dd\lambda = \dd\tau/\Sigma$. 
We focus on bound orbits, characterized by $0 \le \EN < 1$ and $\CAR \ge 0$ \cite{Chandrasekhar:1985kt}.

The radial potential $R(r)$ is a quartic polynomial that factorizes as
\begin{equation}
   R(r)=(1-\EN^2)(r_1 -r)(r-r_2)(r-r_3)(r-r_4)\; ,
\end{equation}
with roots $r_4 \le r_3 \le r_2 \le r_1$. The radial motion is confined between the turning points $r_2 \le r \le r_1$. 
Similarly, the polar potential can be written as
\begin{equation}
   \varTheta(z)=(z^2-z_1^2)\,[a^2(1-\EN^2)z^2-z_2^2]\ ,
\end{equation}
so that the motion in $z$ is bounded between 
$\pm z_1$, corresponding to polar angles 
$\theta_{\min} = \arccos|z_1|$ and 
$\theta_{\max} = \pi - \theta_{\min}$. 
Both radial and polar motions are periodic, with fundamental periods $\Lambda_r$ and $\Lambda_z$ in Mino time such that 
\begin{equation}
    y^{i}(\lambda+\Lambda_i)=y^{i}(\lambda) \ , \quad i \in \{r,z\}\ . 
\end{equation}
Introducing angle variables\footnote{These quantities can be shifted by arbitrary initial phases. We choose $z(0)=z_1$, $z(\pi)=-z_1$, $r(0)=r_2$, and $r(\pi)=r_1$, allowing consistent integration over $q_r\in[0,\pi]$ in the computation of adiabatic fluxes.}
\begin{equation}\label{eq:anglevar}
    q_r=\lambda \Upsilon_r\; , \quad  q_z=\lambda\Upsilon_z+\pi/2\; ,
\end{equation}
the radial and polar coordinates become $2\pi$–periodic functions of $(q_r,q_z)$, with $\Upsilon_r$ and $\Upsilon_z$ denoting the radial and polar frequencies in Mino time.

Explicit solutions can be written in terms of Jacobi 
elliptic functions \cite{vandeMeent:2019cam}:
\begin{align}
    r(q_r) &= \frac{r_3 (r_1 - r_2)\text{sn}^2\left(\frac{\text{K}(k_r)}{\pi} q_r|k_r\right)-r_2(r_1-r_3)}{(r_1 - r_2) \text{sn}^2\left(\frac{\text{K}(k_r)}{\pi} q_r|k_r\right)-(r_1 -r_3)}\; ,\label{eq:r_geo_sol} \\ 
    z(q_z) &= z_1 \text{sn}\left(\text{K}(k_z) \frac{2 q_z}{\pi}|k_z \right)\; ,
    \label{eq:z_geo_sol}
\end{align}
where $\text{K}(k)$ is the complete elliptic integral of the first kind and $\text{sn}(u|k)$ the Jacobi elliptic sine. Their definitions and conventions are summarized in Appendix~\ref{app:ellipticintegrals}.

The additional roots $r_3,r_4$ and constant $z_2$ relate to $r_1,r_2,z_1$ as
\begin{align}
    r_3 =& \frac{1}{1-\EN^2}-\frac{r_1+r_2}{2}  
    +\nonumber\\
    &\sqrt{\left( \frac{r_1+r_2}{2}-\frac{1}{1-\EN^2}\right)^2-\frac{a^2 \CAR}{r_1 r_2(1-\EN^2)}} \; ,\\ 
    r_4 &= \frac{a^2 \CAR}{r_1r_2r_3(1-\EN^2)} \; , \\ 
    z_2 &= \sqrt{a^2(1-\EN^2)+\frac{\ANG^2}{1-z_1^2}}\; ,
\end{align}
with elliptic moduli 
\begin{equation}
    k_r = \frac{(r_1 - r_2)(r_3-r_4)}{(r_1-r_3)(r_2-r_4)}\ ,\quad 
   k_z=a^2(1-\EN^2)\frac{z_1^2}{z_2^2}\ . 
\end{equation}
From these solutions, the frequencies are
\begin{align}
    \Upsilon_r& = \frac{\pi}{2 \text{K}(k_r)}\sqrt{(1-\EN^2)(r_1-r_3)(r_2-r_4)}\; ,\\
    \Upsilon_z &= \frac{\pi z_2}{2 \text{K}(k_z)} \; .
\end{align}

The geodesic equations for $t$ and $\phi$ can be solved analogously. 
Equation~\eqref{eq:t_KerrGeo} can be decomposed as
\begin{equation}
   T(r,z)= T_t+T_r(r)+T_z(z) \; ,
\end{equation}
where $T_t$ is independent of $r$ and $z$. 
Integrating $T(r,z)$ yields
\begin{equation}
  t(\lambda)= \int \frac{T_r(r)}{\dd r/\dd \lambda}\dd r+ \int \frac{T_z(z)}{\dd z/\dd \lambda}\dd z+ T_t \lambda \ , 
\end{equation}
leading to the solution
\begin{equation}
    t(q_t, q_r, q_z) = q_t +t_r(q_r) + t_z (q_z) \; ,
\end{equation}
with $t_r$ and $t_z$ expressed in terms of elliptic integrals:
\begin{subequations}\label{eq:trtzeqs}
\begin{align}
    t_r(q_r)& = \tilde{t}_r\left[\text{am}\left( \text{K}(k_r)\frac{q_r}{\pi}|k_r  \right)\right]- \frac{\tilde{t}_r(\pi)}{2\pi} q_r \; ,\\ 
    t_z(q_z) &= \tilde{t}_z\left[\text{am}\left( \text{K}(k_z)\frac{q_z}{\pi}|k_z  \right)\right]- \frac{\tilde{t}_z(\pi)}{2\pi} q_z \; .
\end{align}
\end{subequations}
where am is the Jacobi amplitude function.
Applying the same procedure to $\phi$, we find
\begin{equation}
    \phi(q_\phi, q_r, q_z) = q_\phi +\phi_r(q_r) + \phi_z (q_z)\; ,
\end{equation}
where
\begin{subequations}\label{eq:phirphizeq}
\begin{align}
    \phi_r(q_r) &= \tilde{\phi}_r\left[\text{am}\left( \text{K}(k_r)\frac{q_r}{\pi}|k_r  \right)\right]- \frac{\tilde{\phi}_r(\pi)}{2\pi} q_r\; ,\\
    \phi_z(q_z) &= \tilde{\phi}_z\left[\text{am}\left( \text{K}(k_z)\frac{q_z}{\pi}|k_z  \right)\right]- \frac{\tilde{\phi}_z(\pi)}{2\pi} q_z \; .
\end{align}
\end{subequations}
The phases $q_t$ and $q_\phi$ represent the secularly growing linear parts of the motion,
\begin{equation}\label{eq:q_i}
    q_i = \Upsilon_i \lambda \; ,
\quad  \Upsilon_i = \tilde{\Upsilon}_{i,r} + \tilde{\Upsilon}_{i,z} \quad\quad i = t, \phi\ ,
\end{equation}
where the functions $\tilde{t}_r, \tilde{t}_z,\tilde{\phi}_r, \tilde{\phi}_z,\tilde{\Upsilon}_{i,r}, \tilde{\Upsilon}_{i,z}$ are listed in Appendix~\ref{app:OrbitalMotion}.

So far we have discussed geodesic motion in terms of $\{\EN,\ANG,\CAR\}$.
It is convenient to parametrize geodesics using orbital parameters $p,\ e$ and $\theta_{\rm inc}$, namely the semi-latus rectum, the eccentricity and the inclination angle measured respect the equatorial plane.
In terms of these quantities, the radial and polar 
turning points $r_1$, $r_2$ and $z_1$ are 
\begin{equation}
    r_1=\frac{p}{1-e} \ ,\quad r_2=\frac{p}{1+e} \ ,\quad z_1=\sin\theta_{\rm inc} \ .
\end{equation}
Given the orbital elements $r_1,\ r_2$ and $z_1$, the constants of motion $\{\EN,\ANG,\CAR\}$ can be computed 
as
\begin{align}
    \EN =& \sqrt{\frac{\kappa \rho + 2 \epsilon \sigma -2 a \sqrt{\frac{\sigma}{a^2}} (\sigma \epsilon^2+ \rho \epsilon \kappa - \eta \kappa^2)}{\rho^2+4 \eta \sigma}}\ ,\\
    \ANG =& -\frac{g(r_2)\EN}{h(r_2)} - \nonumber\\& \frac{\sqrt{(g(r_2)^2+h(r_2)f(r_2))\EN^2-h(r_2)d(r_2)}}{h(r_2)}\ ,\\
    \CAR =& z_1^2\Bigg(a^2(1-\EN^2)+\frac{\ANG^2}{1-z_1^2}\Bigg)\ ,
\end{align}
where the auxiliary coefficients are defined as
\begin{align}
    \kappa &= d(r_2)h(r_1)-d(r_1)h(r_2)\ ,\\
    \epsilon &= d(r_2)g(r_1)-d(r_1)g(r_2)\ ,\\
    \rho &= f(r_2)h(r_1)-f(r_1)h(r_2)\ ,\\
    \eta &= f(r_2)g(r_1)-f(r_1)g(r_2)\ ,\\
    \sigma &= g(r_2)h(r_1)-g(r_1)h(r_2)\ ,
\end{align}
and the functions $d(r)$, $f(r)$, $g(r)$, and $h(r)$ 
read
\begin{align}
    d(r)&=\Delta(r)(r^2+a^2z_1^2)\ ,\\
    f(r)&=r^4+a^2(r(r+2)+z_1^2\Delta(r))\ ,\\
    g(r)&=2ar\ ,\\
    h(r)&=r(r-2)+\frac{z_1^2\Delta(r)}{1-z_1^2}\ .
\end{align}

%
\subsection{Scalar field perturbations}\label{subsec:scalarfieldperts}
%

Since the equations governing the metric perturbations 
coincide with those of GR, we focus here on the computation 
of scalar field perturbations sourced by the secondary 
charged BH orbiting the massive primary. 
We emphasize, however, that our numerical 
framework is fully capable of treating both scalar 
and tensor modes, as discussed in 
Sec.~\ref{sec:resultsscalarspectra}.

The equation for $\psi^{(1)}$, Eq.~\eqref{eq:scalar}, 
can be written explicitly in Boyer–Lindquist 
coordinates as \cite{Teukolsky:1973ha}
\begin{equation}
\begin{aligned}
&\left[\frac{(r^2+a^2)^{2}}{\Delta} - a^2 \sin^2\theta \right] \pdv[2]{\psi^{(1)}}{t}
+ \frac{4 M a r}{\Delta} \pdv{\psi^{(1)}}{t}{\phi} \\
&\quad + \left[ \frac{a^2}{\Delta} - \frac{1}{\sin^2\theta} \right] \pdv[2]{\psi^{(1)}}{\phi}
- \pdv{}{r} \left[ \Delta \pdv{\psi^{(1)}}{r} \right] \\
&\quad - \frac{1}{\sin\theta} \pdv{}{\theta} \left[ \sin\theta \pdv{\psi^{(1)}}{\theta} \right]
= -4\pi \Sigma T \ .
\end{aligned}
\end{equation}
The scalar field $\psi^{(1)}(t,r,\theta,\phi)$ and the source term $T(t,r,\theta,\phi)$ admit expansions in spin-weighted spheroidal harmonics. In the frequency domain, after a Fourier transform, they read
\begin{equation}\label{eq:scalarfielddecomposition}
    \psi^{(1)}(x^\mu)=\int \dd \omega \sum_{\ell m} \tilde{R}_{\ell m}(\omega,r) S_{\ell m}(\omega, \theta)e^{i m \phi} e^{-i \omega t} \ , 
\end{equation}
\begin{equation}\label{eq:scalarfielddecomposition2}
\begin{split}
    J(x^\mu)&= 4\pi\Sigma T(x^\mu)\\
    &=\int \dd \omega \sum_{\ell m}\tilde{J}_{\ell m}(\omega,r)S_{\ell m}(\omega,\theta)e^{i m \phi} e^{-i\omega t}  \ .
\end{split}
\end{equation}
Here $\ell=0,\ldots,\infty$ and $-\ell\le m\le \ell$. The decomposition \eqref{eq:scalarfielddecomposition}–\eqref{eq:scalarfielddecomposition2} decouples the Teukolsky equation into angular and radial parts. The angular equation is
\begin{equation}\label{eq:angularcomponenteq}
\begin{split}
    \Big[\frac{1}{\sin\theta}\diff{}{\theta} \Big(\sin\theta\diff{}{\theta}\Big)&-\gamma^2\sin^2\theta-\frac{m^2}{\sin^2\theta}\\
    &+2m\gamma +\lambda_{\ell m}\Big]S_{\ell m}=0 \ ,
\end{split}
\end{equation}
where $\gamma = a \omega$, and $\lambda_{\ell m}$ is the angular eigenvalue. The solutions $S_{\ell m}(\omega,\theta)$ are spin-weighted spheroidal harmonics, satisfying the orthogonality relation
\begin{equation}\label{eq:orthorelation}
    \int \dd\Omega \, S_{\ell m}( \omega,\theta)e^{im\phi}\,S_{\ell'm'}^*(\omega,\theta)e^{-im'\phi}=\delta_{\ell \ell'}\delta_{mm'} \ .
\end{equation}
The radial functions $\tilde{R}_{\ell m}(\omega, r)$ satisfy the inhomogeneous equation
\begin{equation}\label{eq:Requation}
\begin{split}
    \diff{}{r}\left[\Delta\diff{\tilde{R}_{\ell m}}{r} \right]+\left[\frac{K^2}{\Delta}-\lambda_{\ell m}\right]\tilde{R}_{\ell m}=\tilde{J}_{\ell m} \ ,
\end{split}
\end{equation}
where $K(\omega,r)=(r^2+a^2)\omega-a m$.
The solution to Eq.~\eqref{eq:Requation} can be found with the standard Green’s function method:
\begin{equation}\label{eq:Rsol}
    \tilde{R}_{\ell m}={\tilde{R}}^+_{\ell m}\int_{r_+}^{r}\dd r'\,\frac{\tilde{R}^-_{\ell m}\tilde{J}_{\ell m}}{W_{\ell m}\Delta}
    +\tilde{R}^-_{\ell m}\int_{r}^{\infty}\dd r' \,\frac{\tilde{R}^+_{\ell m}\tilde{J}_{\ell m} }{W_{\ell m}\Delta}\ ,
\end{equation}
where $\tilde R_{\ell m}^-$ ($\tilde R_{\ell m}^+$) denote the purely ingoing (outgoing) solution at the horizon (infinity) of the homogeneous equation and the Wronskian is
\begin{equation}
\label{def_W}
    W_{\ell m}=\frac{\dd \tilde{R}^+_{\ell m}}{\dd r}\tilde{R}^-_{\ell m} - \tilde{R}^+_{\ell m} 
    \frac{\dd\tilde{R}^-_{\ell m}}{\dd r}\ .
\end{equation}
The asymptotic behavior of the solutions $\tilde{R}_{\ell m}^{\pm}$ are easier to find if we introduce the 
auxiliary functions:
\begin{equation}\label{eq:YtoR}
\tilde{Y}^\pm_{\ell m}(\omega, r) =\sqrt{r^2+a^2}\,\tilde{R}^\pm_{\ell m}(\omega, r) \ ,
\end{equation}
which satisfies the homogeneous equation
\begin{equation}\label{eq:eqtosolveinY}
    \diff[2]{}{r_\star}\tilde{Y}^\pm_{\ell m}+V\,\tilde{Y}^\pm_{\ell m}=0\ ,
\end{equation}
where $r_\star$ is the tortoise coordinate defined by
\begin{equation}\label{eq:drs}
    \diff{r_\star}{r}=\frac{r^2+a^2}{\Delta} \ ,
\end{equation} 
and the effective potential is
\begin{equation}
    V(\omega, r)=\frac{K(\omega,r)^2-\lambda_{\ell m}\Delta(r)}{(r^2+a^2)^{2}}-G(r)^2-\diff{G(r)}{r_\star} \ ,
\end{equation}
with $G = \frac{r \Delta}{(r^2+a^2)^2}$.

At spatial infinity ($r \to \infty$, $r_\star \to \infty$) and near the horizon ($r \to r_+$, $r_\star \to -\infty$), Eq.~\eqref{eq:eqtosolveinY} reduces to
\begin{subequations}\label{eq:Y_asymp}
\begin{align}
    \diff[2]{}{r_\star}\tilde{Y}^\pm_{\ell m} + \omega^2 \tilde{Y}^\pm_{\ell m} &= 0  \quad\quad ( r_\star \rightarrow + \infty)\ ,\\
    \diff[2]{}{r_\star}\tilde{Y}^\pm_{\ell m} + k^2  \tilde{Y}^\pm_{\ell m} &= 0 \quad\quad ( r_\star \rightarrow -\infty)\ ,
\end{align}
\end{subequations}
where $k = \omega - m \omega_+$ and $\omega_+ = a/(2 M r_+)$.

From Eqs.~\eqref{eq:Y_asymp} we find the asymptotic behaviour of the functions $\tilde{Y}^\pm_{\ell m}$, i.e.
$\tilde{Y}^-_{\ell m}\sim e^{-ik r_\star}$ for $r\to-\infty$ and $\tilde{Y}^+_{\ell m}\sim e^{+i\omega r_\star}$ for $r\to+\infty$. From these, using Eq.~\eqref{eq:YtoR} we find the asymptotic behaviour of the functions $\tilde R_{\ell m}^\pm$. We choose their normalizations such that
\begin{subequations}\label{eq:Rminusasymptotic}
\begin{align}
    \tilde{R}_{\ell m}^- &\to e^{- i k r_\star} \quad &( r_\star \rightarrow -\infty) \ ,\\
    \tilde{R}_{\ell m}^- &\to A_{\ell m}^{\text{in}} \frac{e^{- i \omega r_\star}}{r} + A_{\ell m}^{\text{out}} \frac{e^{ i \omega r_\star}}{r} \quad &( r_\star \rightarrow +\infty) \ ,
\end{align}
\end{subequations}
and 
\begin{subequations}\label{eq:Rplusasymptotic}
\begin{align}
    \tilde{R}^+_{\ell m} &\to B_{\ell m}^{\text{in}} e^{- i k r_\star} + B_{\ell m}^{\text{out}} e^{ i k r_\star} \quad\quad &( r_\star \rightarrow -\infty)\ ,\\
    \tilde{R}^+_{\ell m} &\to \frac{e^{ i \omega r_\star}}{r} \quad\quad &( r_\star \rightarrow +\infty) \ .
\end{align}
\end{subequations}
The radial solution $\tilde{R}_{\ell m}$ in Eq.~\eqref{eq:Rsol} is ingoing (outgoing) at the horizon (infinity) by construction and its asymptotic amplitude $\delta  \tilde{\psi}^{-}_{\ell m}$ ($\delta  \tilde{\psi}^{+}_{\ell m}$), i.e.,
\begin{equation}
    \delta  \tilde{\psi}^{\mp}_{\ell m}(\omega)=\int_{r_+}^{+\infty}\dd r\,\frac{\tilde{R}_{\ell m}^\pm (\omega,r)\tilde{J}_{\ell m}(\omega,r)}{W_{\ell m}(\omega,r) \Delta(r)}\ ,
    \label{eq:deltaphi_amplitude}
\end{equation}
is the physical relevant quantity for computing GW fluxes. In Eq.~\eqref{eq:deltaphi_amplitude}, the product $W \Delta$ is independent of $r$; 
evaluating it as $r\rightarrow \infty$ gives
\begin{equation}
     w_{\ell m}(\omega) = W_{\ell m} (\omega,r)\Delta(r)\big\vert_{r\rightarrow\infty} = 2 i \omega A^{\text{in}}_{\ell m}(\omega)\ .\label{eq:wronskian}
\end{equation}
%
\subsection{The source term}\label{subsec:sourceterm}
%

We now turn our attention on the frequency-domain 
components, $\tilde{J}_{\ell m}(\omega, r)$, that 
appear as source terms of the scalar field equation 
\eqref{eq:Requation}, 
\begin{equation}\label{eq:Jlm}
\begin{split}
    \tilde{J}_{\ell m}(\omega, r)
    =& \frac{1}{2\pi}
    \int_{-\infty}^{+\infty} \! \dd t
    \int \sin\theta \, \dd\theta \, \dd\phi \;
    S_{\ell m}(\omega, \theta) \, \times \\&
    e^{i\omega t - i m \phi} \,
    J(t, r, \theta, \phi) \ ,
\end{split}
\end{equation}
where the time-domain source reads (see Eq.~\eqref{eq:scalar})
\begin{equation}\label{eq:Jeq_first_form}
    J = 4\pi m_{\rm p} d \,
    \frac{
        \delta(r - r_{\rm p}) \,
        \delta(\theta - \theta_{\rm p}) \,
        \delta(\phi - \phi_{\rm p})
    }{
        |\sin\theta| \, \dot{t}
    }\ ,
\end{equation}
with $\dot{t} = \dd t/\dd \tau$. The functions 
$r_{\rm p}$, $\theta_{\rm p}$, and $\phi_{\rm p}$ denote 
the coordinates of the secondary, and depend 
on the coordinate time, e.g. $r_{\rm p}=r_{\rm p}(t)$.
Swapping the order of the time and spatial 
integrals, replacing Eq.~\eqref{eq:Jeq_first_form} 
into Eq.~\eqref{eq:Jlm} and performing the 
angular integrals, we get
\begin{equation}
    \tilde{J}_{\ell m}
    =\left.2 m_{\rm p} d\int_{-\infty}^{+\infty}  \frac{\dd t}{\dot{t}} S_{\ell m}e^{i(\omega t-m\phi)}\delta(r-r_{\rm p})\right|_{\substack{\phi = \phi_{\rm p} \\ \theta = \theta_{\rm p}}}\ ,\label{eq:emri:generic:J_final}
\end{equation}
where we used $|\sin(\theta)|=\sin\theta$ since $\theta \in [0,\pi]$.

Plugging Eq.~\eqref{eq:emri:generic:J_final} in Eq.~\eqref{eq:deltaphi_amplitude} and performing the integration over $r$, we get the explicit expression of the mode amplitudes $\delta\tilde\varphi^\pm_{\ell m}$:
\begin{equation}
    \delta \tilde{\psi}_{\ell m}^{\mp}=\left. \frac{2 m_{\rm p} d}{w_{\ell m}}\int_{-\infty}^{+ \infty}\frac{\dd t}{\dot{t}} e^{i(\omega t-m\phi)}\tilde{R}_{\ell m}^\pm S_{\ell m}\right|_{\substack{\phi = \phi_{\rm p} \\ \theta = \theta_p\\
    r=r_{\rm p}}}\ .
\end{equation}
Writing the $t$ and $\phi$ 
particle’s coordinates 
in terms of the Mino time 
$\lambda$,
\begin{equation}
\begin{split}
    t(\lambda) &= \Upsilon_t \lambda + t_r(r(\lambda)) + t_\theta(\theta(\lambda))\ , \\
    \phi(\lambda) &= \Upsilon_\phi \lambda + \phi_r(r(\lambda)) + \phi_\theta(\theta(\lambda))\ ,
\end{split}
\end{equation}
where $\Upsilon_i$ are the Mino 
frequencies, the mode coefficients read
\begin{align}\label{eq:modecoefflm}
    \delta \tilde{\psi}_{\ell m}^{\mp}  = \frac{2 m_{\rm p} d}{w_{\ell m }}\int_{-\infty}^{+ \infty}\dd \lambda e^{i(\omega \Upsilon_t - m \Upsilon_\phi) \lambda} \mathcal{I}_{\ell m}^\pm \ ,
\end{align}
with
\begin{equation}\label{eq:Iplusminu}
\begin{split}
     \mathcal{I}_{\ell m}^\pm(\omega, r(\lambda),\theta(\lambda))  = & e^{i\left[\omega (t_r(r) + t_\theta(\theta))-  m (\phi_r(r) + \phi_\theta(\theta))\right]} \\ 
     &\left. \Sigma(r,\theta) \tilde{R}_{\ell m}^\pm(\omega,r) S_{\ell m}(\omega,\theta)\right|_{\substack{r = r_{\rm p}(\lambda) \\ \theta = \theta_{\rm p}(\lambda)}}\ .
    \end{split}
\end{equation}
Since $\mathcal{I}^{\pm}$ is a periodic function 
of $\lambda$, it admits a Fourier expansion
\begin{equation}\label{eq:mathcalIasseries}
    \mathcal{I}^{\pm}_{\ell m}=\sum_{kn} \hat{\mathcal{I}}_{\ell mkn}^{\pm} e^{-i(k\Upsilon_\theta + n \Upsilon_r)\lambda}\ ,
\end{equation}
where $k$ and $n$ are the  angular and radial harmonic index respectively,
with coefficients
\begin{equation}\label{eq:coefficientsofmathcalIseries}
\begin{split}
    \hat{\mathcal{I}}_{\ell mkn}^{\pm}(\omega)=&\frac{1}{(2 \pi)^2}\int_0^{2\pi}\dd q_\theta\int_0^{2\pi}\dd q_r e^{i(k q_\theta+n q_r)} \\
    &\mathcal{I}_{\ell m}^{\pm}[\omega,r(q_r),\theta(q_\theta)]\ ,
\end{split}
\end{equation}
where we used the angle variables $q_r$ and 
$q_\theta$ defined in Eq.~\eqref{eq:anglevar}. 
Inserting Eq.~\eqref{eq:Iplusminu} into 
Eq.~\eqref{eq:modecoefflm} we obtain:
\begin{equation}\label{eq:SolutionDeltaVarphi}
    \delta \tilde{\psi}_{\ell m}^{\pm}(\omega)=\sum_{kn}\delta \hat{\psi}_{\ell mkn}^{\pm}\delta(\omega-\omega_{mkn})\ ,
\end{equation}
with
\begin{equation}
\omega_{mkn}=k\Omega_\theta+n\Omega_r+m\Omega_\phi\ , \quad\quad  \Omega_i=\frac{\Upsilon_i}{\Upsilon_t} \ ,
\end{equation}
and
\begin{equation}\label{eq:deltaphi_lmnk}
\begin{split}
    \delta \hat{\psi}_{\ell mkn}^{\mp}=&\frac{m_{\rm p} d}{\pi w \Upsilon_t} \int_{0}^{2\pi}\dd q_\theta \int_0^{2\pi} \dd q_r e^{i(k q_\theta+n q_r)}\\
    &\mathcal{I}^{\pm}_{\ell m}(\omega_{mkn},r(q_r),\theta(q_\theta))\ .
\end{split}
\end{equation}
Noting that (with $i = r,\theta$) 
\begin{equation}
    \begin{split}
        &r(q_i)=r(- q_i)\ ,\\
        &t_i(q_i)=-t_i(- q_i)\ ,\\
        &\phi_i(q_i)=-\phi_i(- q_i)\ ,\\
    \end{split}
\end{equation}
we can further simplify the domain of integration.
Indeed, applying these symmetries to the integrand in Eq.~\eqref{eq:deltaphi_lmnk}, the amplitudes of the radial Teukolsky function become
\begin{widetext}
\begin{align}
    \delta \hat{\psi}_{\ell m kn}^{\mp}
    &= \frac{4 m_{\rm p} d}{w \pi \Upsilon_t} \int_{0}^{\pi}\dd q_\theta \int_0^{\pi} \dd q_r \Sigma\tilde{R}_{\ell m}^\pm(\omega, r_{\rm p}(q_r)) S_{\ell m} (\theta_{\rm p}(q_\theta), \omega)\cos[n q_r + \omega t_r (q_r)- m \phi_r(q_r)]\cos[n q_\theta + \omega t_\theta (q_\theta)- m \phi_\theta(q_\theta)]\nonumber\\
    &=\frac{4 m_{\rm p} d}{w \pi\Upsilon_t} ( I_\theta^{(1)} I_r^{\pm(2)}+I_\theta^{(2)} I_r^{\pm(1)}) \ ,
\end{align}
where 
\begin{equation}
    \begin{split}
        I_\theta^{(1)} = \int_0^\pi &\dd q_\theta \  S_{\ell m} (\theta_{\rm p}(q_\theta), \omega)\cos[k q_\theta + \omega t_\theta (q_\theta)- m \phi_\theta(q_\theta)]\ , \\
        I_\theta^{(2)} = \int_0^\pi &\dd q_\theta \   a^2 \cos^2(\theta_{\rm p}(q_\theta))  S_{\ell m} (\theta_{\rm p}(q_\theta), \omega)\cos[k q_\theta + \omega t_\theta (q_\theta)-  m \phi_\theta(q_\theta)]\ , \\
        I_r^{\pm(1)} = \int_0^\pi &\dd q_r \ \tilde{R}_{\ell m}^\pm(\omega, r_{\rm p}(q_r))\cos[n q_r + \omega t_r (q_r)- m \phi_r(q_r)]\ ,\\
        I_r^{\pm(2)} = \int_0^\pi &\dd q_r \ r^2_p(q_r) \tilde{R}_{\ell m}^\pm(\omega, r_{\rm p}(q_r))\cos[n q_r + \omega t_r (q_r)- m \phi_r(q_r)]\ .
    \end{split}
\end{equation}
\end{widetext}

%
\subsection{Energy and angular momentum scalar fluxes}\label{subsec:energyangmomscalarfluxes}
%
The energy and angular momentum 
fluxes associated to the scalar field perturbation, 
carried away at the horizon $(-)$ and at infinity 
$(+)$ can be computed as 
\begin{equation}
    \dot{E}^\pm = \mp \Delta \int T_{tr}^{\text{scal}}\dd \Omega \ ,\quad
    \dot{L}^\pm = \mp \Delta \int T_{ \phi r}^{\text{scal}}\dd \Omega \ ,
\end{equation}
where the components of the stress-energy tensor needed to determine 
$\dot{E}^{\pm}$ and $\dot{L}^{\pm}$ 
are
\begin{equation}
    T_{jr}^\text{scal} =\frac{1}{16\pi} \partial_j\psi\partial_r\psi = \frac{1}{16\pi} \partial_j\psi(\partial_r\psi)^* \ , \quad j=\{t,\phi\}\ ,
\end{equation}
since the scalar field is real. From the spheroidal 
harmonics expansion \eqref{eq:scalarfielddecomposition}, 
and the asymptotic behavior of the radial function 
$\tilde{R}_{\ell m}(\omega,r)$ at the horizon and at spatial infinity, Eqns.~\eqref{eq:Rminusasymptotic}-\eqref{eq:Rplusasymptotic}, we find 
\begin{equation}
    \begin{split}
        &\psi_{\ell m } \sim e^{-i\omega t} e^{-i k r_\star} S_{\ell m } e^{i m \phi}\quad\quad r\rightarrow r_+ \\
        &\psi_{\ell m } \sim e^{-i\omega t} \frac{e^{i \omega r_\star}}{r}S_{\ell m } e^{i m \phi} \quad\quad r\rightarrow + \infty\ .
    \end{split}
\end{equation}
From this, we find the $t$ and $\phi$ derivatives of the scalar field in both limits $r\to r_+$, $r\to+\infty$:
\begin{equation}
        \partial_t \psi = -i \omega \psi\ ,\quad
         \partial_\phi \psi = i m \psi\ ,
\end{equation}
while the derivative respect to the radial coordinate shows different asymptotic behaviours: 
\begin{equation}
    \begin{split}
        &(\partial_r\psi)^* \sim e^{i\omega t} \frac{r^2+a^2}{\Delta}(-i k )e^{i k r_\star} S_{\ell m } e^{-i m \phi}\quad\quad r\rightarrow r_+ 
        \ , \\
        &(\partial_r\psi)^* \sim e^{i\omega t} (- i \omega)\frac{e^{-i \omega r_\star}}{r} S_{\ell m } e^{-i m \phi}\quad\quad~~~~~~ r\rightarrow + \infty\ .
    \end{split}
\end{equation}
Using the orthogonality properties of the spheroidal harmonics, 
Eq.~\eqref{eq:orthorelation}, we 
obtain the final expressions for the energy fluxes 
at infinity and at the horizon:\,\footnote{
Notice that the expression for the horizon flux differs from that in~\cite{Maselli:2021men} due to a different normalization of the ingoing solution for the homogeneous equation that was adopted there.}
\begin{equation}\label{eq:final_expression_scalar_fluxes_E}
\begin{split}
\dot{E}^+ &= \sum_{\ell m n k } \dot{E}^+_{\ell m n k }\\&= \frac{1}{16\pi}\sum_{\ell m n k} \, \omega_{mnk}^2 \left| \delta\psi^+_{\ell m n k } \right|^2 \ ,\\
\dot{E}^- &= \sum_{\ell mnk} \dot{E}^-_{\ell m n k }\\&=\frac{1}{16\pi} \sum_{\ell m n k} \, \omega_{mnk} k_{mnk} \left| \delta\psi^-_{\ell m n k } \right|^2 \left( r_+^2 + a^2 \right)\ ,
\end{split}
\end{equation}
and for the angular momentum fluxes at infinity and at horizon:
\begin{equation}\label{eq:final_expression_scalar_fluxes_L}
\begin{split}
\dot{L}^+ &= \sum_{\ell m n k } \dot{L}^+_{\ell m n k }\\& =\frac{1}{16\pi}\sum_{\ell m n k} \, m \omega_{mnk} \left| \delta\psi^+_{\ell m n k } \right|^2 \ ,\\
\dot{L}^- &= \sum_{\ell m n k } \dot{L}^-_{\ell m n k }\\& =\frac{1}{16\pi} \sum_{\ell m n k} \, m  k_{mnk} \left| \delta\psi^-_{\ell m n k } \right|^2 \left( r_+^2 + a^2 \right)\ ,
\end{split}
\end{equation}
with $k_{mnk} = \omega_{mnk} - m \omega_+$. Note that, since each flux mode satisfies $\dot L^{\pm}_{\ell mnk}=(m/\omega_{mnk})\dot E^{\pm}_{\ell mnk}$, the angular momentum fluxes can be straightforwardly obtained from the 
energy fluxes (and viceversa).
The same applies to the rate of change of the Carter constant, that can be obtained from the energy and angular momentum fluxes~\cite{Drasco:2005is,Isoyama:2013yor,Zi:2025lio}.

In Eqs.~\eqref{eq:final_expression_scalar_fluxes_E} and ~\eqref{eq:final_expression_scalar_fluxes_L}, we have neglected possible couplings among different $(n,k)$ modes. Indeed, some harmonics can give rise to \emph{resonances}~\cite{Ruangsri:2013hra}: their averaged crossing-term is not vanishing and it contributes to the energy and angular momentum fluxes. It occurs if   
\begin{equation}
    \exists k,n \in \mathbb{Z} \ : \ k\Omega_\phi+n\Omega_\theta=0\ . 
\end{equation}
Resonances contribute at order $\epsilon^{-1/2}$ in the waveform 
frequency and they can be consistently neglected at leading order $\epsilon$~\cite{Lynch:2024ohd}. In this regard, a resonant 
scalar self-force model was presented in~\cite{Nasipak:2021qfu}.
%

%
\section{Results}\label{sec:resultsscalarspectra}
%

To compute scalar fluxes we use \texttt{STORM (Scalar Tensor Orbital Radiation from EMRIs)}, a new \texttt{C++} framework designed to model the evolution of asymmetric binaries in generic extensions of GR~\cite{STORM}.
The code supports arbitrary-precision arithmetic through the \texttt{Boost} multiprecision library~\cite{boost}, and implements all numerical ingredients required to evaluate perturbations of spinning BHs. 
Its structure is organized into six main modules:

\begin{itemize}
    \item \texttt{SpinWeightedSpheroidalHarmonics} evaluates the homogeneous solutions of the angular Teukolsky equation for arbitrary spin weight $s$, providing the angular basis functions used in the mode decomposition;
    \item \texttt{RadialHomogeneousSolution} computes the solutions of the homogeneous radial Teukolsky equation for arbitrary field spin $s$, which are required to construct the radial Green’s function;
    \item \texttt{GeodesicOrbitalMotion} integrates the geodesic motion of the secondary along eccentric and inclined Kerr orbits (see Sec.~\ref{subsec:orbitalmotiton}), supplying the source trajectories for the perturbation equations;  
    \item \texttt{ScalarFluxMode} evaluates the energy flux of individual scalar modes (c.~f.~Eq.~\eqref{eq:final_expression_scalar_fluxes_E});
    \item \texttt{GravitationalFluxMode} evaluates the energy flux of individual gravitational modes;
    \item \texttt{special\_functions} provides high-accuracy implementations of complex special functions, including the Gamma function and hypergeometric functions, which appear in the analytic solutions of the Teukolsky equations;
    \item \texttt{config}: a configuration file which allows the user to select the numerical data type (e.g., double or multiprecision) and to set the tolerance parameters controlling all iterative procedures. In this way, the user can balance computational cost and numerical accuracy according to the requirements of the specific application.
\end{itemize}

A detailed description of the numerical methods used 
to solve the angular and radial Teukolsky equations 
is provided in Appendix~\ref{app:codeimplementation}.\\

With this computational infrastructure in hand, we 
carry out a systematic exploration of the scalar 
radiation emitted by generic binaries, mapping out 
the full four-index mode structure $(\ell,m,n,k)$ 
associated with general trajectories, both inclined 
and eccentric.

Spectra of the scalar fluxes shown hereafter, 
are generated by fixing the orbital parameters $(a,e,p,\theta_{\rm inc})$ 
and the angular mode numbers $(\ell,m)$, while 
performing a systematic scan in the $(k,n)$ harmonic 
plane.

For a given value of $k$, the Fourier index $n$ is 
first set to $n=0$ and then scanned toward both positive 
and negative values. All mode contributions to the energy flux with 
$|\dot{E}^{\pm}_{\ell m k n}| < \bar{\dot{E}}^{\pm}_{\ell m k n}=10^{-30}$ 
are set to zero. Here we take the absolute value since $\dot{E}^{-}_{\ell m k n}$ 
can take negative values. The scan in a given direction 
is terminated once three consecutive values of $n$ yield 
vanishing contributions. This procedure is repeated for 
successive values of $k$, starting from $k=0$ and scanning 
toward both positive and negative $k$. In each direction, 
the scan over $k$ is stopped when the sum over $n$ 
of the corresponding modes falls below the numerical 
threshold, $\sum_n \dot{E}^{\pm}_{\ell m k n}<\bar{\dot{E}}^{\pm}_{\ell m k n}$, for 
three consecutive values of $k$. The scan over $k$ 
can be simplified by observing that, for a fixed value 
of $\ell + m$, only $k$-modes with matching parity 
contribute. Specifically, when $\ell + m$ is even (odd), 
only even (odd) values of $k$ appear in the sum. This 
selection rule follows from the parity properties of the 
angular functions and is discussed in Appendix~B of 
Ref.~\cite{DellaRocca:2024pnm} for the case of inclined, 
circular orbits. We apply the same criterion to fluxes
of angular momentum, finding that the sum over $n$ and 
$k$ saturates the threshold at the same values as for the 
energy fluxes (see 
Tables~\ref{tab:dominant_modes}-\ref{tab:subdominant_modes} 
and discussion below).

\begin{figure}
    \centering

    \begin{subfigure}{0.238\textwidth}
        \centering
        \includegraphics[width=\textwidth]{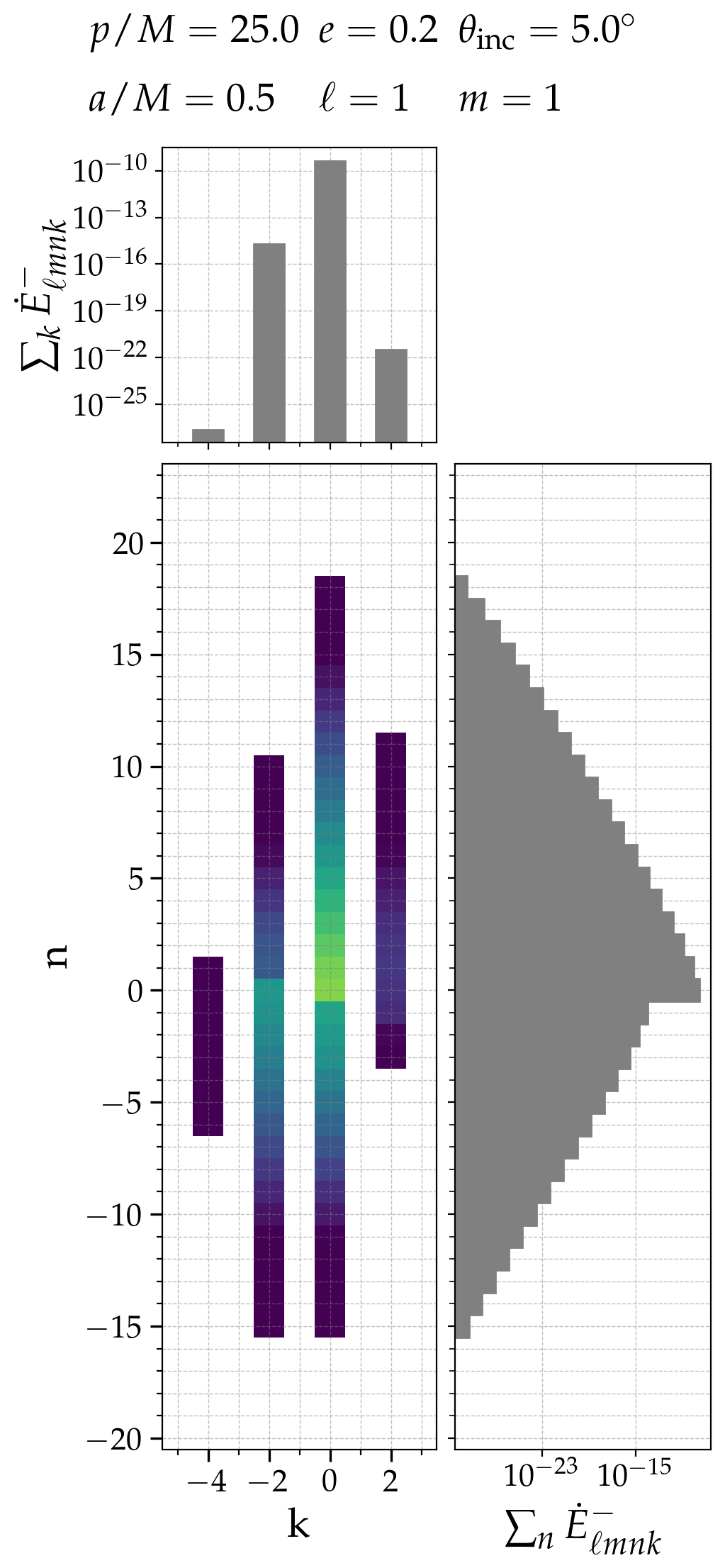}
        \caption{Energy scalar flux at the horizon.}
    \end{subfigure}
    \hfill
    \begin{subfigure}{0.238\textwidth}
        \centering
        \includegraphics[width=\textwidth]{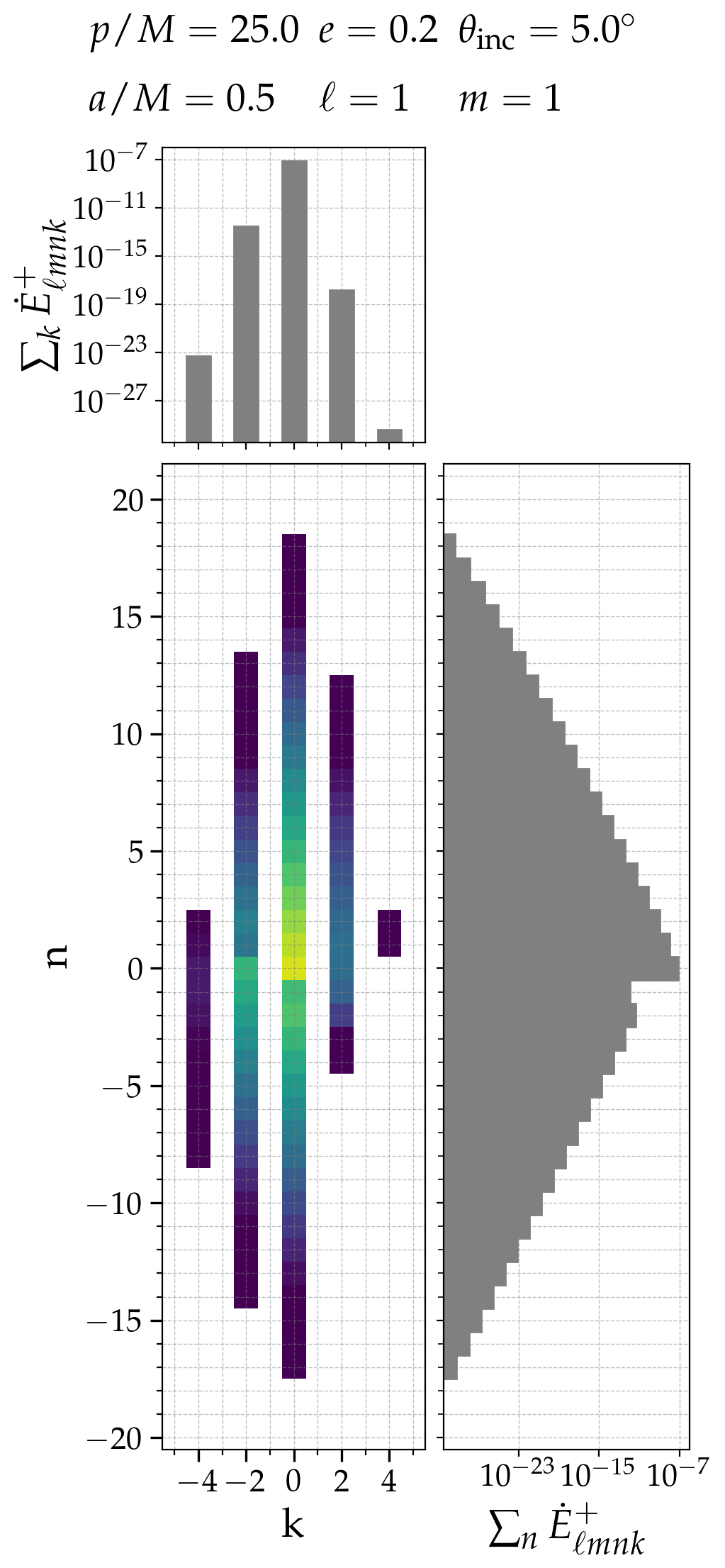}
        \caption{Energy scalar flux at infinity.}
    \end{subfigure}

    \vspace{0.3cm} 

    \begin{subfigure}{0.238\textwidth}
        \centering
        \includegraphics[width=\textwidth]{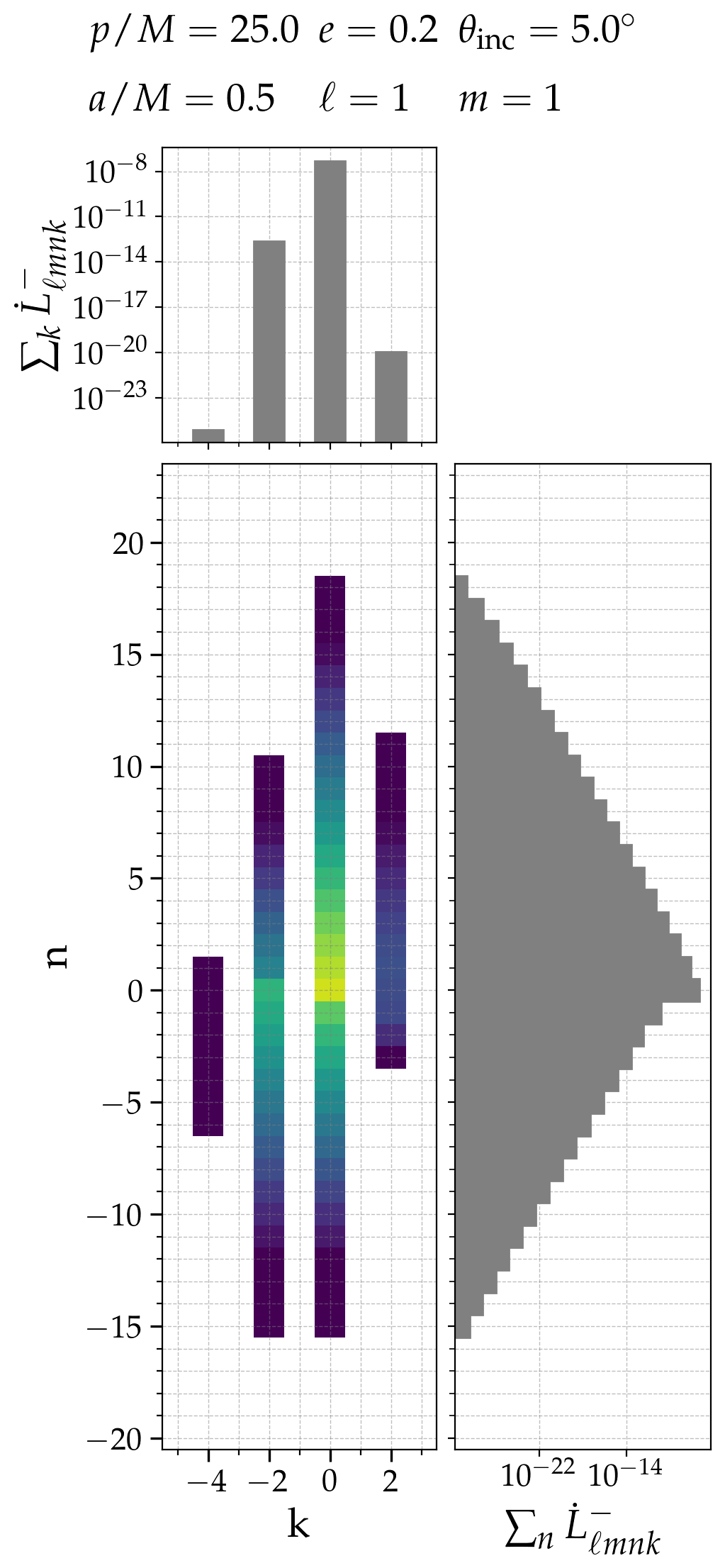}
        \caption{Angular momentum scalar flux at horizon.}
    \end{subfigure}
    \hfill
    \begin{subfigure}{0.238\textwidth}
        \centering
        \includegraphics[width=\textwidth]{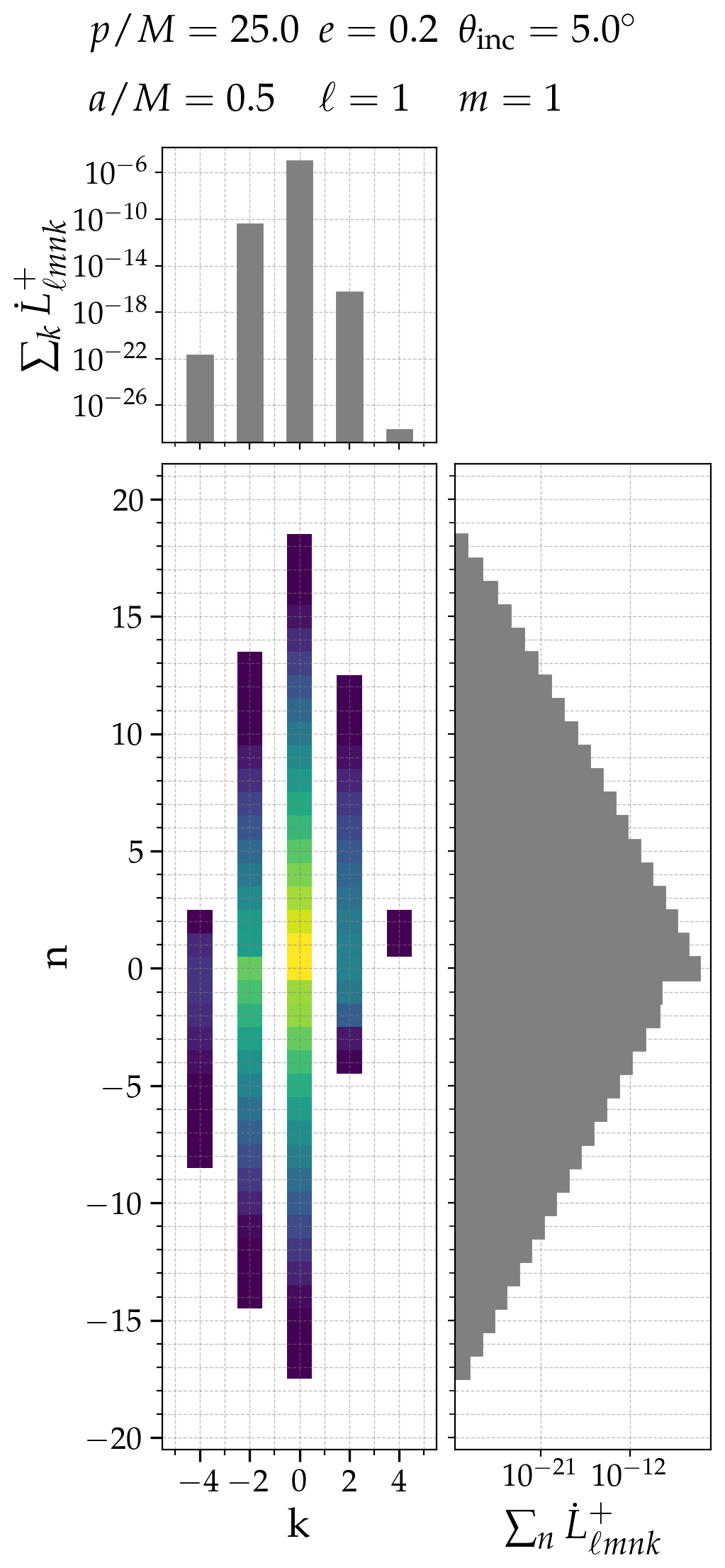}
        \caption{Angular momentum scalar flux at infinity.}
    \end{subfigure}

    \caption{Spectrum of the scalar energy (top panels) and angular momentum (bottom panels)
    fluxes emitted at the horizon (left panels) and at 
    infinity (right panels) for the $(\ell,m)=(1,1)$ 
    harmonic, corresponding to a binary 
    configuration with $(a/M,e,p/M,\theta_{\rm inc})=(0.5,0.2,25,5^\circ)$. The central panel in each 
    plot shows the mode contributions in the $(n,k)$ plane, with the color scale indicating the 
    magnitude of the single-mode flux. The top (right) 
    panels display the {\it marginalized} energy 
    flux contributions, obtained for a given 
    $k$ ($n$) by summing over the index $n$ ($k$).}
    \label{fig:flux_horizon_infinity}
    \end{figure}

Figure~\ref{fig:flux_horizon_infinity} shows a 
representative example of the scalar energy and angular 
momentum fluxes at the horizon and at infinity for a 
binary configuration with 
$(a/M,e,p/M,\theta_{\rm inc})=(0.5,0.2,25,5^\circ)$. 
Each panel displays a two-dimensional distribution 
of mode contributions in the 
$(n,k)$ plane, together with the corresponding 
marginal distributions, which are constructed 
by summing the computed mode contributions 
over $n$ or $k$, respectively. Finally, the 
color scale indicates the magnitude of the 
single-mode flux.

%
\subsection{Mode hierarchy across \texorpdfstring{$\ell$}{l} and \texorpdfstring{$m$}{m}}
%

We begin by analyzing the dependence of the scalar 
flux on the harmonic indices $(\ell,m)$. To this end, 
we fix the orbital configuration to 
$(a/M,e,p/M,\theta_{\rm inc})=(0.3,0.5,7,30^\circ)$, 
and have verified that the qualitative features 
described below persist across different setups.

\begin{figure}[hbpt!]
\centering
\includegraphics[scale=0.32]{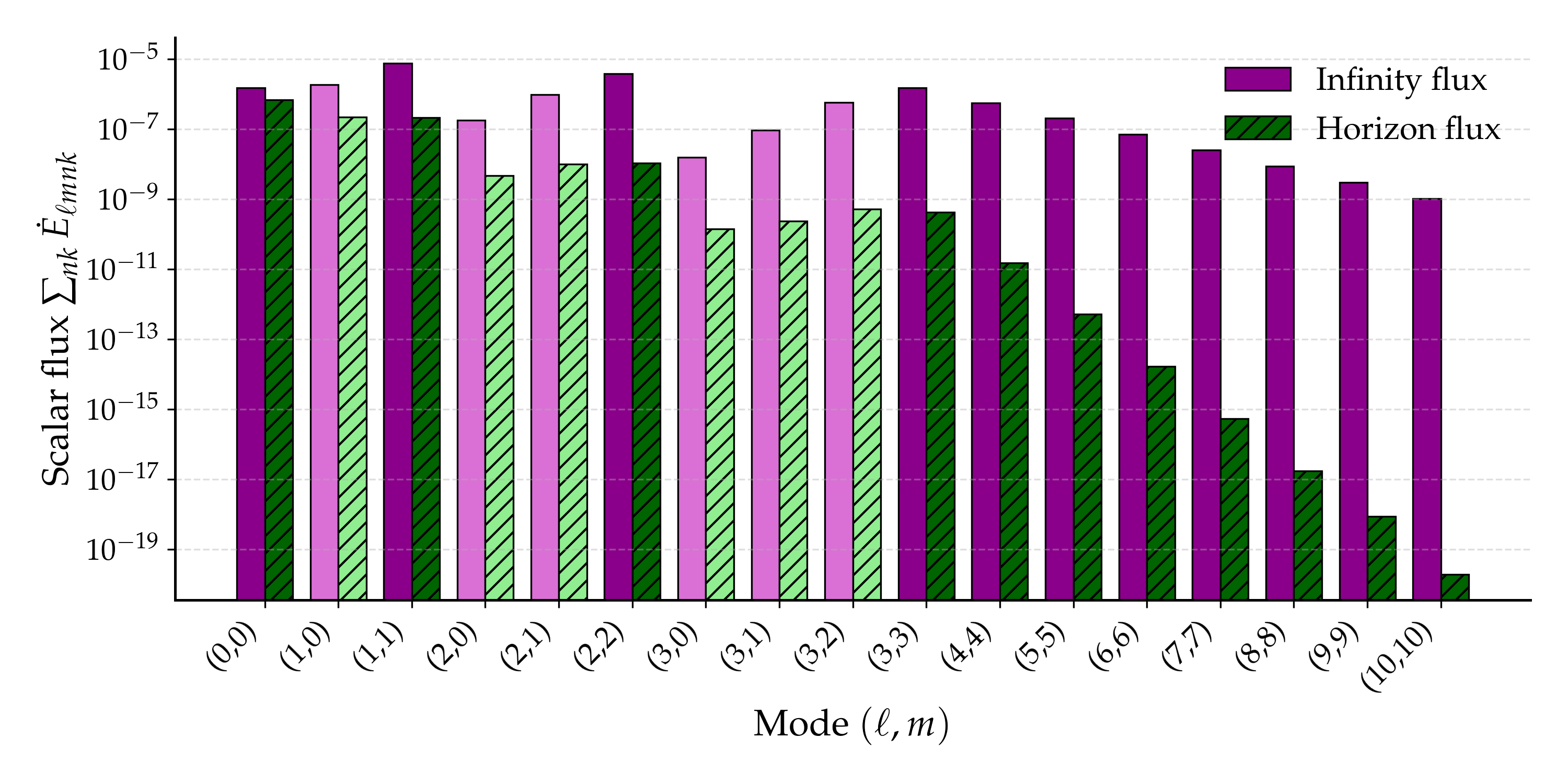}
\includegraphics[scale=0.32]{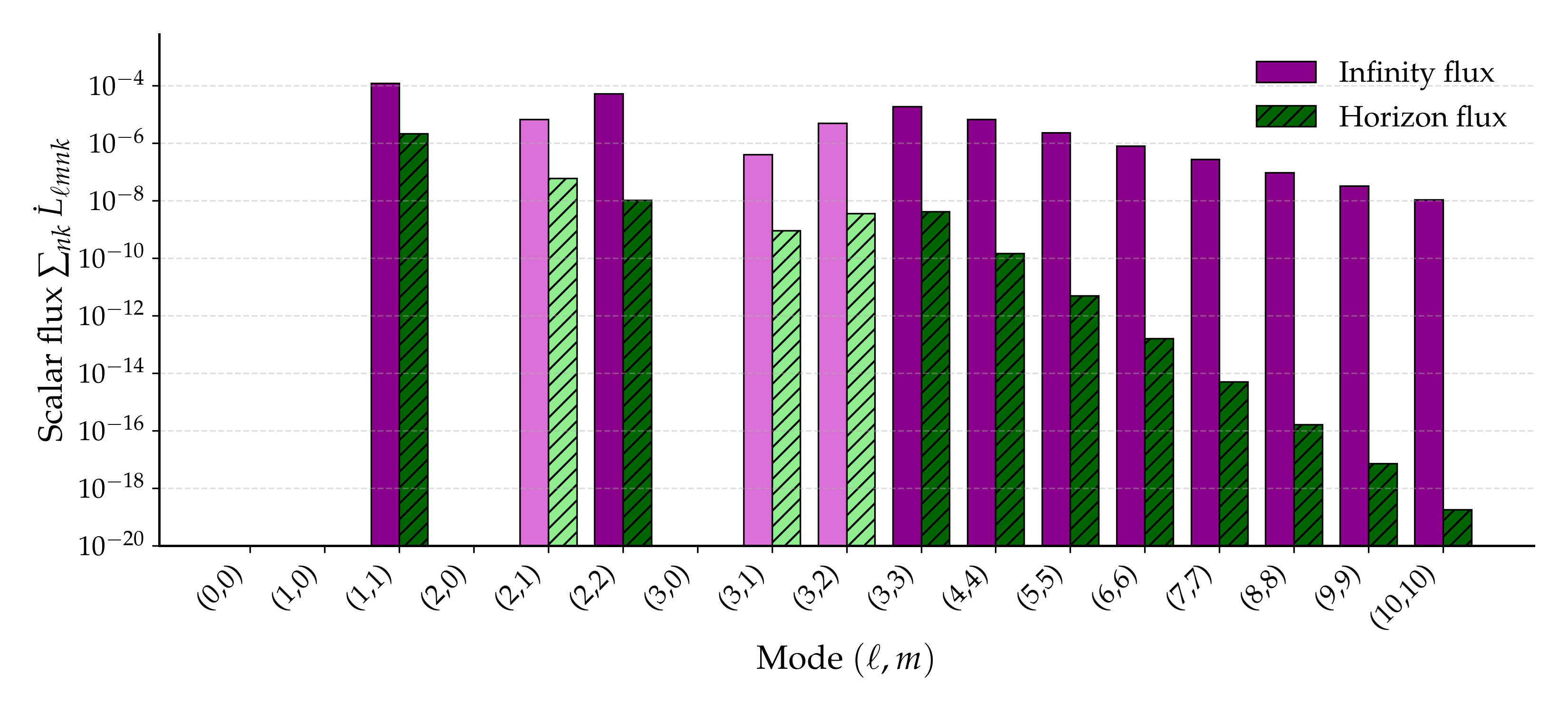}
\caption{Bar plot of the energy (top panel) and angular momentum 
(bottom panel) fluxes at infinity and at the horizon, 
for scalar modes with different angular indices $(\ell,m)$, 
summed over the $(n,k)$ harmonics. Darker (lighter) shades 
denote modes with $\ell = m$ ($\ell \neq m$), respectively.}\label{fig:histogramfluxes}
\end{figure}

Figure~\ref{fig:histogramfluxes} presents bar plots of 
the energy and angular momentum fluxes, summed over the 
harmonic indices $n$ and $k$, for selected $(\ell,m)$ modes. 
The corresponding numerical values are listed in 
Tables~\ref{tab:scalar_spetra_l_m}-\ref{tab:angular_momentum_scalar_spetra_l_m}. 
Representative spectra for these modes are shown in 
Figs.~\ref{fig:flux_grid_a_0.9_horizon_Edot}-\ref{fig:flux_grid_a_0.9_infinity_Edot} 
and \ref{fig:flux_grid_a_0.9_horizon_Ldot}-\ref{fig:flux_grid_a_0.9_infinity_Ldot} 
of Appendix~\ref{app:scalar_spectra}.

The fluxes for $(\ell,m)$ and $(\ell,-m)$ are identical, 
as a consequence of the invariance under the transformation~\cite{Hughes:2021exa}
\begin{equation}
(\ell,m,n,k)\longrightarrow(\ell,-m,-n,-k)\ ,
\end{equation}
which ensures symmetry under $m \to -m$ after summation 
over $n$ and $k$. Accordingly, Table~\ref{tab:scalar_spetra_l_m} 
includes only modes with $m \ge 0$.

\begin{table}[]
\begin{tabular}{c|c|c}
\hline
\hline
$(\ell, m)$ & $\sum_{n,k}\dot{E}^+_{\ell m n k}$  & $\sum_{n,k}\dot{E}^-_{\ell m n k}$       \\ 
\hline
\hline
(0,0)       & $1.519 \times 10^{-6}$    & $6.829 \times 10^{-7}$     \\ \hline
(1,0)       & $1.851 \times 10^{-6}$    & $2.236 \times 10^{-7}$     \\ \hline
(1,1)       & $7.682 \times 10^{-6}$    & $2.119 \times 10^{-7}$     \\ \hline
(2,0)       & $1.810 \times 10^{-7}$    & $4.789 \times 10^{-9}$     \\ \hline
(2,1)       & $9.658 \times 10^{-7}$    & $9.943 \times 10^{-9}$     \\ \hline
(2,2)       & $3.836 \times 10^{-6}$    & $1.071 \times 10^{-8}$     \\ \hline
(3,0)       & $1.572 \times 10^{-8}$    & $1.418 \times 10^{-10}$     \\ \hline
(3,1)       & $9.291 \times 10^{-8}$    & $2.363 \times 10^{-10}$     \\ \hline
(3,2)       & $5.832 \times 10^{-7}$    & $5.216 \times 10^{-10}$     \\ \hline
(3,3)       & $1.531 \times 10^{-6}$    & $4.312 \times 10^{-10}$     \\ \hline
(4,4)       & $5.689 \times 10^{-7}$    & $1.543 \times 10^{-11}$     \\ \hline
(5,5)       & $2.045 \times 10^{-7}$    & $5.195 \times 10^{-13}$     \\ \hline
(6,6)       & $7.218 \times 10^{-8}$    & $1.695 \times 10^{-14}$     \\ \hline
(7,7)       & $2.518 \times 10^{-8}$    & $5.444 \times 10^{-16}$     \\ \hline
(8,8)       & $8.720 \times 10^{-9}$    & $1.739 \times 10^{-17}$     \\ \hline
(9,9)       & $3.027 \times 10^{-9}$    & $8.686 \times 10^{-19}$     \\ \hline
(10,10)     & $1.031 \times 10^{-9}$    & $1.938 \times 10^{-20}$     \\ 
\hline
\hline
\end{tabular}
\caption{Scalar energy flux at infinity and at the 
horizon summed over $n$ and $k$ for different $(\ell, m)$ 
modes. Orbital parameters are fixed to $a/M = 0.3$, $e = 0.5$, $p/M = 7$, 
and $\theta_\text{inc}= 30^{\circ}$.}
\label{tab:scalar_spetra_l_m}
\end{table}

\begin{table}[]
\begin{tabular}{c|c|c}
\hline
\hline
$(\ell, m)$ & $\sum_{n,k}\dot{L}^+_{\ell m n k}$  & $\sum_{n,k}\dot{L}^-_{\ell m n k}$       \\ 
\hline
\hline
(0,0)       & $0$                       & $0$     \\ \hline
(1,0)       & $0$                       & $0$     \\ \hline
(1,1)       & $1.237 \times 10^{-4}$    & $2.201 \times 10^{-6}$     \\ \hline
(2,0)       & $0$                       & $0$     \\ \hline
(2,1)       & $6.793 \times 10^{-6}$    & $6.014 \times 10^{-8}$     \\ \hline
(2,2)       & $5.205 \times 10^{-5}$    & $1.059 \times 10^{-8}$     \\ \hline
(3,0)       & $0$                       & $0$     \\ \hline
(3,1)       & $4.019 \times 10^{-7}$    & $9.244 \times 10^{-10}$     \\ \hline
(3,2)       & $4.926 \times 10^{-6}$    & $3.616 \times 10^{-9}$     \\ \hline
(3,3)       & $1.895 \times 10^{-5}$    & $4.149 \times 10^{-9}$     \\ \hline
(4,4)       & $6.688 \times 10^{-6}$    & $1.468 \times 10^{-10}$     \\ \hline
(5,5)       & $2.330 \times 10^{-6}$    & $4.905 \times 10^{-12}$     \\ \hline
(6,6)       & $8.055 \times 10^{-7}$    & $1.591 \times 10^{-13}$     \\ \hline
(7,7)       & $2.770 \times 10^{-7}$    & $5.091 \times 10^{-15}$     \\ \hline
(8,8)       & $9.487 \times 10^{-8}$    & $1.621 \times 10^{-16}$     \\ \hline
(9,9)       & $3.255 \times 10^{-8}$    & $7.222 \times 10^{-18}$     \\ \hline
(10,10)     & $1.105 \times 10^{-8}$    & $1.762 \times 10^{-19}$     \\ 
\hline
\hline
\end{tabular}
\caption{ 
Scalar angular momentum flux at infinity and at the horizon 
summed over $n$ and $k$ for different $(\ell, m)$ modes. 
Orbital parameters are fixed to $a/M = 0.3$, $e = 0.5$, $p/M = 7$, 
and $\theta_\text{inc}= 30^{\circ}$.}
\label{tab:angular_momentum_scalar_spetra_l_m}
\end{table}

The total flux decreases with increasing multipole order 
$\ell$, with the dominant contribution arising from 
$\ell=1$. Modes with $m=0$ do not carry angular momentum, 
and therefore satisfy $\dot{L}_{\ell m n k}^{\pm}=0$.

Consistent with the behavior observed for eccentric orbits 
in Ref.~\cite{Barsanti:2022ana}, we find that for dominant 
modes with $\ell=m$ the location of the spectral peak 
depends strongly on $\ell$. The peak in the $k$ spectrum 
occurs at $k=0$, while the peak in the $n$ spectrum shifts 
toward higher values as $\ell$ increases. 
For clarity, we define $n_{\rm peak}^{\pm}$ and
$k_{\rm peak}^{\pm}$ as the harmonic indices at which
$\dot{E}_{\ell m n k}^{\pm}$ (and equivalently
$\dot{L}_{\ell m n k}^{\pm}$) attain their maximum values.
Table~\ref{tab:dominant_modes} lists the harmonic index 
$n$ corresponding to the maximum flux for 
$\ell=m=0,1,\dots,10$, at both infinity and the horizon.

For subdominant modes with $\ell \ne m$, the spectral 
structure is more intricate. 
Table~\ref{tab:subdominant_modes} summarizes the peak 
locations of $\dot{E}_{\ell m n k}^{\pm}$ and 
$\dot{L}_{\ell m n k}^{\pm}$. The maximum along $n$ is 
primarily determined by $\ell$ and depends only weakly 
on $m$, whereas the peak along $k$ is controlled by $m$. 
For all $(\ell,m)$ pairs analyzed, the maximum in $k$ 
occurs at $k=\ell-m$, providing a practical guideline 
for selecting the initial $k$ range when constructing 
scalar-flux grids for waveform generation.

Finally, the values of $n_{\rm peak}^{\pm}$
are nearly identical for the energy and angular momentum 
fluxes, for both $\ell=m$ and $\ell \ne m$, across all 
configurations analyzed.

\begin{table}[t]
\centering
\begin{tabular}{c|c c}
\hline
\hline
$\ell=m$ & $n_{\rm peak}^{+}$ & $n_{\rm peak}^{-}$ \\ \hline
\hline
0  &  1 & 1  \\
1  &  1(1) & 4(3) \\
2  &  4(4)& 6(6)  \\
3  &  6(6) & 10(9) \\
4  &  9(9) & 13(12) \\
5  & 12(12) & 16(16) \\
6  & 15(14) & 19(19) \\
7  & 17(17) & 22(22) \\
8  & 20(20) & 26(25) \\
9  & 23(23) & 49(49) \\
10 & 26(25) & 32(32) \\
\hline
\hline
\end{tabular}
\caption{Location of the maximum in the $n$ spectrum 
for dominant modes with $\ell=m$ at $k=0$, for fluxes 
of energy at infinity ($n^+_\tn{peak}$) and at horizon ($n^-_\tn{peak}$). Values within 
round brackets correspond to the location of maxima 
for the fluxes of angular momentum. We assume a binary 
system with $a/M = 0.3$, $e = 0.5$, $p/M = 7$, and $\theta_\text{inc}= 30^{\circ}$.}
\label{tab:dominant_modes}
\end{table}

\begin{table}[t]
\centering
\begin{tabular}{c c|c c|c c}
\hline
\hline
\multicolumn{2}{c|}{} &
\multicolumn{2}{c|}{Infinity flux} &
\multicolumn{2}{c}{Horizon flux} \\
$\ell$ & $m$ & $k^+_{\rm peak}$ & $n^+_{\rm peak}$ & $k^-_{\rm peak}$ & $n^-_{\rm peak}$ \\ \hline\hline
1 & 0 & 1 & 1 & 1 & 3 \\
2 & 0 & 2 & 3 & 2 & 5 \\
2 & 1 & 1(1) & 4(3) & 1(1) & 6(5) \\
3 & 0 & 3 & 6 & 3 & 8 \\
3 & 1 & 2(2) & 6(6) & 2(2) & 9(8) \\
3 & 2 & 1(1) & 6(6) & 1(1) & 9(9) \\
\hline
\hline
\end{tabular}
\caption{Peak locations $(k^\pm_{\rm peak},n^{\pm}_{\rm peak})$ 
of modes with $\ell\ne m$ for fluxes of energy at 
infinity and at the horizon. Values within round 
brackets refer to peaks for fluxes of angular momentum.
Binary parameters are fixed to $a/M = 0.3$, $e = 0.5$, 
$p/M = 7$, and $\theta_\text{inc}= 30^{\circ}$.}
\label{tab:subdominant_modes}
\end{table}

%
\subsection{Systematic exploration of the orbital-parameter dependence}
%

\begin{figure}[t]
    \centering

    \begin{subfigure}{0.238\textwidth}
        \centering
        \includegraphics[width=\textwidth]{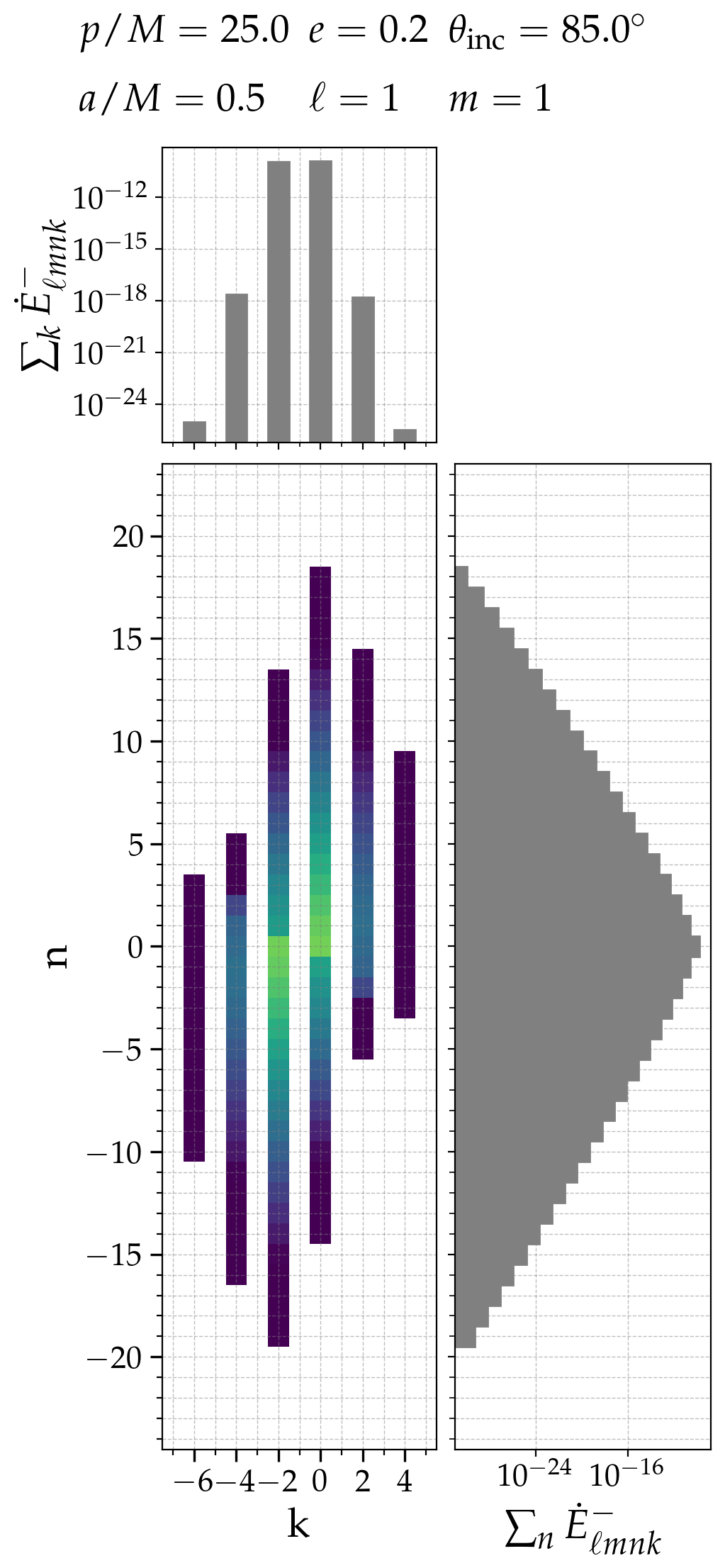}
        \caption{Energy scalar flux at the horizon.}
        \label{fig:flux_horizon_theta}
    \end{subfigure}
    \hfill
    \begin{subfigure}{0.238\textwidth}
        \centering
        \includegraphics[width=\textwidth]{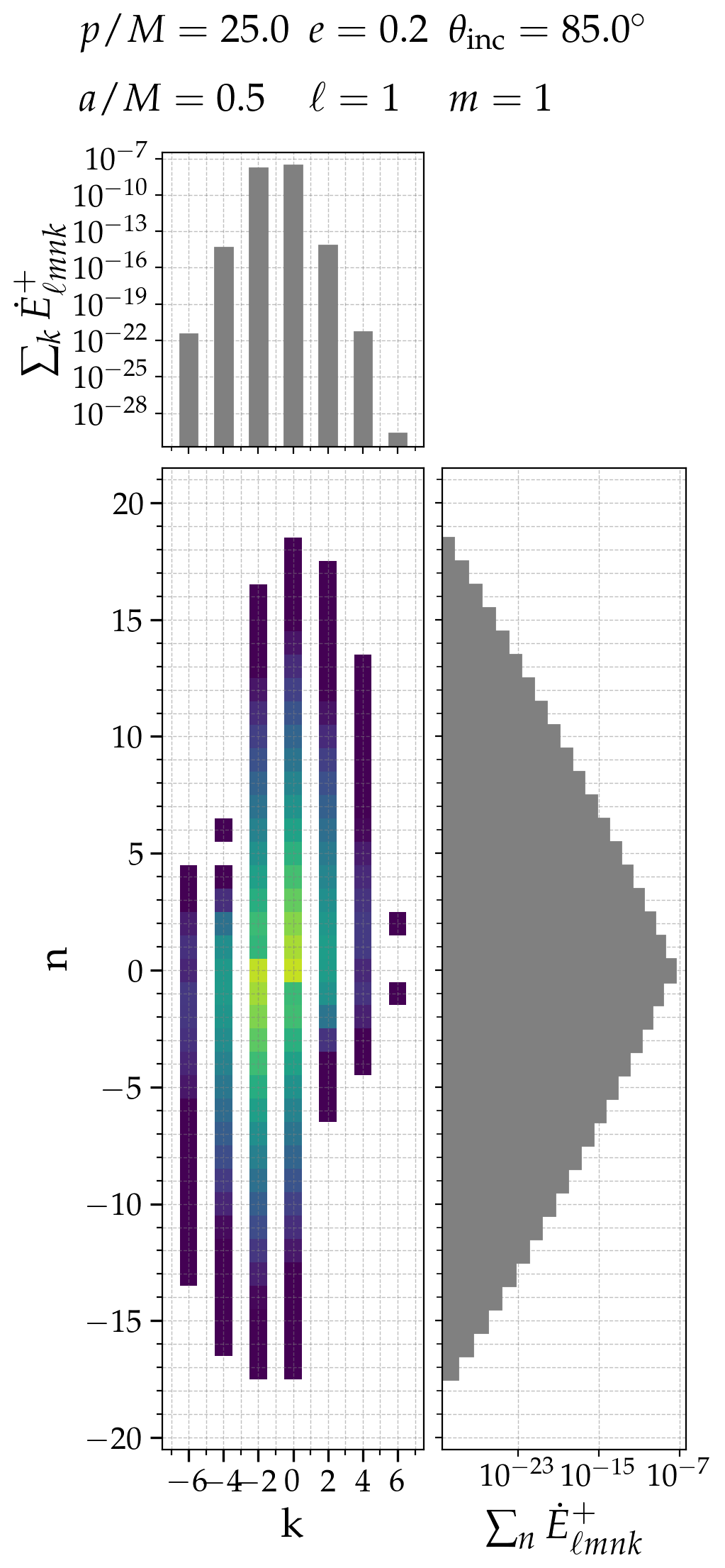}
        \caption{Energy scalar flux at infinity.}
        \label{fig:flux_infinity_theta}
    \end{subfigure}
    \caption{Same as panels (a) and (b) of Fig.~\ref{fig:flux_horizon_infinity}, 
    but for $\theta_{\rm inc}=85^\circ$.}
    \label{fig:flux_horizon_infinity_theta}
\end{figure}

\begin{figure}[ht]
    \centering

    \begin{subfigure}{0.238\textwidth}
        \centering
        \includegraphics[width=\textwidth]{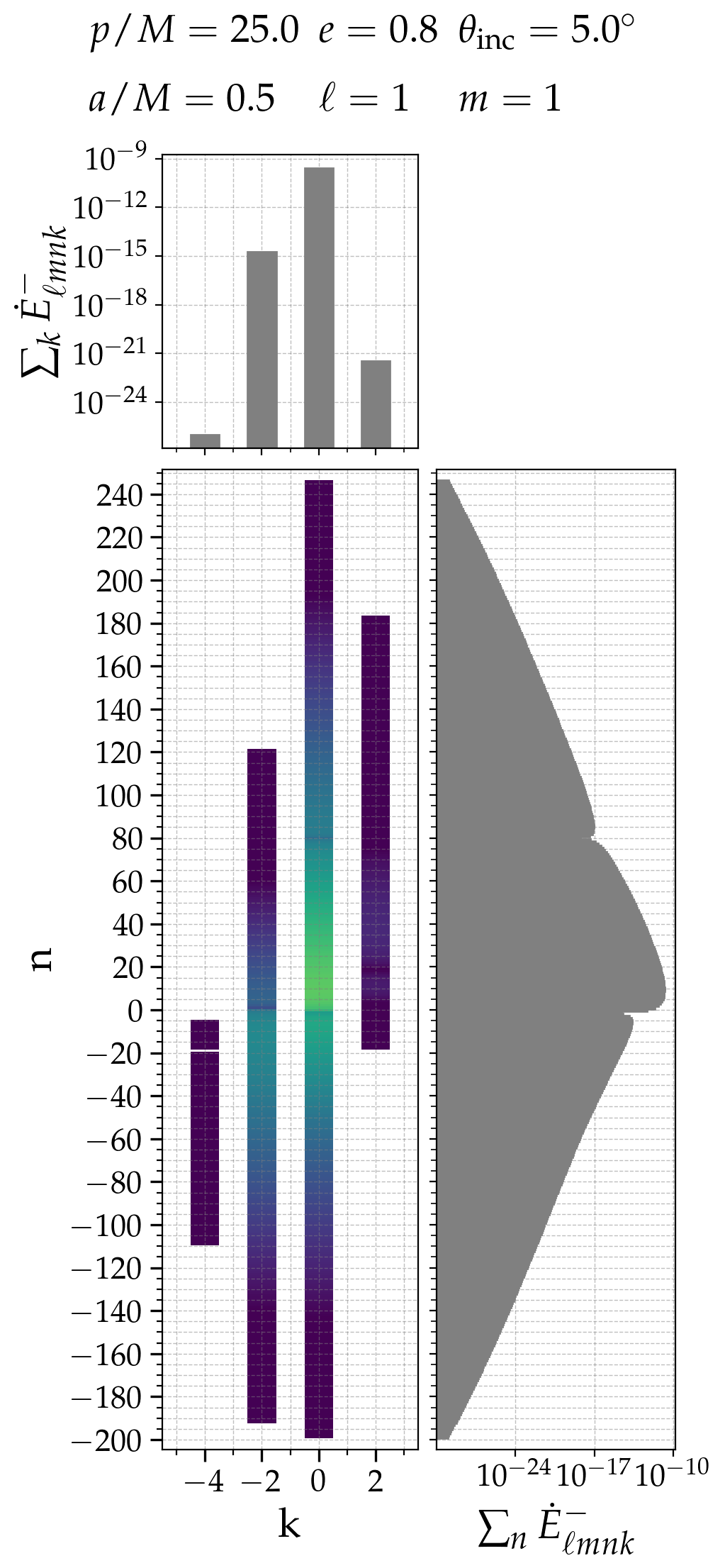}
        \caption{Energy scalar flux at the horizon.}
        \label{fig:flux_horizon_ecc}
    \end{subfigure}
    \hfill
    \begin{subfigure}{0.238\textwidth}
        \centering
        \includegraphics[width=\textwidth]{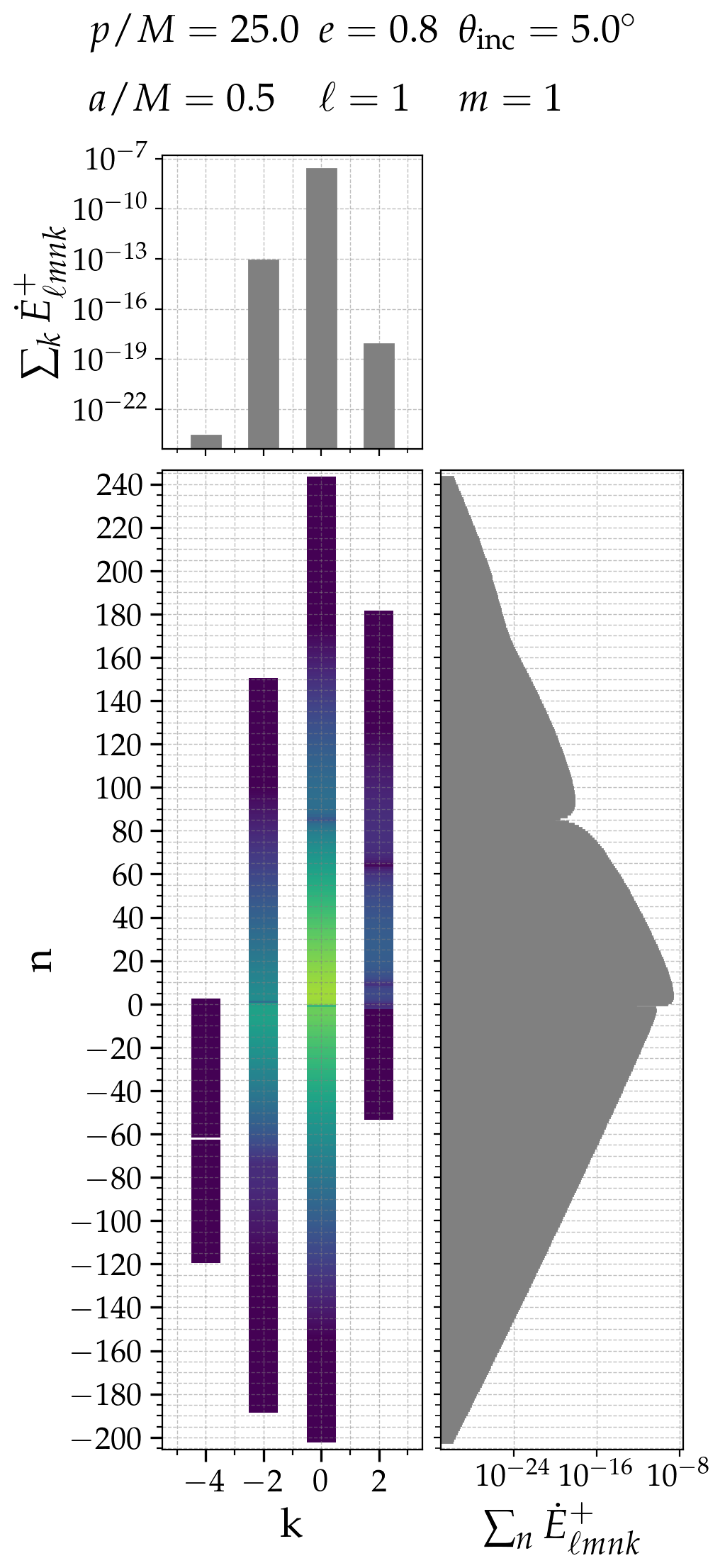}
        \caption{Energy scalar flux at infinity.}
        \label{fig:flux_infinity_ecc}
    \end{subfigure}

  \caption{Same as panels (a) and (b) of Fig.~\ref{fig:flux_horizon_infinity}, 
    but for $e = 0.8$.}
    \label{fig:flux_horizon_infinity_ecc}
\end{figure}

As we have verified that the radiation is predominantly 
dipolar, we begin by constructing two–dimensional mode 
grids in the $(n,k)$ plane for the $(\ell,m)=(1,1)$ 
harmonic across a representative set of orbital 
configurations. We extract the scalar spectra for 
all possible combinations of the following orbital 
parameters:
\begin{equation}
    \begin{split}
        &a/M=\{0.1,0.5,0.9\} \ ,\qquad 
        e=\{0.2,0.4,0.8\}\ ,\\
        &\theta_{\rm inc}=\{5^\circ,40^\circ,85^\circ\} \ ,\qquad
        p/M=\{8,25\} \ .
    \end{split}
\end{equation}
For illustrative purposes, we present in the main 
text figures corresponding to a subset of these configurations, 
focusing in particular on the impact of variations 
in inclination and eccentricity on the resulting flux 
spectra. The trends discussed below are observed 
throughout the entire parameter space explored 
(see Appendix~\ref{app:scalar_spectra}). 
The behavior of the energy and angular momentum 
spectra is also remarkably similar (cf.\ top and 
bottom rows of Fig.~\ref{fig:flux_horizon_infinity}). 
For this reason, hereafter we only discuss the 
dependence of $\dot{E}$ on the binary parameters:
\begin{itemize}
    \item \textbf{Inclination angle.} 
    By comparing Fig.~\ref{fig:flux_horizon_infinity} and Fig.~\ref{fig:flux_horizon_infinity_theta}, 
    we observe that increasing the inclination angle $\theta_{\rm inc}$ leads to a broadening of the distribution 
    in the $k$ direction, reflecting the richer harmonic structure of strongly inclined orbits. 
    Despite this broadening, the flux remains strongly peaked at $k=0$, 
    with only a narrow range of $k$ modes contributing significantly.
    \item \textbf{Eccentricity.}  
    Increasing the eccentricity $e$ substantially broadens the spectrum along the $n$ direction 
    (see Fig.~\ref{fig:flux_horizon_infinity_ecc}). 
    For low eccentricity, the dominant contribution arises from the $n=0$ mode, 
    whereas at high eccentricity the peak shifts to $|n|\neq 0$, 
    often accompanied by secondary maxima. 
    This behavior becomes increasingly pronounced in relativistically strong–field configurations, 
    namely for smaller orbital separations or highly inclined orbits.
    \item \textbf{Orbital radius.}  
    For larger values of $p$, the total flux summed over $(n,k)$ decreases, 
    as expected for orbits located further from the primary BH. 
    This trend is quantified in Table~\ref{tab:scalar_spectra_p_e_theta}, 
    which reports the total fluxes $\dot{E}^{\pm}$ 
    for all configurations analyzed.
\end{itemize}

As discussed in previous studies~\cite{Barsanti:2022ana,DellaRocca:2024pnm}, 
the structure of the scalar flux spectra becomes progressively richer as the complexity of the orbital motion increases. 
Overall, the simultaneous presence of orbital inclination 
and eccentricity does not qualitatively alter 
the main spectral features reported in the literature for 
eccentric equatorial orbits and circular inclined orbits.

For instance, at $a/M=0.5$, the dominant harmonic in $k$ 
is $k=0$ for all values of $n$ when $\theta_{\rm inc}=5^\circ$ 
and $40^\circ$. For highly inclined orbits 
($\theta_{\rm inc}=85^\circ$), the dominant contribution 
shifts to $k=2$ for $n \lesssim 30$.

At fixed $k$, the location of the peak in the $n$ spectrum 
exhibits only a weak dependence on inclination. 
At $a/M=0.5$ and moderate eccentricity ($e=0.2$), 
the dominant contribution to the energy flux at infinity 
always arises from the $(n,k)=(0,0)$ mode, independently 
of the inclination angle. At higher eccentricity ($e=0.8$), 
the position of the dominant peak depends on inclination, 
with the leading contribution corresponding to $k=0$ 
and $n=5$, $6$, and $9$ for $\theta_{\rm inc}=5^\circ$, $40^\circ$, and $85^\circ$, respectively. 
For the $k=2$ harmonic at $e=0.8$, the dominant contribution 
in $n$ varies only mildly -- from $n=34$ to $n=36$ -- despite 
the large change in inclination. A similar weak dependence 
on inclination is observed for the energy flux absorbed 
at the horizon.

These features can be exploited in the construction of 
flux grids for the evolution of asymmetric binaries, 
providing practical guidance for the choice of summation 
strategies over the harmonic indices.

\begin{table*}[!htb]
\footnotesize
\centering
\begin{tabular}{c|@{\hskip 3pt}c@{\hskip 5pt}|c@{\hskip 5pt}|c@{\hskip 5pt}|c}
\hline\hline
\multicolumn{5}{c}{$a/M = 0.1$} \\ 
\hline
$p/M$ & $\theta_{\rm inc}$ & $e=0.2$ & $e=0.4$ & $e=0.8$ \\ 
\hline
8 & 5$^\circ$ & 6.090$\times 10^{-6}$ (7.374$\times 10^{-8}$) & 5.804$\times 10^{-6}$ (1.710$\times 10^{-7}$) & 3.087$\times 10^{-6}$ (3.944$\times 10^{-7}$) \\
  & 40$^\circ$ & 4.911$\times 10^{-6}$ (6.131$\times 10^{-8}$) & 4.711$\times 10^{-6}$ (1.406$\times 10^{-7}$) & 2.583$\times 10^{-6}$ (3.277$\times 10^{-7}$) \\
  & 85$^\circ$ & 3.159$\times 10^{-6}$ (7.040$\times 10^{-8}$) & 3.083$\times 10^{-6}$ (1.377$\times 10^{-7}$) & 1.870$\times 10^{-6}$ (3.020$\times 10^{-7}$) \\
\hline
25 & 5$^\circ$ & 8.592$\times 10^{-8}$ (5.985$\times 10^{-11}$) & 7.473$\times 10^{-8}$ (5.243$\times 10^{-11}$) & 2.639$\times 10^{-8}$ (3.159$\times 10^{-11}$) \\
   & 40$^\circ$ & 6.858$\times 10^{-8}$ (4.877$\times 10^{-11}$) & 5.969$\times 10^{-8}$ (4.352$\times 10^{-11}$) & 2.113$\times 10^{-8}$ (2.693$\times 10^{-11}$) \\
   & 85$^\circ$ & 4.362$\times 10^{-8}$ (4.689$\times 10^{-11}$) & 3.800$\times 10^{-8}$ (5.321$\times 10^{-11}$) & 1.349$\times 10^{-8}$ (4.292$\times 10^{-11}$) \\
\hline\hline
\multicolumn{5}{c}{$a/M = 0.5$} \\ 
\hline
8 & 5$^\circ$ & 5.610$\times 10^{-6}$ (1.626$\times 10^{-7}$) & 5.014$\times 10^{-6}$ (1.577$\times 10^{-7}$) & 2.049$\times 10^{-6}$ (9.286$\times 10^{-8}$) \\
  & 40$^\circ$ & 4.698$\times 10^{-6}$ (1.535$\times 10^{-7}$) & 4.275$\times 10^{-6}$ (1.363$\times 10^{-7}$) & 1.856$\times 10^{-6}$ (8.634$\times 10^{-8}$) \\
  & 85$^\circ$ & 3.159$\times 10^{-6}$ (1.659$\times 10^{-7}$) & 3.038$\times 10^{-6}$ (2.326$\times 10^{-7}$) & 1.653$\times 10^{-6}$ (2.877$\times 10^{-7}$) \\
\hline
25 & 5$^\circ$ & 8.472$\times 10^{-8}$ (4.352$\times 10^{-10}$) & 7.328$\times 10^{-8}$ (4.683$\times 10^{-10}$) & 2.537$\times 10^{-8}$ (2.663$\times 10^{-10}$) \\
   & 40$^\circ$ & 6.815$\times 10^{-8}$ (3.490$\times 10^{-10}$) & 5.912$\times 10^{-8}$ (3.768$\times 10^{-10}$) & 2.067$\times 10^{-8}$ (2.165$\times 10^{-10}$) \\
   & 85$^\circ$ & 4.364$\times 10^{-8}$ (2.460$\times 10^{-10}$) & 3.801$\times 10^{-8}$ (2.795$\times 10^{-10}$) & 1.349$\times 10^{-8}$ (1.872$\times 10^{-10}$) \\
\hline\hline
\multicolumn{5}{c}{$a/M = 0.9$} \\ 
\hline
8 & 5$^\circ$ & 5.242$\times 10^{-6}$ (3.108$\times 10^{-7}$) & 4.509$\times 10^{-6}$ (3.171$\times 10^{-7}$) & 1.605$\times 10^{-6}$ (1.673$\times 10^{-7}$) \\
  & 40$^\circ$ & 4.498$\times 10^{-6}$ (2.570$\times 10^{-7}$) & 3.954$\times 10^{-6}$ (2.691$\times 10^{-7}$) & 1.509$\times 10^{-6}$ (1.548$\times 10^{-7}$) \\
  & 85$^\circ$ & 3.145$\times 10^{-6}$ (2.784$\times 10^{-7}$) & 2.967$\times 10^{-6}$ (3.706$\times 10^{-7}$) & 1.455$\times 10^{-6}$ (3.746$\times 10^{-7}$) \\
\hline
25 & 5$^\circ$ & 8.360$\times 10^{-8}$ (7.850$\times 10^{-10}$) & 7.193$\times 10^{-8}$ (8.461$\times 10^{-10}$) & 2.444$\times 10^{-8}$ (4.846$\times 10^{-10}$) \\
   & 40$^\circ$ & 6.769$\times 10^{-8}$ (6.302$\times 10^{-10}$) & 5.852$\times 10^{-8}$ (6.816$\times 10^{-10}$) & 2.021$\times 10^{-8}$ (3.946$\times 10^{-10}$) \\
   & 85$^\circ$ & 4.361$\times 10^{-8}$ (4.416$\times 10^{-10}$) & 3.798$\times 10^{-8}$ (5.012$\times 10^{-10}$) & 1.347$\times 10^{-8}$ (3.345$\times 10^{-10}$) \\
\hline\hline
\end{tabular}
\caption{Values of the scalar flux at infinity (horizon) 
summed over the $n$ and $k$ harmonics for the $(\ell, m) = (1,1)$ mode. For each spin we consider different values of the 
binary semi-latus rectum $p/M$, inclination $\theta_{\rm inc}$, 
and eccentricity $e$.}
\label{tab:scalar_spectra_p_e_theta}
\end{table*}

\begin{figure}[htbp]
    \centering
    \begin{subfigure}{0.238\textwidth}
        \centering
        \includegraphics[width=\textwidth]{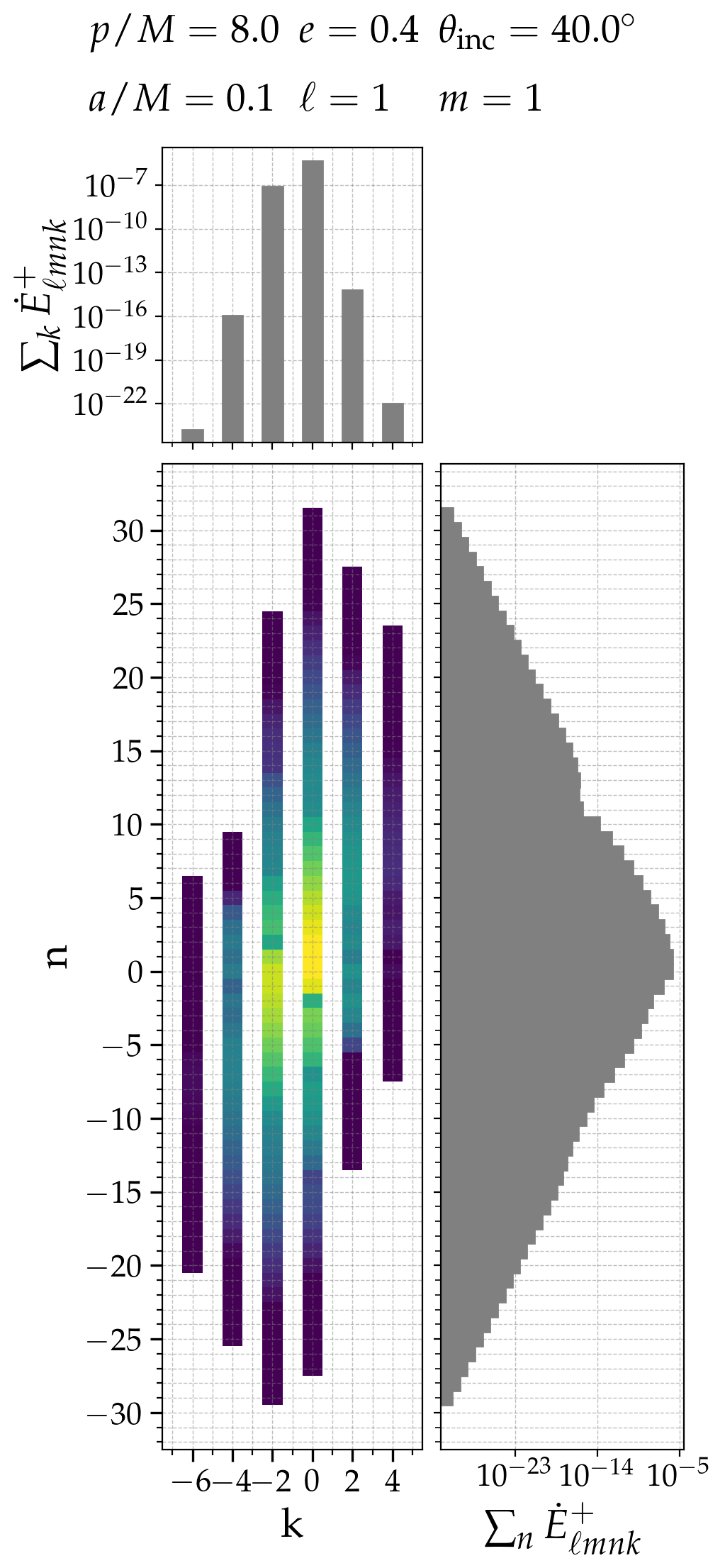}
        \caption{Scalar flux at infinity.}
        \label{fig:flux_horizon_spin}
    \end{subfigure}
    \hfill
   \begin{subfigure}{0.238\textwidth}
        \centering
        \includegraphics[width=\textwidth]{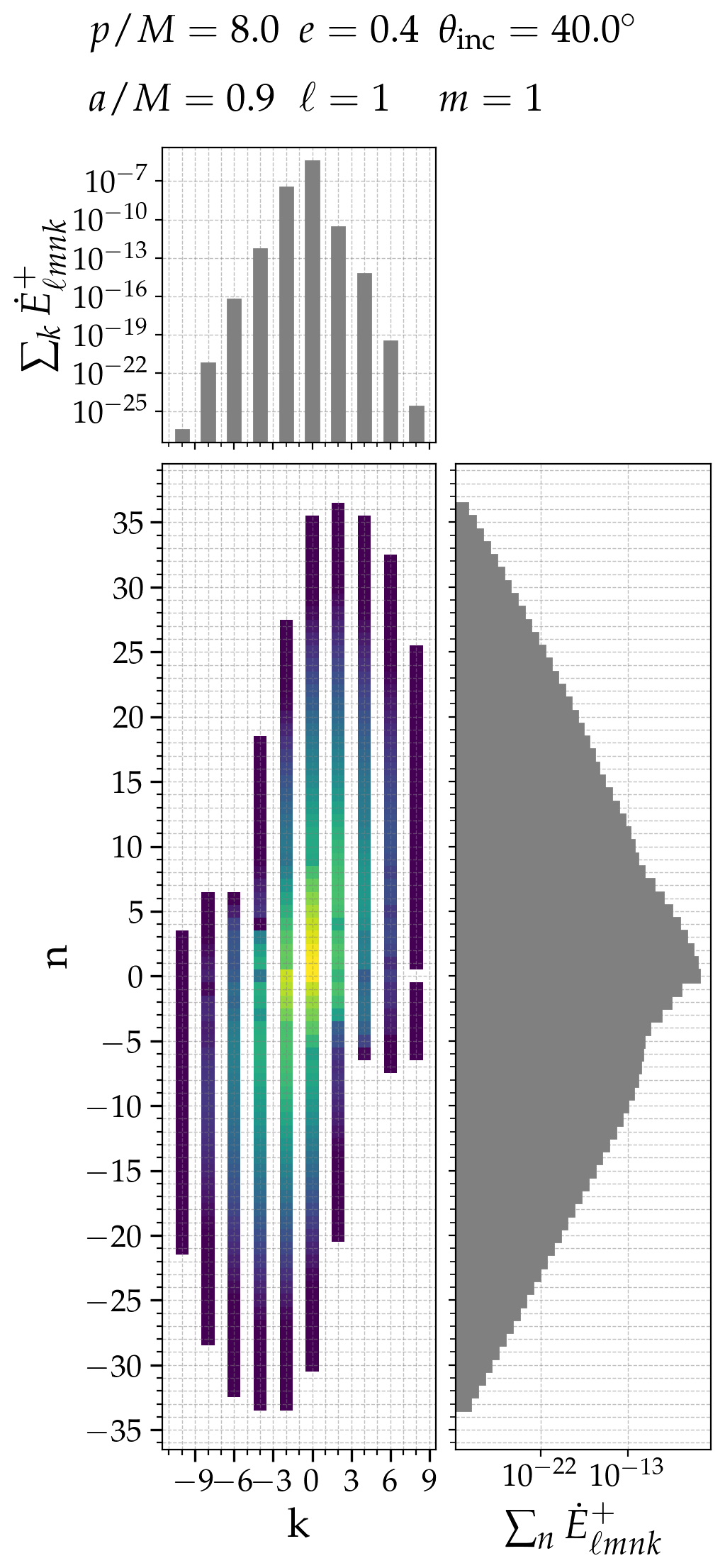}
        \caption{Scalar flux at infinity.}
        \label{fig:flux_infinity_spin}
    \end{subfigure}
    \caption{Spectrum of the scalar energy flux 
    emitted at infinity for the $(\ell,m)=(1,1)$ harmonic. 
    The binary parameters are $(e,p/M,\theta_{\rm inc})=(0.4,8,40^\circ)$. The left and right panels correspond 
    to primary spins $a/M=0.1$ and $a/M=0.9$, respectively. 
    The layout of the panels follows the same format as Fig.~\ref{fig:flux_horizon_infinity_ecc}.}
    \label{fig:flux_horizon_infinity_spin}
\end{figure}

\subsection{Dependence on black-hole spin}
%
To characterize the dependence of the energy flux on the 
BH spin, we compare scalar energy spectra 
computed at different values of $a/M$, while keeping 
$(p/M,e,\theta_{\rm inc})$ fixed. We find that the modal 
structure is remarkably insensitive to the spin. 
As a general feature, the distribution in $n$ remains 
nearly unchanged across the sampled values of $a$, 
while the width of the spectrum in $k$ increases only 
mildly at high spin.

As a representative example, 
Fig.~\ref{fig:flux_horizon_infinity_spin} shows the flux 
at infinity for $(p/M,e,\theta_{\rm inc})=(8,0.4,40^\circ)$ 
and two values of the spin, $a/M=0.1$ and $a/M=0.9$. 
In both cases, the emission is dominated by the 
$k=0$ modes with $n=0,1$, while all other contributions 
are suppressed by at least one order of magnitude. 

A broader exploration of the parameter space 
(see Appendix~\ref{app:scalar_spectra}) indicates that 
the spin can induce a modest shift in the location of 
the spectral peak at higher eccentricity. 
For instance, at $(p/M,e,\theta_{\rm inc})=(8,0.8,40^\circ)$, 
the dominant contribution corresponds to $(k,n)=(0,8)$ 
for $a/M=0.1$, while it shifts to $(k,n)=(0,5)$ for 
$a/M=0.9$.

Overall, the harmonic content of the scalar flux is far 
more strongly influenced by the orbital eccentricity 
and inclination than by the BH spin.

%
\section{Conclusion and outlook}\label{sec:concl}
%
In this study, we have taken a concrete step toward 
the systematic modeling of asymmetric binaries 
in theories of gravity endowed with a massless 
scalar degree of freedom. Considering fully generic 
bound orbits, we have computed the scalar energy 
and angular-momentum fluxes both at infinity and 
across the horizon, resolving their complete 
harmonic structure over a broad region of parameter 
space.

\begin{figure*}[ht!]
\centering
\includegraphics[scale=0.5]{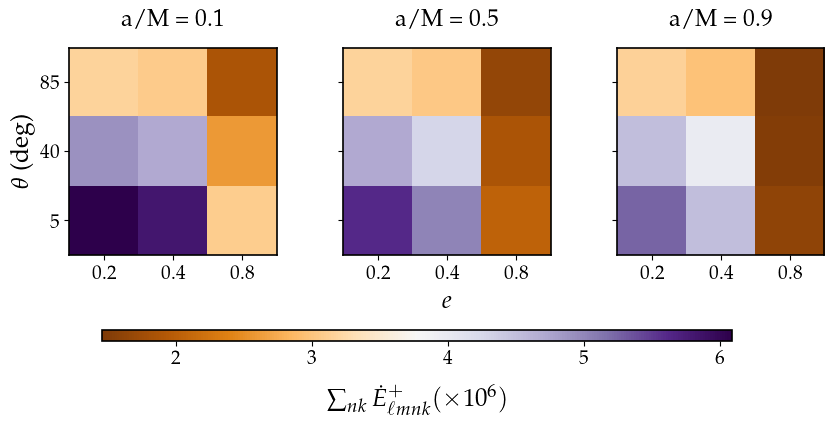}
\caption{Scalar energy flux 
(in units of $10^{-6}$) emitted at infinity 
by binaries at fixed orbital radius, $p/M=8$
shown as a function 
of eccentricity e (horizontal axis) and inclination $\theta_{\rm inc}$ (vertical axis), for three values of 
the spin parameter $a/M$, for $\ell=m=1$. Each panel corresponds 
to a different spin. The color scale indicates the magnitude of the flux, with a common normalization 
across all panels to allow direct comparison.}\label{fig:finalflux}
\end{figure*}

By decomposing the scalar field into frequency-domain modes labeled by the indices $(\ell,m,n,k)$, 
we have characterized the distribution of power 
across the scalar spectrum and identified the 
hierarchy of contributions that governs the emission. 
The radiation is predominantly dipolar, with the 
$\ell = m = 1$ mode providing the leading contribution 
to the total flux. Higher multipoles are progressively suppressed with increasing $\ell$, yet they 
remain non-negligible, especially for eccentric 
and inclined configurations where the spectral 
content broadens and subdominant harmonics become comparatively enhanced. Accurate flux modeling 
therefore requires retaining these additional 
modes, as illustrated in Fig.~\ref{fig:histogramfluxes}.

Focusing on the $\ell=m=1$ component, we have mapped the 
structure of the spectrum in the $(n,k)$ plane. 
Eccentricity leads to a broadening of the distribution of  
the radial harmonic index $n$ in the flux, shifting the spectral 
peak away from $n=0$ at high $e$, while inclination affects the distribution in the polar index $k$, 
increasing the number of contributing harmonics without 
qualitatively altering the dominance of low-$k$ 
modes. The BH spin, by contrast, affects the 
harmonic structure only mildly, inducing at most 
modest shifts in the location of the dominant 
peaks. These systematic trends provide practical 
guidance for constructing flux grids and optimizing 
harmonic summations in waveform models.

Figure~\ref{fig:finalflux} offers a global view of the total scalar energy flux emitted at infinity for $p/M=8$, 
as a function of eccentricity, inclination, and spin. 
As also quantified in Table~\ref{tab:scalar_spectra_p_e_theta}, the 
overall magnitude of the flux varies only moderately 
across the sampled configurations. While the detailed 
spectral distribution is significantly affected even by small changes 
in the orbital geometry, the total emitted power remains 
within a relatively narrow range, and the same qualitative 
behavior is observed in the fluxes absorbed at the horizon.

Since the instantaneous flux at a given orbital 
radius exhibits limited variation across the 
geometric parameter space, the impact of scalar 
charge constraints on the binary geometry will then 
depend 
on the cumulative effect of scalar radiation over the 
full inspiral, and on the coupled evolution of 
eccentricity, inclination, and orbital radius.\\

The results presented here provide the essential 
building blocks for investigating the long-term 
evolution and waveform generation of scalar-charged 
EMRIs and IMRIs. 
In our framework, the resulting 
waveforms are exact at the adiabatic order, \emph{i.e.}, 
at first order in the perturbative expansion. 
Achieving this level of accuracy requires the 
construction of dense grids of scalar amplitudes 
and fluxes as functions of the orbital parameters, 
in close analogy with the infrastructure underlying 
state-of-the-art models in GR. 
The analysis carried out in this work delivers precisely 
the information needed to build such grids in the 
scalar sector.

More broadly, although our primary motivation is 
precision modeling of asymmetric binaries beyond 
GR, the numerical framework developed 
here can also serve as a validated tool for 
high-accuracy flux computations within GR itself, 
thereby contributing to the refinement of 
perturbative modeling in both contexts.\\

The next stage of our research program using EMRIs 
and IMRIs to probe fundamental physics is the 
incorporation of the scalar amplitude and flux grids into inspiral 
and waveform-generation frameworks, such as \texttt{FEW}~\cite{Chua:2020stf,Katz:2021yft,Speri:2023jte}. 
In a forthcoming companion work, we will implement the 
grids within an adiabatic inspiral scheme, enabling 
consistent orbital evolution and parameter-estimation 
studies for scalar-charged EMRIs and IMRIs. 

While the fluxes computed here capture the instantaneous emission for given orbital configurations, assessing in full LISA's ability to constrain the scalar charge  
requires modeling the cumulative effect of scalar 
radiation over the complete inspiral. Embedding these 
results into an inspiral framework will allow us to 
explore the parameter space of asymmetric binaries and 
identify configurations that maximize sensitivity to 
scalar charge, including potential “golden” systems 
arising from specific combinations of spin, eccentricity, 
and inclination. Extending this analysis to larger mass 
ratios will also be of interest, particularly to assess 
whether the merger phase may contribute non-negligibly 
to the observable signal~\cite{Roy:2025kra}.

\acknowledgments
M.D.R.~is supported by the MUR FIS2 Advanced Grant ET-NOW 
(CUP:~B53C25001080001) and by the INFN TEONGRAV initiative.
A.M.~acknowledges financial support from MUR PRIN 
Grants No.~2022-Z9X4XS and No.~2020KB33TP.
TPS acknowledges partial support from the STFC Consolidated Grant nos. ST/V005596/1,  ST/X000672/1, and UKRI2492.

\bibliography{sources}

\appendix
%
\section{Elliptic integrals}\label{app:ellipticintegrals}
%
We show here the form of the elliptic integrals used in Sec.~\ref{subsec:orbitalmotiton} to determine 
a solution for Kerr geodesics. The integral of the first 
kind is given by:
\begin{align}
    \text{F}(\phi|m)=\int_0^\phi (1-m \sin^2\theta)^{-1/2}\dd\theta\ ,
\end{align}
for real $m$, $-\pi/2<\phi<\pi/2<$, $m \sin^2\phi < 1$.
This function is the inverse of the Jacobi Amplitude
\begin{align}
    \phi = \text{am}(u|\phi) \quad \text{with} \quad u=F(\phi|m), m<1 \ .
\end{align}
When $\phi = \pi/2$ we get the complete elliptic integral of the first kind
\begin{align}
    \text{K}(m) = F(\pi/2|m) \ .
\end{align}

The elliptic integral of the second kind is
\begin{align}
    \text{E}(\phi|m)=\int_0^\phi (1-m \sin^2\theta)^{1/2}\dd\theta\ ,
\end{align}
for $m<1$, $-\pi/2<\phi<\pi/2<$.
When $\phi = \pi/2$ we get the complete elliptic integral of the second kind
\begin{align}
    \text{E}(m) = E(\pi/2|m)\ .
\end{align}

Finally, the elliptic integral of the third kind is
\begin{align}
    \Pi(n;\phi|m)=\int_0^\phi (1-n \sin^2\theta)^{-1}(1-m \sin ^2 \theta)^{-1/2}\dd\theta\ ,
\end{align}
for real $m,n$, $-\pi/2<\phi<\pi/2<$ and $m\sin^2\phi < 1$.
When $\phi = \pi/2$ we get the complete elliptic integral of the third kind
\begin{align}
    \Pi(n|m) = \Pi(n;\pi/2|m) \ .
\end{align}
%

%
\section{Functions describing the orbital motion }\label{app:OrbitalMotion}
%
We collect here the explicit expressions for the orbital 
motion functions introduced in Section \ref{subsec:orbitalmotiton}.
The expressions defined in Eq.~\eqref{eq:trtzeqs} are
\begin{align}
\tilde{t}_r(\xi_r) &= \frac{\EN(r_2-r_3)}{\sqrt{(1-\EN^2)(r_1-r_3)(r_2-r_4)}}\nonumber\\
&\times\Bigg\{(4+r_1+r_2+r_3+r_4)\Pi(h_r;\xi_r|k_r)\nonumber\\
&
-\frac{4}{r_{+}-r_{-}}\Bigg(\frac{r_{+}(4-a \ANG/\EN)-2a^2}{(r_2-r_{+})(r_3-r_{+})}\nonumber\\
&\quad\Pi(h_{+};\xi_r |k_r)- (+\leftrightarrow -)\Bigg) 
	+\frac{(r_1-r_3)(r_2-r_4)}{r_2-r_3}\nonumber\\
	&\quad \left(\text{E}(\xi_r |k_r)- h_r \frac{\sqrt{1-k_r\sin^2\xi_r}\sin\xi_r\cos\xi_r}{1-h_r\sin^2\xi_r}\right)\Bigg\}\ ,
\\
\tilde{t}_z(\xi_z) &= -\frac{\EN}{1-\EN^2}z_2 \text{E}(\xi_z|k_z)\ ,
\end{align}
where $r_1$, $r_2$, $r_3$, $r_4$, $z_1$, $z_2$, $r_+$, $r_-$ and $h_r$, $k_r$, 
The expressions defined in Eq.~\eqref{eq:phirphizeq} are
\begin{align}
\tilde{\phi}_r(\xi_r) &= -\frac{2a\EN(r_2-r_3)}{(r_{+}-r_{-})\sqrt{(1-\EN^2)(r_1-r_3)(r_2-r_4)}}\nonumber\\
& \times\left(\frac{2r_{+}-a \ANG/\EN}{(r_2-r_{+})(r_3-r_{+})}\Pi(h_{+};\xi_r|k_r)- (+\leftrightarrow -)
\right)\ ,\\
\tilde{\phi}_z(\xi_z) &=\frac{\ANG}{z_2} \Pi(z_1^2;\xi_z|k_z) \; .
\end{align}
where,
\begin{equation}
     h_r = \frac{r_1-r_2}{r_1-r_3}\; ,\quad  h_\pm = h_r \frac{r_3 - r_\pm}{r_2 -r_\pm}\ .
\end{equation}

The expressions defined in Eq.~\eqref{eq:q_i} are
\begin{align}
    \tilde{\Upsilon}_{\phi,r}= &\frac{a}{r_+-r_-}\left(\frac{2\EN r_+-a\ANG}{r_3-r_+}\right.\nonumber\\
    &\left.\times\left(1-\frac{r_2-r_3}{r_2-r_+}\frac{\Pi(h_+|k_r)}{\text{K}(k_r)}\right)-(+\leftrightarrow -)\right)\; ,\\
    \tilde{\Upsilon}_{\phi,z}=&\frac{\ANG}{\text{K}(k_z)}\Pi(z_1^2|k_z)\ .
\end{align} 
\begin{align}
    \tilde{\Upsilon}_{t,r}= &\ (4+a^2)\EN + \EN \Bigg(\frac{1}{2}\Bigg((4+r_1+r_2+r_3)r_3-\nonumber
    \\ 
    & r_1r_2+(r_1-r_3)(r_2-r_4)\frac{\text{E}(k_r)}{\text{K}(k_r)} + (4+r_1\nonumber\\
    &+r_2+r_3+r_4)(r_2-r_3)\frac{\Pi(h_r|k_r)}{\text{K}(k_r)}\Bigg)+\nonumber\\
    &\frac{2}{r_+ - r_-}\Bigg(\frac{(4-a \ANG/\EN)r_+-2a^2}{r_3-r_+}\times \nonumber\\
    &\Bigg( 1-\frac{r_2-r_3}{r_2-r_+}\frac{\Pi(h_+|k_r)}{\text{K}(k_r)} \Bigg) - (+\leftrightarrow -)\Bigg)\Bigg),\\
    \tilde{\Upsilon}_{t,z}=& - a^2 \EN + \frac{\EN \CAR}{(1-\EN^2)z_1^2}\left(1- \frac{\text{E}(k_z)}{\text{K}(k_r)}\right)\ .
\end{align} 
Here $\mathrm{E}$ and $\Pi$ are the elliptic functions defined in Appendix~\ref{app:ellipticintegrals}.

%
\section{Numerical implementation of the code}\label{app:codeimplementation}
%
We describe the mathematical techniques 
implemented in \texttt{STORM} \cite{STORM} to solve the angular 
and radial Teukolsky equations.

\subsection{Spin-Weighted Spheroidal Harmonics}
The computation of the spin-weighted spheroidal harmonics 
\[
\mathcal{S}_{\ell m}(\omega,\theta,\phi) = S_{\ell m}(\omega,\theta)e^{im\phi}\ ,
\]
is implemented in the \texttt{SpinWeightedSpheroidalHarmonics} class. These functions satisfy the angular Teukolsky equation for a generic spin weight $s$:
\begin{align}
    \Biggl[
    &\frac{1}{\sin\theta}\frac{\mathrm{d}}{\mathrm{d}\theta}
    \left(\sin\theta\frac{\mathrm{d}}{\mathrm{d}\theta}\right)
    + \zeta^2 \cos^2\theta 
    - 2 s \zeta \cos\theta
    - \frac{(m + s \cos\theta)^2}{\sin^2\theta} \nonumber \\
    &+ E_{\ell m}(\zeta) - s^2
    \Biggr] 
    S_{\ell m}(\omega,\theta) = 0 \, ,
    \label{eq:spheroid}
\end{align}
where $\zeta = a \omega$ is the oblateness parameter. 
In our implementation, the continued-fraction procedure determines the angular separation constant $\lambda_{\ell m}$, which is related to the quantity $E_{\ell m}(\zeta)$ appearing in Eq.~\eqref{eq:spheroid} through
\begin{equation}\label{eq:lambdaseparationconstant}
    E_{\ell m}(\zeta)
    = \lambda_{\ell m}
    + s(s+1) - \zeta^2 + 2 m \zeta \, .
\end{equation}

The class therefore solves Eq.~\eqref{eq:spheroid} by computing the separation constant $\lambda_{\ell m}$ and constructing the corresponding eigenfunctions for arbitrary spin weight $s$, multipole indices $(\ell,m)$, and oblateness parameter $\zeta = a \omega$.

\subsubsection{Eigenvalue computation}

To determine the eigenvalue, the class employs a \emph{continued-fraction} approach, following Ref.~\cite{Fujita_2004}. The solution of the angular Teukolsky equation is expanded in terms of Jacobi polynomials, leading to a three-term recurrence relation for the series coefficients $a_n$:
\begin{equation}
\alpha_n a_{n+1} + \beta_n a_n + \gamma_n a_{n-1} = 0 \, ,
\end{equation}
where the coefficients $\alpha_n$, $\beta_n$, and 
$\gamma_n$ depend on $(s,\ell,m,\zeta)$ and on 
the eigenvalue ${}_sE_{\ell m}(\zeta)$ (we refer the 
reader to Ref.~\cite{Fujita_2004} for their explicit 
expressions). Defining the ratios  
\begin{equation}
R_n = \frac{a_n}{a_{n-1}}, \qquad 
L_n = \frac{a_n}{a_{n+1}} \, ,
\end{equation}
the recurrence relation can be recast as a continued-fraction condition,
\begin{equation}
\beta_n + \alpha_n R_{n+1} + \gamma_n L_{n-1} = 0 \, ,
\end{equation}
which is solved using a root-finding algorithm, yielding the eigenvalue
\begin{equation}
\lambda_{\ell m} 
= E_{\ell m}(\zeta) 
- s(s+1) + \zeta^2 - 2 m \zeta \, ,
\end{equation}
to the desired precision. 

In the limits of low and high oblateness, i.e. $\zeta \ll 1$ and $\zeta \gg 1$, respectively, perturbative analytic expansions provide sufficiently accurate initial guesses~\cite{Fujita_2004,Casals:2018cgx}. However, these expansions lose accuracy in the intermediate regime.
In contrast, the spectral method based on spherical-harmonic expansions provides a robust initial estimate at any $\zeta$~\cite{PhysRevD.61.084004}. 
In this approach, the spin-weighted spheroidal harmonics are expanded in a truncated basis of spin-weighted spherical harmonics, 
and the angular Teukolsky equation is projected onto this basis, yielding a finite-dimensional symmetric matrix eigenvalue problem. 
The eigenvalues of this matrix are computed using the classical Jacobi rotation method for symmetric matrices 
(see, e.g., \cite{golub2013matrix}). The eigenvalue corresponding to the desired $\ell$ is then selected as an accurate initial guess~\cite{PhysRevD.61.084004}. 
Finally, a Newton–Raphson algorithm is used to refine the eigenvalue iteratively to the target precision.
 
\subsubsection{Angular function evaluation}

Once the eigenvalue $E_{\ell m}(\zeta)$ is determined, 
the spin-weighted spheroidal harmonic is constructed 
using Leaver’s method \cite{Leaver:1985ax}. In this 
approach, the recurrence relation is expressed in 
powers of $(1+\cos\theta)$ rather than Jacobi 
polynomials, so that the recurrence coefficients 
$\tilde\alpha_n$, $\tilde\beta_n$, and $\tilde\gamma_n$ differ from 
those in the continued-fraction method~\cite{Leaver:1985ax}.

The spheroidal harmonic is then given by
\begin{equation}
\begin{split}
\mathcal{S}_{\ell m}(\omega,\theta) =& \,\mathcal{N}\, e^{a \omega \cos\theta}\,(1+\cos\theta)^{k_1}(1-\cos\theta)^{k_2} \\
&\times \sum_{n=0}^{N_{\rm max}} \tilde a_n (1+\cos\theta)^n \ ,
\end{split}
\end{equation}
where $k_1 = |m-s|/2$, $k_2 = |m+s|/2$, and $\mathcal{N}$ 
is a normalization constant.

In practice, the coefficients $\tilde a_n$ are generated 
iteratively using the three-term recurrence relation.  
The normalization constant is then determined by enforcing 
the orthonormality condition in Eq.~\eqref{eq:orthorelation}, 
and the overall sign is fixed to ensure continuity with the 
spherical-harmonic limit. This class also computes the first 
and second derivatives of $\mathcal{S}_{\ell m}(\omega,\theta,\phi)$ with respect to 
$\theta$, needed for gravitational energy-flux calculations. 
This procedure yields a fully normalized, numerically stable 
representation of the spin-weighted spheroidal harmonics, 
suitable for both scalar and gravitational perturbations.

\subsection{Homogeneous Radial Function}
The class \texttt{RadialHomogeneousSolution} computes the two linearly independent homogeneous solutions of the radial Teukolsky equation, which are required to construct the corresponding Green-function solution and to evaluate the fluxes at the horizon and at infinity for arbitrary spin weight $s$:
\begin{equation}
    \Delta^{-s}\frac{\dd}{\dd r}\left(\Delta^{s+1}\frac{\dd R_{\ell m}}{\dd r}\right)+V R_{\ell m}=0\ ,
\end{equation}
with potential 
\begin{equation}
    V = \frac{K^2-2is (r- M)K }{\Delta} + 4is \omega r - \lambda_{\ell m}\ ,
\end{equation}
where
\begin{equation}
K = (r^2 + a^2)\omega - a m\ ,
\end{equation}
and $\lambda_{\ell m}$ is the angular separation constant defined in Eq.~\eqref{eq:lambdaseparationconstant}.

The class computes the purely ingoing (outgoing) solution $\tilde{R}^-_{\ell m}$ ($\tilde{R}^+_{\ell m}$) at the horizon (infinity) using a hybrid numerical–analytical scheme based on the Mano–Suzuki–Takasugi (MST) formalism \cite{Mano:1996vt}. 
In this work, we adopt the normalization convention defined in 
Eqs.~\eqref{eq:Rminusasymptotic} and \eqref{eq:Rplusasymptotic}, 
such that $\tilde{R}^-_{\ell m}$ is purely ingoing at the horizon 
with unit amplitude, while $\tilde{R}^+_{\ell m}$ is purely outgoing 
at infinity with unit amplitude. 

With this choice, the normalized solutions $\tilde{R}^\pm_{\ell m}$ 
are related to the standard MST solutions $R^{\rm in}_{\ell m}$ and 
$R^{\rm up}_{\ell m}$ via~\cite{Sasaki:2003xr}
\begin{equation}
\tilde{R}^+ = \frac{R^\nu_{\text{up}}}{C^{\rm trans}}\ , 
\qquad 
\tilde{R}^- = \frac{R^\nu_{\text{in}}}{B^{\rm trans}} \ .
\end{equation}
where $C^{\rm trans}$ and $B^{\rm trans}$ are normalization constants and $\nu$ is the so-called \emph{renormalized angular momentum}~\cite{Mano:1996vt}, i.e., the extra parameter introduced in the MST method to allow for a convergent series representation of the functions $R^{\rm in,up}$.
Here and thereafter, we suppress the $(\ell,m)$ indices when no ambiguity arises.

The class also provides a routine to compute the Wronskian defined in Eq.~\eqref{eq:wronskian} via
\begin{equation}
    A^{\rm in} = \frac{B^{\rm inc}}{B^{\rm trans}}\ ,
\end{equation}    
where $B^{\rm inc}$ is the incident amplitude at spatial infinity of $R^{\rm in}$~\cite{Sasaki:2003xr}.

The class also provides routines to compute the first and second
radial derivatives of $R^\pm$. These derivatives are obtained
analytically from the MST series representations, using the known
derivative relations of the hypergeometric and confluent
hypergeometric functions. They are required for the evaluation of
gravitational fluxes at the horizon and at infinity, ensuring that the fluxes are computed with the same high accuracy as the radial functions themselves.

\subsubsection{MST method}
The MST method provides a series representation of the amplitudes $B^{\rm{trans}}$, $C^{\rm{trans}}$ and $B^{\rm inc}$, i.e.~\cite{Sasaki:2003xr},
\begin{eqnarray}
&&B^{\rm trans}=\left(\frac{\epsilon\kappa}{\omega}\right)^{2s}
e^{i\kappa\epsilon_{+}(1 + \frac{2\ln\kappa}{1 + \kappa})}\sum_{n=-\infty}^{\infty}f_{n}^{\nu}\,,\nonumber
\\
&&C^{\rm trans}=\omega^{-1-2s}e^{-i(\epsilon\ln\epsilon -\frac{1-\kappa}{2}\epsilon)}A_{-}^\nu\ ,\\
&&B^{\rm inc}
=\omega^{-1}\left[{K}_{\nu}-
ie^{-i\pi\nu} \frac{\sin \pi(\nu-s+i\epsilon)}
{\sin \pi(\nu+s-i\epsilon)}
{K}_{-\nu-1}\right]\nonumber\\
&&\quad \quad \quad \times A_{+}^{\nu} e^{-i(\epsilon\ln\epsilon -\frac{1-\kappa}{2}\epsilon)}\ ,\nonumber
\end{eqnarray}
and the solutions $R^\nu_\text{up}$ and $R^\nu_\text{in}$ as
\begin{eqnarray}
&&R_{\text{up}}^{\nu}= 2^{\nu}e^{-\pi \epsilon}e^{-i\pi(\nu+1+s)}
e^{i\hat{z}}\hat{z}^{\nu+i\epsilon_+}(\hat{z}-\epsilon\kappa)^{-s-i\epsilon_+}
\nonumber\\
&&\times\sum_{n=-\infty}^{\infty}i^n
\frac{(\nu+1+s-i\epsilon)_n}{(\nu+1-s+i\epsilon)_n}
f_n^{\nu}(2\hat{z})^n \nonumber\\
&&\times
\Psi(n+\nu+1+s-i\epsilon,2n+2\nu+2;-2i\hat{z})\ ,
\end{eqnarray}
and
\begin{equation}
R_{\text{in}}^\nu = R_0^{\nu} + R_0^{-\nu - 1} \ ,
\end{equation}
with
\begin{eqnarray}
&&R_0^{\nu}=e^{i\epsilon \kappa x}(-x)^{-s-{\frac{i}{2}}(\epsilon+\tau)}
(1-x)^{{\frac{i}{2}}(\epsilon+\tau)+\nu}
\nonumber\\
&&
\times\sum_{n=-\infty}
^{\infty}f_n^{\nu}\,\frac{\Gamma(1-s-i\epsilon-i\tau)\Gamma(2n+2\nu+1)
}{{\Gamma(n+\nu+1-i\tau)\Gamma(n+\nu+1-s-i\epsilon)}}
\nonumber\\
&&\times
F\left(-n-\nu-i\tau,-n-\nu-s-i\epsilon;-2n-2\nu;{\frac{1}{1-x}}\right)\nonumber\\
&&\times (1-x)^{n}\,.
\end{eqnarray}
Here $\Psi$ is the irregular confluent hypergeometric function, $F$ is regular hypergeometric function and
\begin{eqnarray}
&&x=\frac{{\omega(r_+-r)}}{\epsilon\kappa}\ ,\quad 
\epsilon = 2 M \omega\ ,\nonumber\\
&&\kappa = \sqrt{1 - \left({\frac{a}{M}}\right)^2},\quad \tau = \frac{\epsilon - m (a/M)}{\kappa}\nonumber\ ,\\
&& \hat{z} = \omega(r-r_-),\quad \epsilon_+ = (\epsilon + \tau)/2\ ,\\
&A_{-}^\nu&=2^{-1-s+i\epsilon}e^{-{\frac{\pi}{2}}i(\nu+1+s)}e^{-{\frac{\pi}{2}}\epsilon}\nonumber\\
&&\sum_{n=-\infty}^{\infty}(-1)^n\frac{(\nu+1+s-i\epsilon)_n}{
(\nu+1-s+i\epsilon)_n}f_n^\nu\ ,\\
&A_{+}^\nu&=e^{-{\frac{\pi}{2}}\epsilon}e^{{\frac{\pi}{2}}i(\nu+1-s)}
2^{-1+s-i\epsilon}\nonumber\\
&&{\frac{\Gamma(\nu+1-s+i\epsilon)}{
\Gamma(\nu+1+s-i\epsilon)}}\sum_{n=-\infty}^{\infty}f_n^\nu,\\
&{K}_{\nu}=&
\frac{e^{i\epsilon\kappa}(2\epsilon \kappa )^{s-\nu-r}2^{-s}i^{r}
\Gamma(1-s-2i\epsilon_+)}
{\Gamma(r+\nu+1-s+i\epsilon)
\Gamma(r+\nu+1+i\tau)}
\nonumber\\
&&\times\frac{\Gamma(r+2\nu+2)}{\Gamma(r+\nu+1+s+i\epsilon)}\nonumber
\\
&&\times \left ( \sum_{n=r}^{\infty}
(-1)^n\, \frac{\Gamma(n+r+2\nu+1)}{(n-r)!}
\right.\nonumber \\
&&\left.\times\frac{\Gamma(n+\nu+1+s+i\epsilon)}{\Gamma(n+\nu+1-s-i\epsilon)} \frac{\Gamma(n+\nu+1+i\tau)}{\Gamma(n+\nu+1-i\tau)}
\,f_n^{\nu}\right)
\nonumber\\
&&\times \left(\sum_{n=-\infty}^{\infty}{r}
\frac{(-1)^n}{(r-n)!
(r+2\nu+2)_n}\right.\nonumber\\
&&\left.\times \frac{(\nu+1+s-i\epsilon)_n}{(\nu+1-s+i\epsilon)_n}
f_n^{\nu}\right)^{-1}\ .
\label{eq:Knu}
\end{eqnarray}

Finally, the coefficients $f_n^\nu$ are determined by the three-term recurrence relation
\begin{equation}
\label{eq:fn_eq}
\alpha_n^{\nu} f^\nu_{n+1} + \beta_n^{\nu} f_n^\nu + \gamma_n^{\nu} f^\nu_{n-1} = 0\ ,
\end{equation}
where the coefficients $\alpha_n^{\nu}$, $\beta_n^{\nu}$, and $\gamma_n^{\nu}$ depend on the mode parameters $(\ell, m, \omega, s)$ and are given explicitly in Ref.~\cite{Sasaki:2003xr}.

The physically relevant sequence $\{f_n^\nu\}$ corresponds to the minimal solution of the recurrence relation in both the $n \to +\infty$ and $n \to -\infty$ directions. This requirement ensures convergence of the hypergeometric and Coulomb series representations. The condition for minimality can be written as
\begin{equation}
\label{eq:nu_eq}
g_n(\nu) = \beta_n^{\nu} + \alpha_n^{\nu} R_{n+1} + \gamma_n^{\nu} L_{n-1} = 0\ ,
\end{equation}
where $R_n = f_n/f_{n-1}$ and $L_n = f_n/f_{n+1}$ are expressed as continued fractions derived from the recurrence relation. Since the choice of $n$ is arbitrary, we set $n=0$ in practice.
The continued fractions entering $R_n$ and $L_n$ are evaluated using 
the continued-fraction routines provided by the \texttt{Boost C++ library}.

The class \texttt{RadialHomogeneousSolution} provides two strategies for computing $\nu$:
\begin{enumerate}
\item A Newton–Raphson iteration in the complex plane 
applied to the function $g_n(\nu)$ defined in Eq.~\eqref{eq:nu_eq}.
\item The \emph{monodromy method} \cite{Nasipak_2025}, 
which expresses $\nu$ in terms of the monodromy data of 
the Teukolsky radial solution.
\end{enumerate}
The Newton–Raphson approach is generally more efficient 
at low frequencies, whereas the monodromy method is required 
at high frequencies. In both cases, the iterative procedures 
are controlled by user-defined tolerance parameters, 
allowing the desired numerical accuracy to be specified at runtime. 
This ensures convergence of the continued-fraction solution 
and gives the user full control over the trade-off between 
accuracy and computational cost.

After determining the renormalized angular momentum, the coefficients $f_n^\nu$ can be computed.

From a technical standpoint, all infinite sums over $n$ appearing in the
MST series representations are evaluated using a symmetric truncation
strategy about $n=0$. At each iteration, the contributions from the
positive and negative indices $+n$ and $-n$ are computed simultaneously
and added to the partial sum. 

The summation is terminated once the relative magnitude of the newly
added terms, measured with respect to the accumulated sum, falls below
a prescribed numerical tolerance for a fixed number of consecutive
iterations. Requiring several consecutive sub-threshold contributions
prevents premature termination due to accidental cancellations or
oscillatory behavior of the series. 

This procedure ensures controlled convergence and uniform accuracy
across the parameter space, providing analytically accurate boundary
data that are subsequently used in the hybrid numerical–analytical
integration scheme described below.

\subsubsection{Hybrid Numerical–Analytical Scheme}
Although the MST series provides an exact analytic representation, 
its direct evaluation at every radial point is computationally expensive. 
To improve efficiency, the class implements a hybrid numerical–analytical 
scheme following Ref.~\cite{Zengino_lu_2011}: the MST formalism is used 
only to set accurate boundary conditions near the horizon and at infinity, 
while the radial Teukolsky equation is integrated numerically in the 
interior region.

To ensure that this numerical integration remains 
stable and efficient across the full domain between 
the horizon and infinity, it is convenient to 
introduce a new radial function $\varphi$, related to 
the Teukolsky solution $R$ by
\begin{equation}\label{eq:psitransnum}
\varphi = r\Delta^{-s} e^{-i m \tilde{\phi}} e^{-i \omega h} R\ ,
\end{equation}
and to redefine the angular variable $\phi$ via
\begin{equation}
  \dd\tilde{\phi} =  \dd\phi + \frac{a}{r^2+a^2} \,\dd r_\star\ .
\end{equation}
These transformations cast the Teukolsky equation 
into a form with a short-ranged effective potential:
\begin{equation}\label{eq:shortrangepoteq}
\frac{(r^2+a^2)^2}{r^4} \frac{\dd^2\varphi}{\dd r_\star^2} - \frac{2 \widetilde{G} }{r^5}\,\frac{\dd\varphi}{\dd r_\star}  + \frac{\tilde{U}}{r^6} \varphi = 0\ ,
\end{equation}
where
\begin{equation}
\widetilde{G} = a^2 \Delta + (r^2+a^2) \left\{r s(r-M)- i r \big[(r^2+a^2)\omega H + m a\big]\right\}\ ,
\end{equation}
and
\begin{eqnarray}
\tilde{U} &=& 2 i s\omega r^2 \left(  r \Delta (1-H) - M(r^2-a^2)(1+H) \right) \nonumber\\ 
&& - 2 i a r \Delta(m+a \omega H) + \Delta \left(2a^2 - r^2\lambda - 2Mr(s+1) \right) \nonumber\\ 
&& - 2 m a \omega r^2(r^2+a^2)(1+H)\nonumber\\
&& + r^2 (r^2+a^2)^2 \left( \omega^2(1-H^2) + i \omega H' \right)\ .
\end{eqnarray}
Here $H = \dd h / \dd r_\star$ satisfies
\begin{equation}
    |H|\leq 1\ ,\qquad \lim_{r_\star\rightarrow\pm \infty}  H = \pm 1\ ,\qquad \lim_{r_\star\rightarrow\pm \infty} \frac{\dd H}{\dd r_\star}= 0\ .
\end{equation}

In practice, the numerical integration is performed with user-controlled tolerance parameters that determine the step size and convergence criteria. 
This allows the user to balance numerical accuracy and computational efficiency, ensuring that the solution remains stable across the full radial domain while retaining high precision near the boundaries set by the MST series.

\section{Fluxes} \label{app:scalar_spectra}

In this Appendix, we present the scalar spectra 
analyzed in Sec.~\ref{sec:resultsscalarspectra}.

\begin{figure*}[p]
    \centering
    \includegraphics[width=\textwidth]{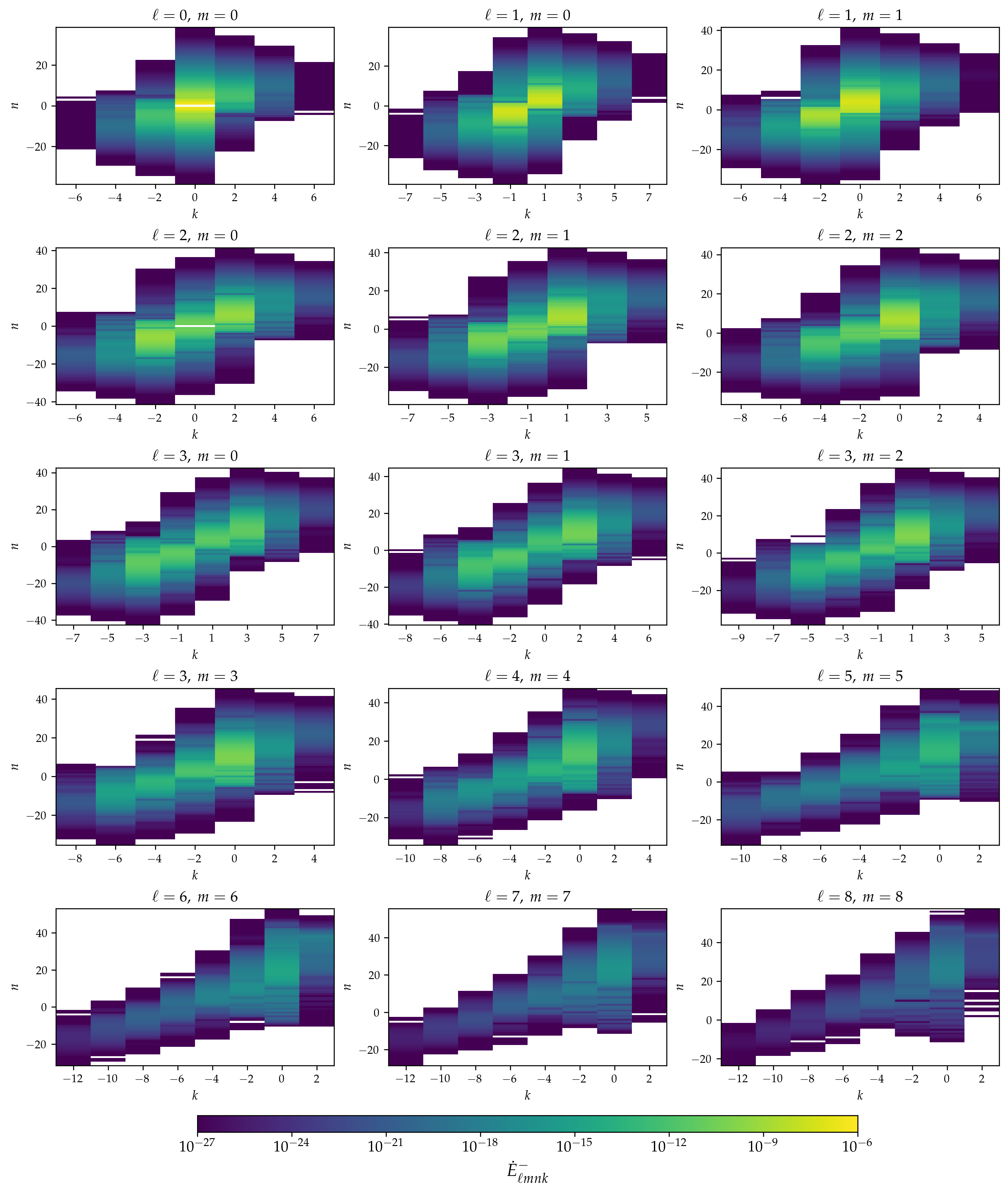}
    \caption{Multipolar $(n,k)$ components of the scalar spectra, 
    $\dot{E}^{-}_{\ell m n k}$, at the horizon for varying $\ell$ and $m$, with fixed $a/M = 0.3$, $p/M = 7$, $e = 0.5$, $\theta_{\rm inc} = 30^\circ$. 
    The color scale indicates the magnitude of the single-mode flux.}
    \label{fig:flux_grid_fixed_apetheta_lm_horizon_Edot}
\end{figure*}

\begin{figure*}[p]
    \centering
    \includegraphics[width=\textwidth]{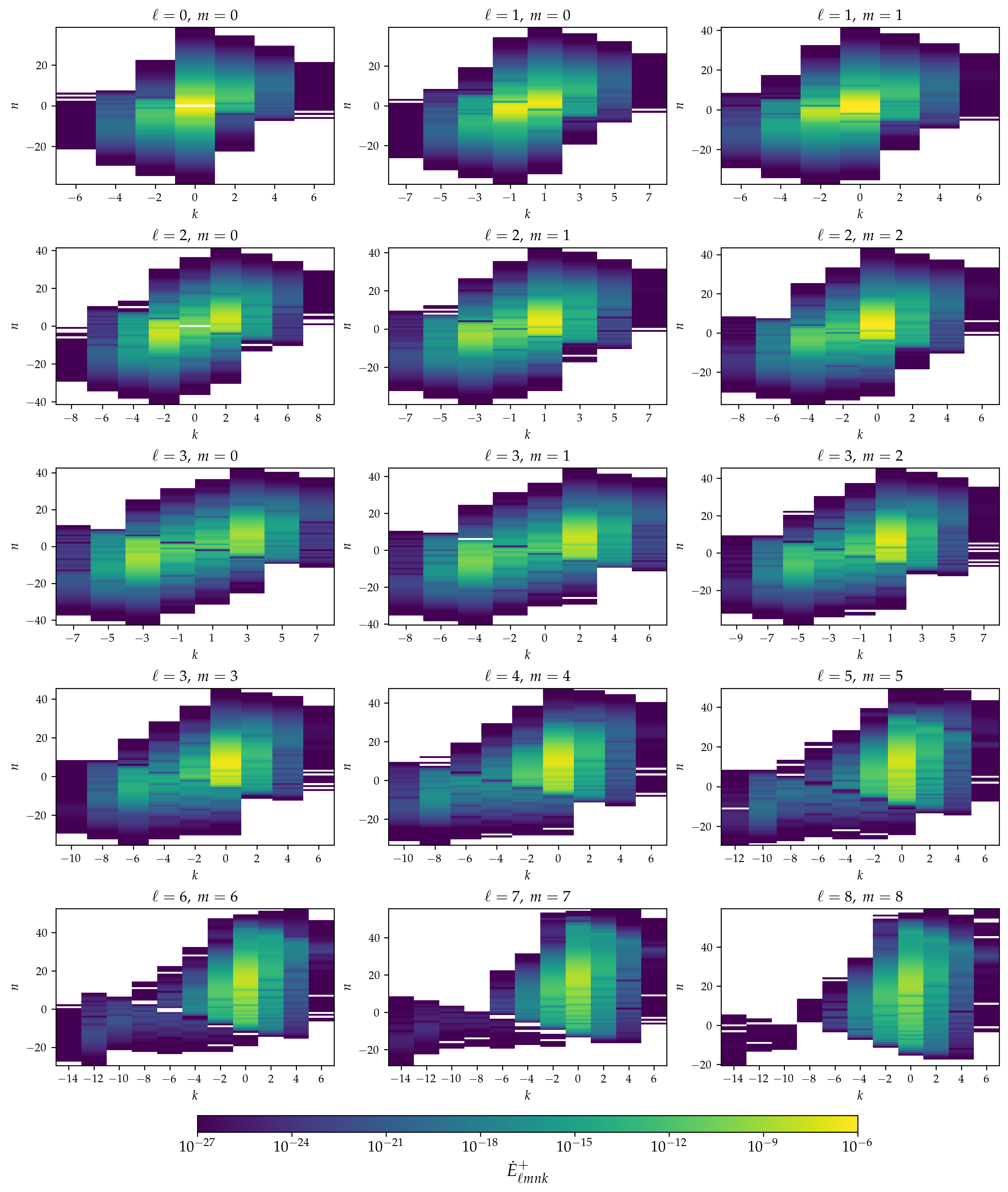}
    \caption{Multipolar $(n,k)$ components of the scalar spectra, $\dot{E}^{+}_{\ell m n k}$, at infinity for varying $\ell$ and $m$, with fixed $a/M = 0.3$, $p/M = 7$, $e = 0.5$, $\theta_{\rm inc} = 30^\circ$. 
    The color scale indicates the magnitude of the single-mode flux.}
    \label{fig:flux_grid_fixed_apetheta_lm_infinity_Edot}
\end{figure*}

\begin{figure*}[p]
    \centering
    \includegraphics[width=\textwidth]{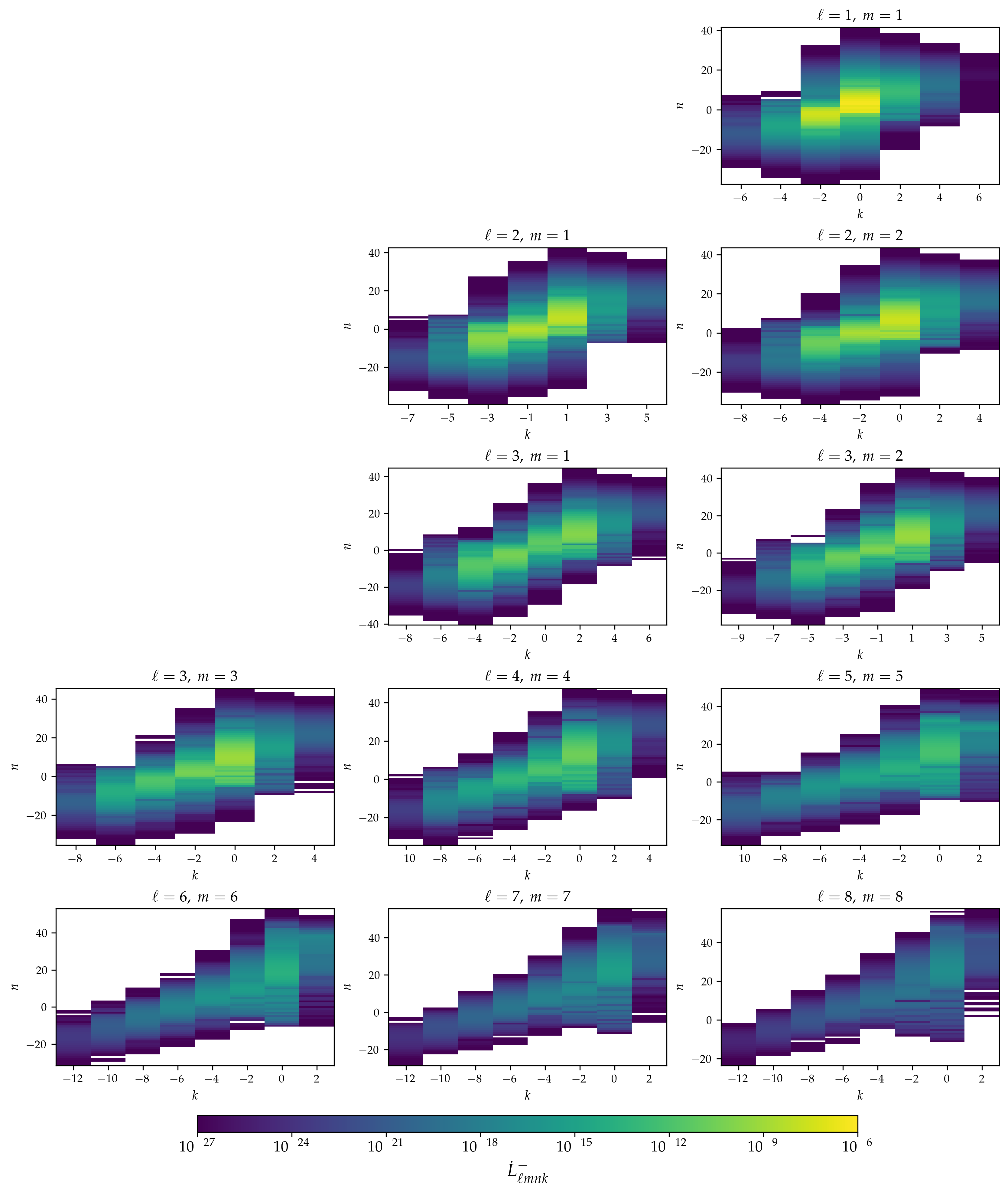}
    \caption{Multipolar $(n,k)$ components of the scalar spectra, 
    $\dot{L}^{-}_{\ell m n k}$, at the horizon for varying $\ell$ and $m$, with fixed $a/M = 0.3$, $p/M = 7$, $e = 0.5$, $\theta_{\rm inc} = 30^\circ$. The 
    color scale indicates the magnitude of the single-mode flux.}
    \label{fig:flux_grid_fixed_apetheta_lm_horizon_Ldot}
\end{figure*}

\begin{figure*}[p]
    \centering
    \includegraphics[width=\textwidth]{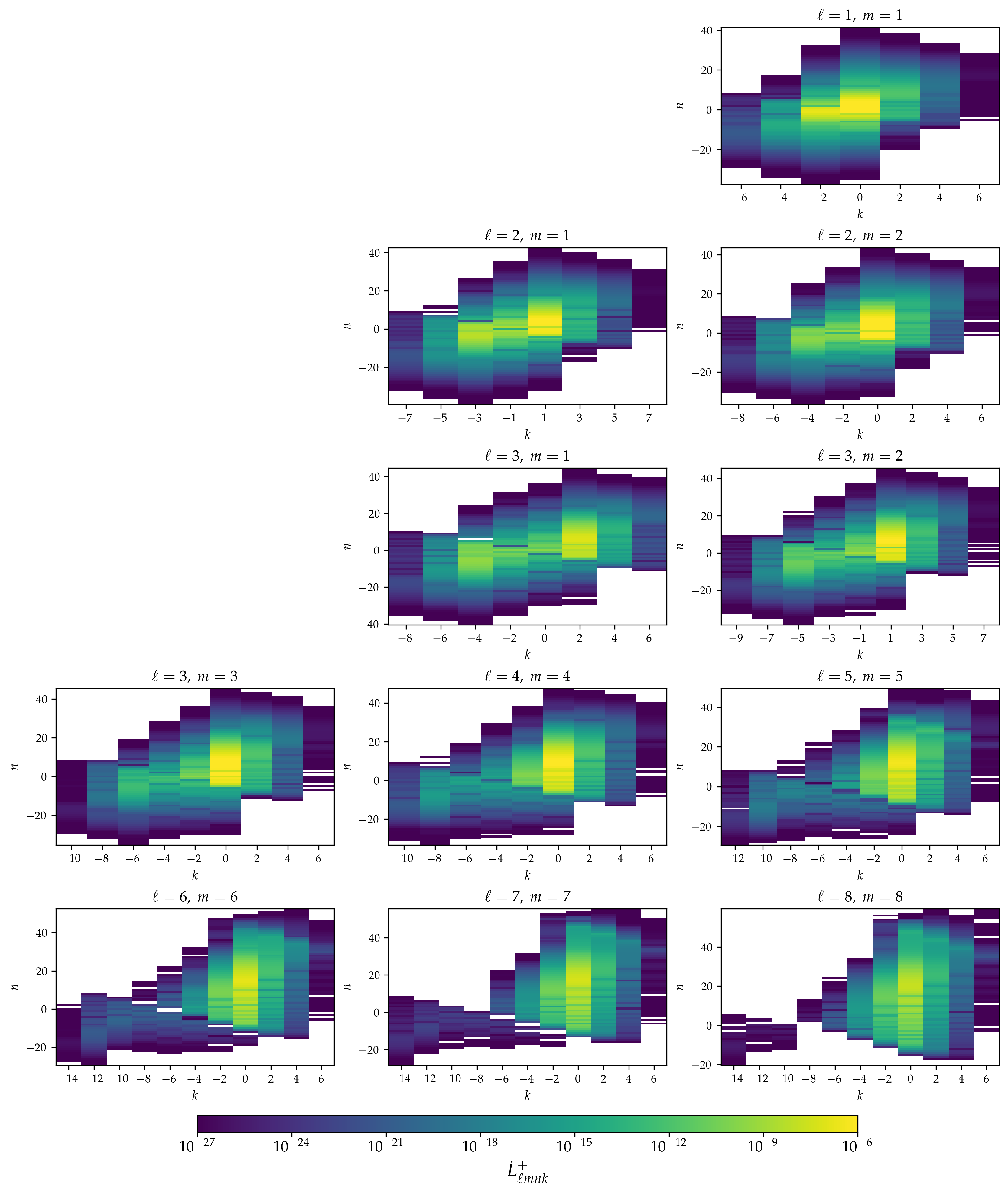}
    \caption{Multipolar $(n,k)$ components of the scalar spectra, $\dot{L}^{+}_{\ell m n k}$, at infinity for varying $\ell$ and $m$, with fixed $a/M = 0.3$, $p/M = 7$, $e = 0.5$, $\theta_{\rm inc} = 30^\circ$. The color scale indicates the magnitude of the single-mode flux.}
    \label{fig:flux_grid_fixed_apetheta_lm_infinity_Ldot}
\end{figure*}

Fig.~\ref{fig:flux_grid_fixed_apetheta_lm_horizon_Edot},~\ref{fig:flux_grid_fixed_apetheta_lm_infinity_Edot},~\ref{fig:flux_grid_fixed_apetheta_lm_horizon_Ldot} and ~\ref{fig:flux_grid_fixed_apetheta_lm_infinity_Ldot} present 
the scalar spectra at the horizon and at infinity for 
different values of $(\ell,m)$, for a fixed orbital 
configuration with parameters $(a/M, p/M, e ,\theta_{\rm inc}) = (0.3, 0.5, 7, 30^\circ)$.

\begin{figure*}[p]
    \centering
    \includegraphics[width=\textwidth]{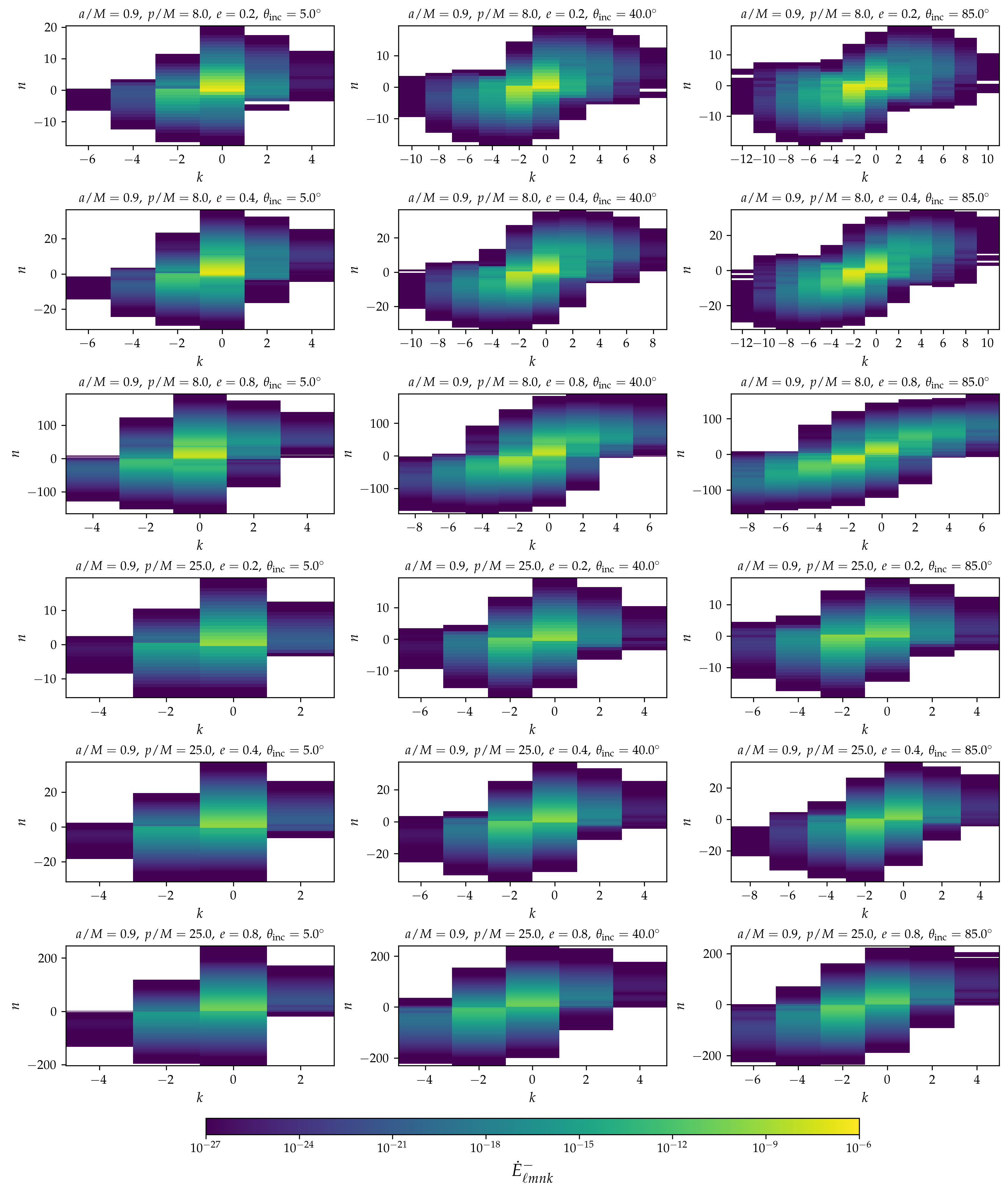}
    \caption{Multipolar $(n,k)$ components of the scalar spectra, 
    $\dot{E}^{-}_{\ell m n k}$, at the horizon for varying $p/M$, $e$ and $\theta_{\rm inc}$, with fixed $a/M = 0.9$, $\ell = m = 1$. The 
    color scale indicates the magnitude of the single-mode flux.}
    \label{fig:flux_grid_a_0.9_horizon_Edot}
\end{figure*}

\begin{figure*}[p]
    \centering
    \includegraphics[width=\textwidth]{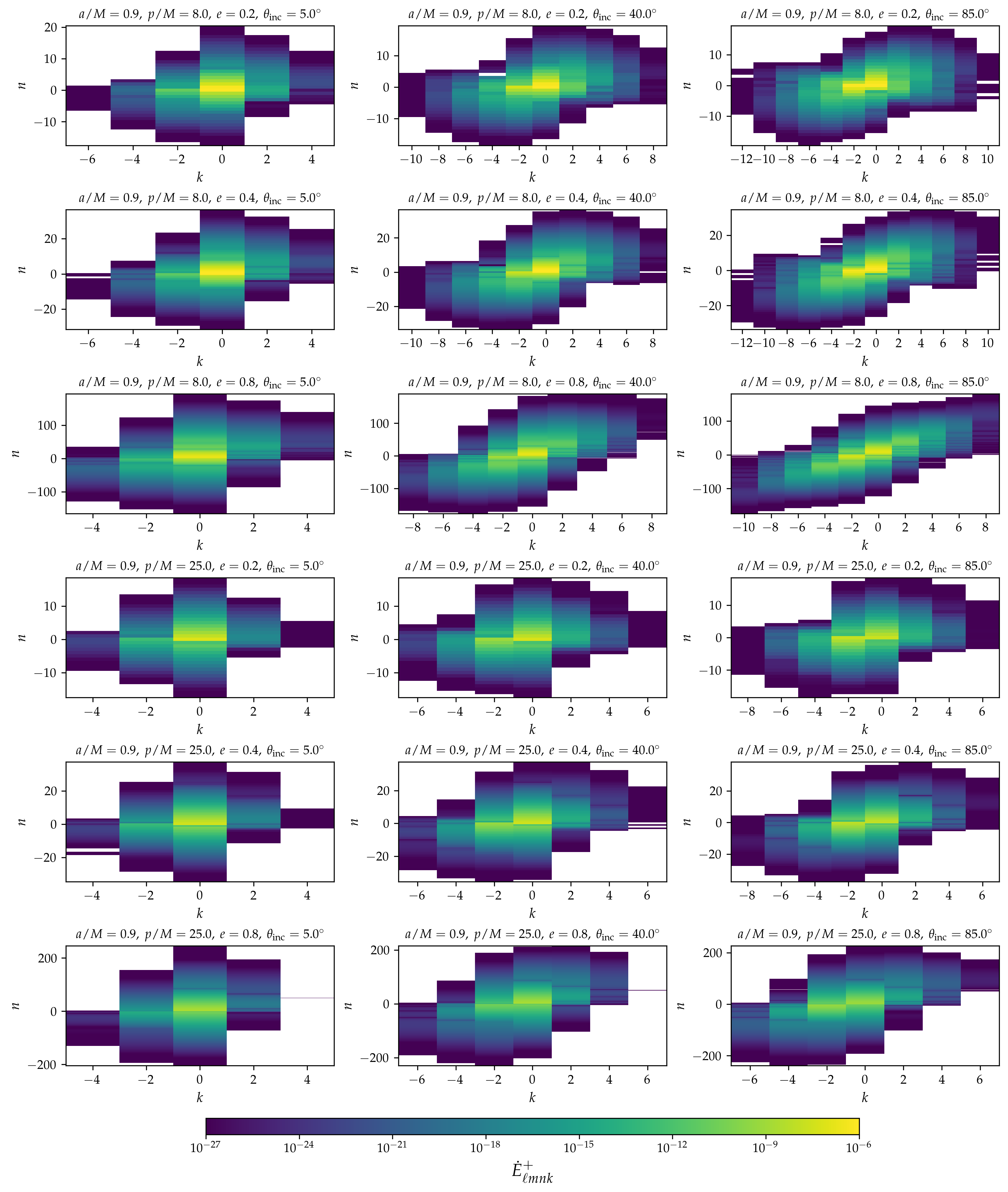}
    \caption{Multipolar $(n,k)$ components of the scalar spectra, $\dot{E}^{+}_{\ell m n k}$, at infinity for varying $p/M$, $e$ and $\theta_{\rm inc}$, with fixed $a/M = 0.9$, $\ell = m = 1$. 
    The color scale indicates the magnitude of the single-mode flux.}
    \label{fig:flux_grid_a_0.9_infinity_Edot}
\end{figure*}

\begin{figure*}[p]
    \centering
    \includegraphics[width=\textwidth]{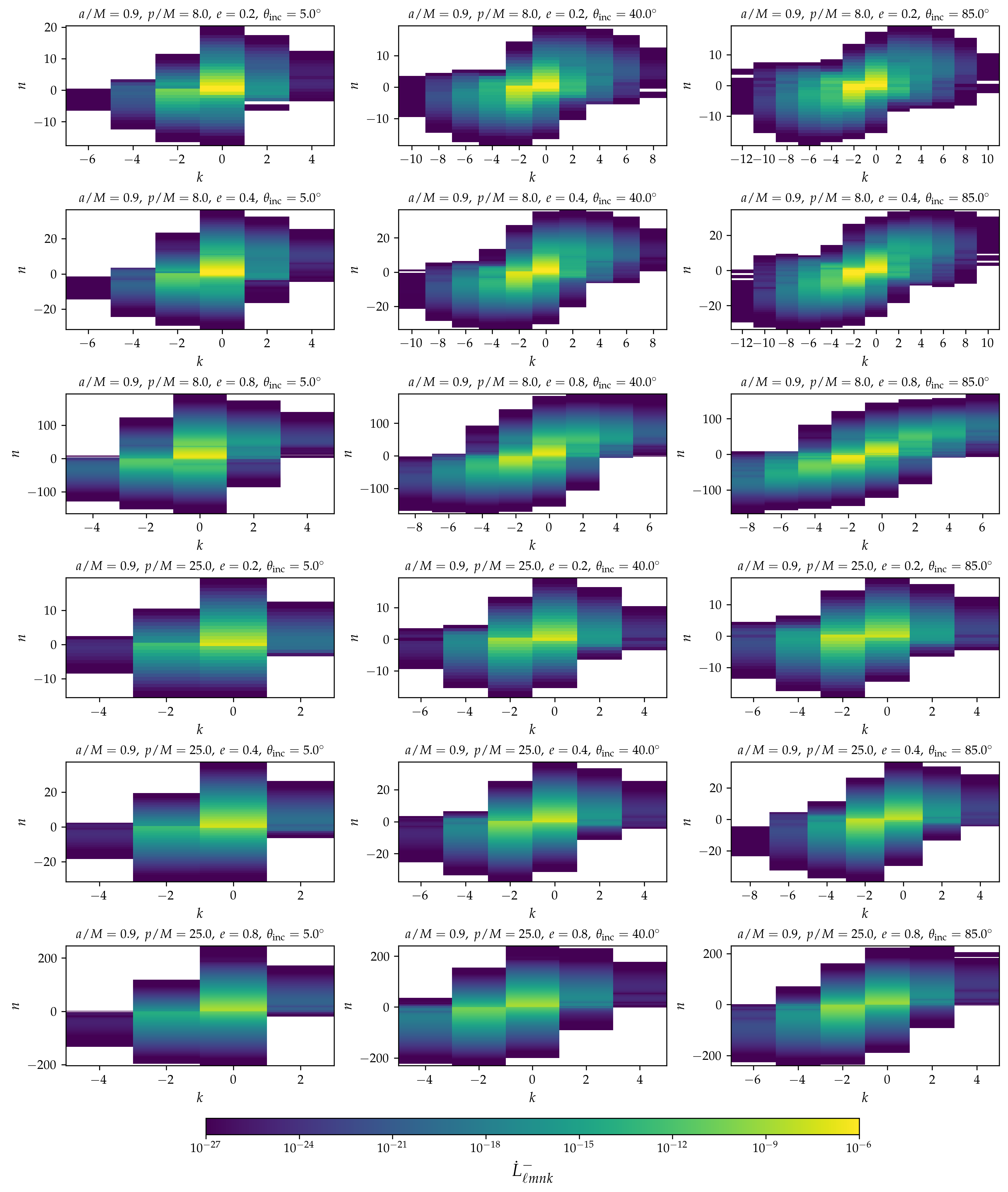}
    \caption{Multipolar $(n,k)$ components of the scalar spectra, $\dot{L}^{-}_{\ell m n k}$, at the horizon for varying $p/M$, $e$ and $\theta_{\rm inc}$, with fixed $a/M = 0.9$, $\ell = m = 1$. The 
    color scale indicates the magnitude of the single-mode flux.}
    \label{fig:flux_grid_a_0.9_horizon_Ldot}
\end{figure*}

\begin{figure*}[p]
    \centering
    \includegraphics[width=\textwidth]{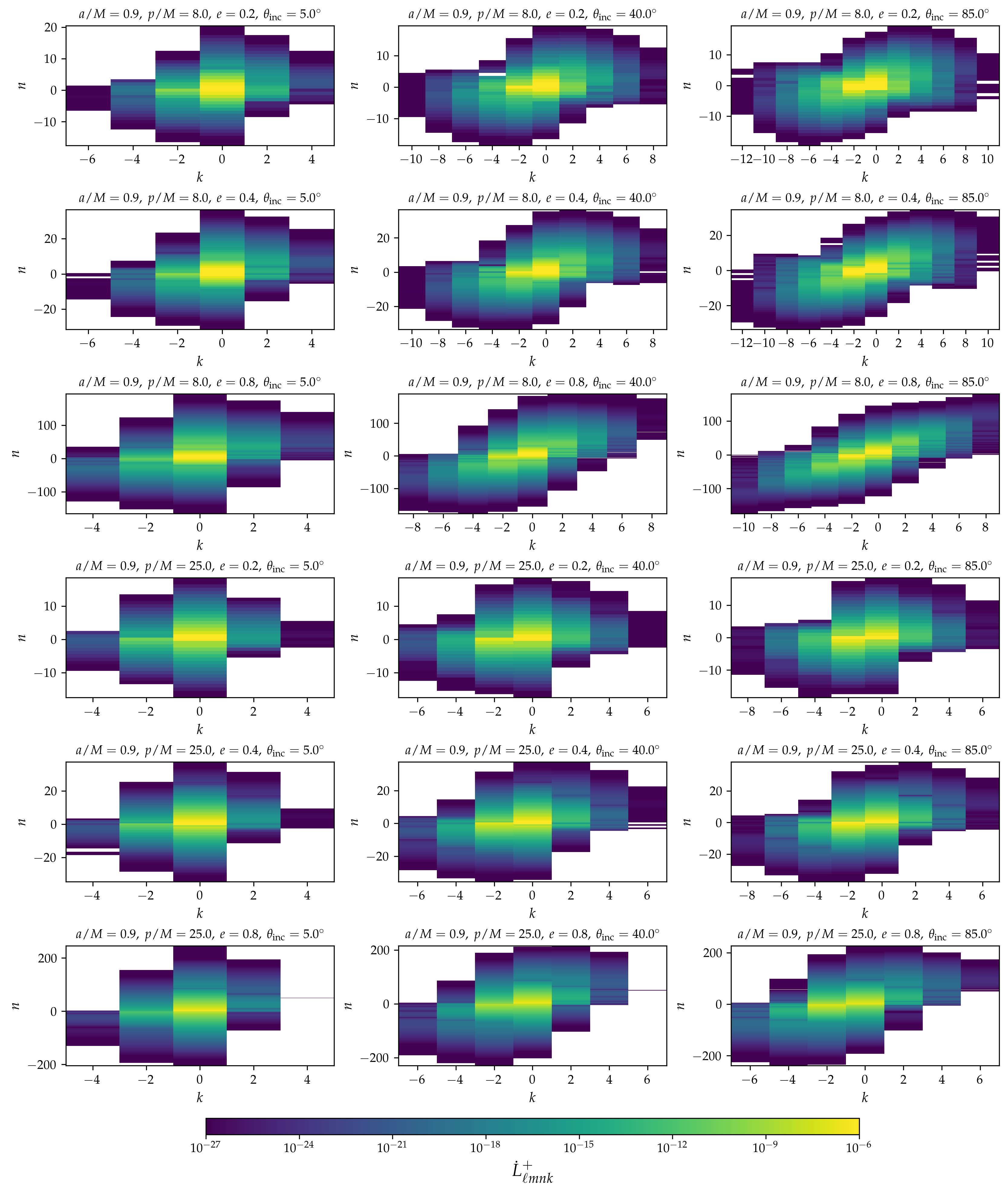}
    \caption{Multipolar $(n,k)$ components of the scalar spectra, $\dot{L}^{+}_{\ell m n k}$, at infinity for varying $p/M$, $e$ and $\theta_{\rm inc}$, with fixed $a/M = 0.9$, $\ell = m = 1$. The 
    color scale indicates the magnitude of the single-mode flux.}
    \label{fig:flux_grid_a_0.9_infinity_Ldot}
\end{figure*}

Fig.~\ref{fig:flux_grid_a_0.9_horizon_Edot},~\ref{fig:flux_grid_a_0.9_infinity_Edot},~\ref{fig:flux_grid_a_0.9_horizon_Ldot} and ~\ref{fig:flux_grid_a_0.9_infinity_Ldot} display 18 
scalar spectra for the energy and angular momentum flux at the horizon and at 
infinity, respectively, illustrating how the $(n,k)$ mode 
structure depends on $(p/M, e, \theta_{\rm inc})$ for 
fixed $(\ell,m)=(1,1)$ and black-hole spin $a/M=0.9$.
Each figure is organized into two $3\times3$ blocks: the 
upper block corresponds to $p/M=8$, while the lower 
block corresponds to $p/M=25$.
Within each block, the rows represent eccentricities $e={0.2,0.4,0.8}$, and the columns represent orbital 
inclinations $\theta_{\rm inc}={5^\circ,40^\circ,85^\circ}$.

\begin{figure*}[p]
    \centering
    \includegraphics[width=\textwidth]{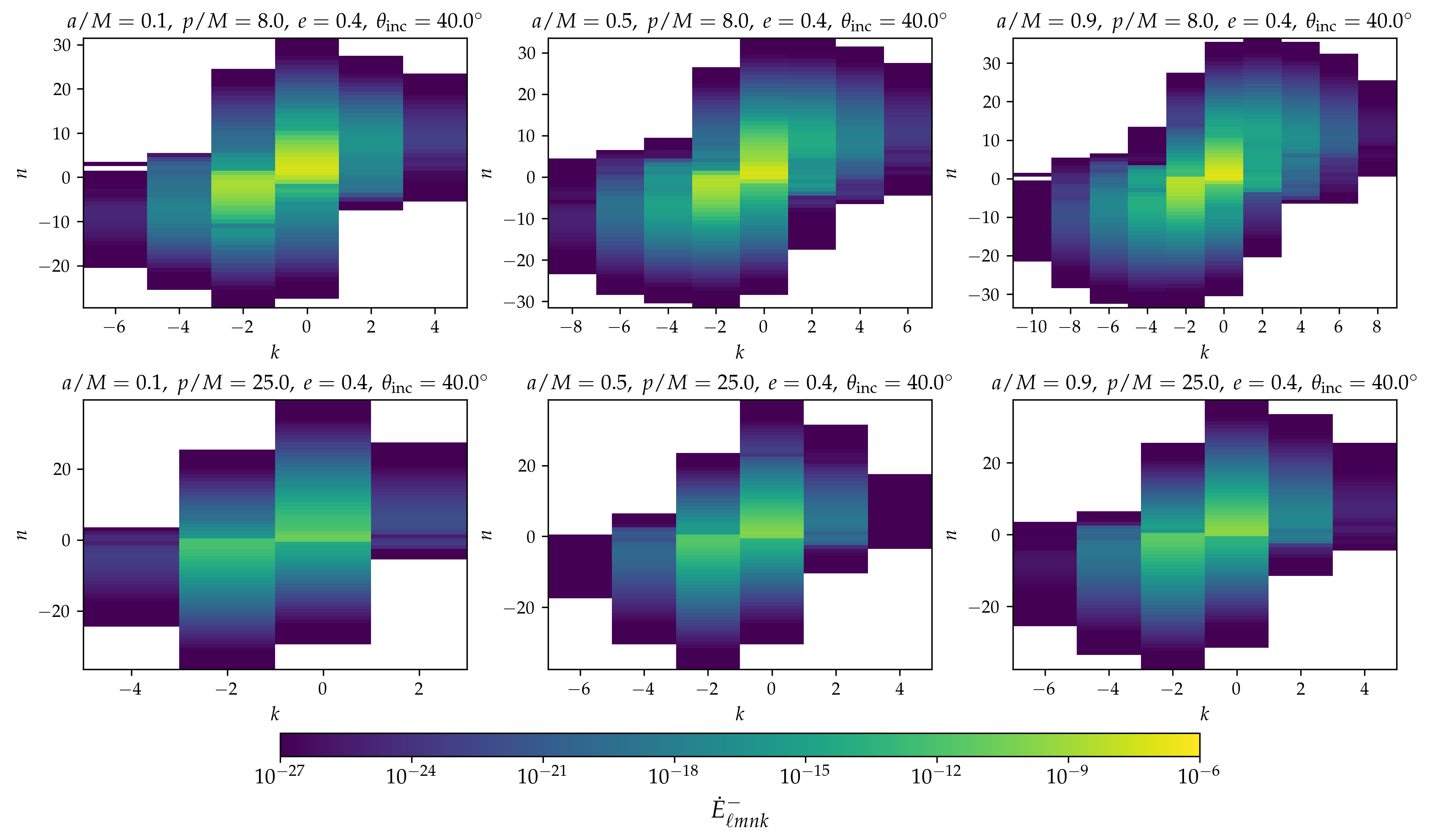}
    \caption{Multipolar $(n,k)$ components of the scalar spectra, 
    $\dot{E}^{-}_{\ell m n k}$, at the horizon for varying $a/M$ and $p/M$, with fixed $e = 0.4$, $\theta_{\rm inc} = 40^\circ$, $\ell = m = 1$. The 
    color scale indicates the magnitude of the single-mode flux.}
    \label{fig:flux_grid_e_0.4_theta_0.40_horizon_Edot}
\end{figure*}

\begin{figure*}[p]
    \centering
    \includegraphics[width=\textwidth]{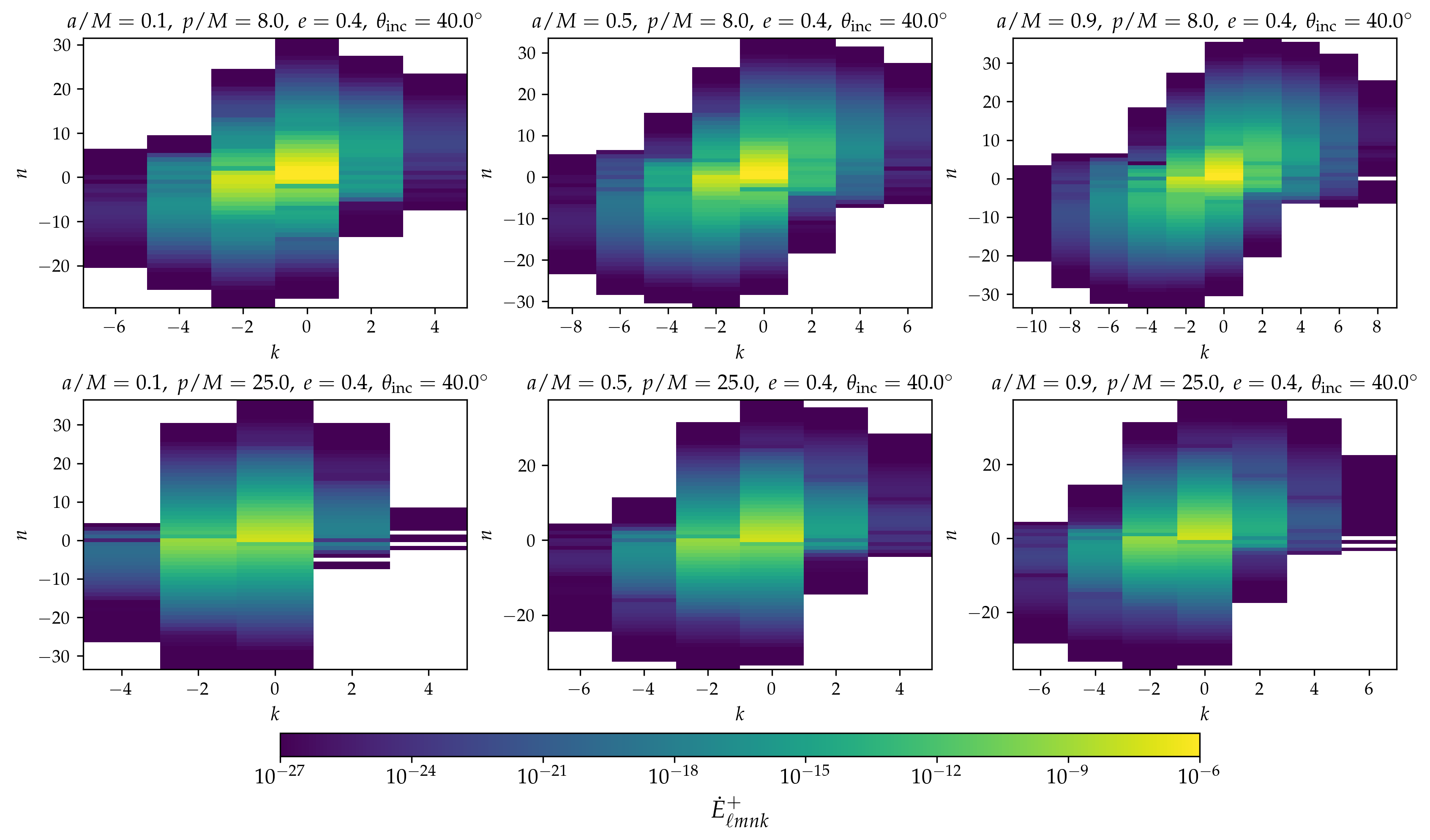}
    \caption{Multipolar $(n,k)$ components of the scalar spectra, $\dot{E}^{+}_{\ell m n k}$, at infinity for varying $a/M$ and $p/M$, with fixed $e = 0.4$, $\theta_{\rm inc} = 40^\circ$, $\ell = m = 1$. 
    The color scale indicates the magnitude of the single-mode flux.}
    \label{fig:flux_grid_e_0.4_theta_0.40_infinity_Edot}
\end{figure*}

\begin{figure*}[p]
    \centering
    \includegraphics[width=\textwidth]{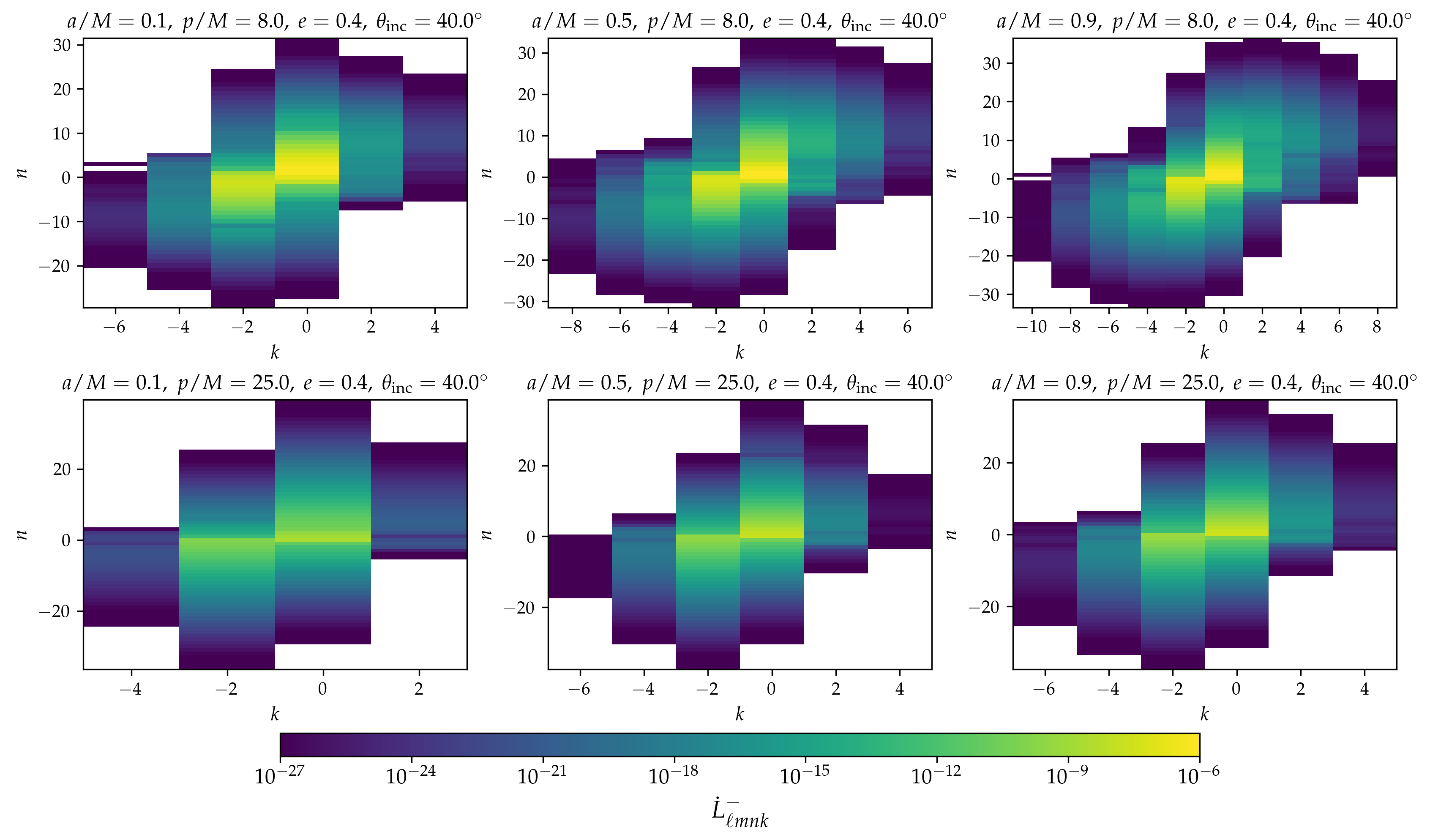}
    \caption{Multipolar $(n,k)$ components of the scalar spectra, 
    $\dot{L}^{-}_{\ell m n k}$, at the horizon for varying $a/M$ and $p/M$, with fixed $e = 0.4$, $\theta_{\rm inc} = 40^\circ$, $\ell = m = 1$. The 
    color scale indicates the magnitude of the single-mode flux.}
    \label{fig:flux_grid_e_0.4_theta_0.40_horizon_Ldot}
\end{figure*}

\begin{figure*}[p]
    \centering
    \includegraphics[width=\textwidth]{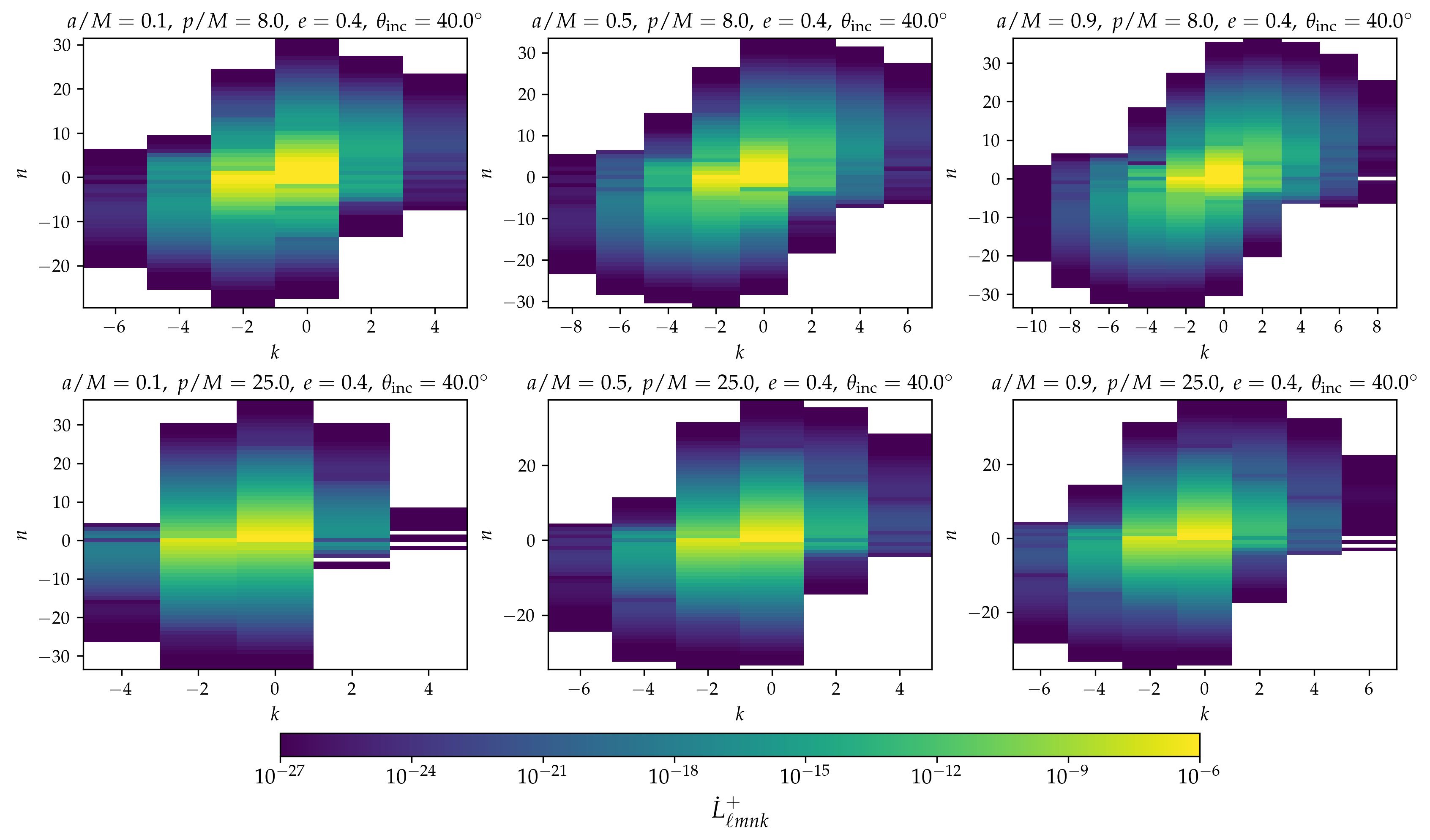}
    \caption{Multipolar $(n,k)$ components of the scalar spectra, $\dot{L}^{+}_{\ell m n k}$, at infinity for varying $a/M$ and $p/M$, with fixed $e = 0.4$, $\theta_{\rm inc} = 40^\circ$, $\ell = m = 1$. The 
    color scale indicates the magnitude of the single-mode flux.}
    \label{fig:flux_grid_e_0.4_theta_0.40_infinity_Ldot}
\end{figure*}

Finally, Fig.~\ref{fig:flux_grid_e_0.4_theta_0.40_horizon_Edot},~\ref{fig:flux_grid_e_0.4_theta_0.40_infinity_Edot},~\ref{fig:flux_grid_e_0.4_theta_0.40_horizon_Ldot} and ~\ref{fig:flux_grid_e_0.4_theta_0.40_infinity_Ldot} 
illustrate the dependence of the scalar spectra on the 
spin $a$ of the primary BH.
In these figures, we fix $(e,\theta_{\rm inc})=(0.4,40^\circ)$ 
and consider two orbital radii, $p/M={8,25}$.
The rows correspond to the two values of $p/M$, while the 
three columns show results for spins $a/M={0.1,0.5,0.9}$.

\end{document}